\documentclass[twoside]{style/cernyrep}

\usepackage[utf8]{inputenc}
\DeclareUnicodeCharacter{2212}{-}
\counterwithin{table}{section}
\counterwithin{figure}{section}
\counterwithin{equation}{section}

\usepackage{comment} 
\usepackage{amsmath}
\usepackage{color}

\usepackage{graphicx}
\usepackage{caption}
\usepackage{subfig}
\usepackage{subfloat}
\usepackage{multirow}
\usepackage{booktabs}
\usepackage{adjustbox}
\usepackage{rotating}
\usepackage{siunitx}
\DeclareSIUnit\clight{\text{c}}
\DeclareSIUnit\barn{b}

\usepackage[backend=biber,sorting=none]{biblatex}

\usepackage{pdfpages}

\usepackage{epstopdf}
\usepackage{acronym}
\usepackage{balance}
\usepackage{authblk}
\usepackage{fancyhdr}
\usepackage{afterpage}
\usepackage{csvsimple}
\usepackage{pgfsys}

\usepackage{todonotes}

\usepackage{tcolorbox}

\usepackage{emptypage}
\usepackage{lipsum}

\usepackage{texnames}
\usepackage{ctable}
\usepackage[T1]{fontenc}

\usepackage{blindtext}

\usepackage[bookmarks, colorlinks=true, linktoc=page, pdftex, linkcolor=blue, citecolor=blue, urlcolor=blue]{hyperref}
\usepackage{float}
\usepackage{xcolor}
\usepackage[toc,page]{appendix}
\usepackage[bookmarks, colorlinks=true, pdftex, linkcolor=blue, citecolor=blue, urlcolor=blue]{hyperref}

\usepackage{times,lipsum}
\usepackage[margin=1in]{geometry}
\usepackage[onehalfspacing]{setspace}

\usepackage{arydshln} 
\let\cleardoublepage=\clearpage

\usepackage[hang,flushmargin]{footmisc}
\fancypagestyle{plain}{}
\newlength{\oddmarginwidth}
\newlength{\evenmarginwidth}
\fancyhfoffset[LO,RE]{\oddmarginwidth}
\fancyhfoffset[LE,RO]{\evenmarginwidth}
\bibliography{main} 
\begin{document}
\pagenumbering{roman}
\setcounter{page}{1}
\thispagestyle{empty}
\setlength{\unitlength}{1mm}

\includepdf[pages={1,2}]{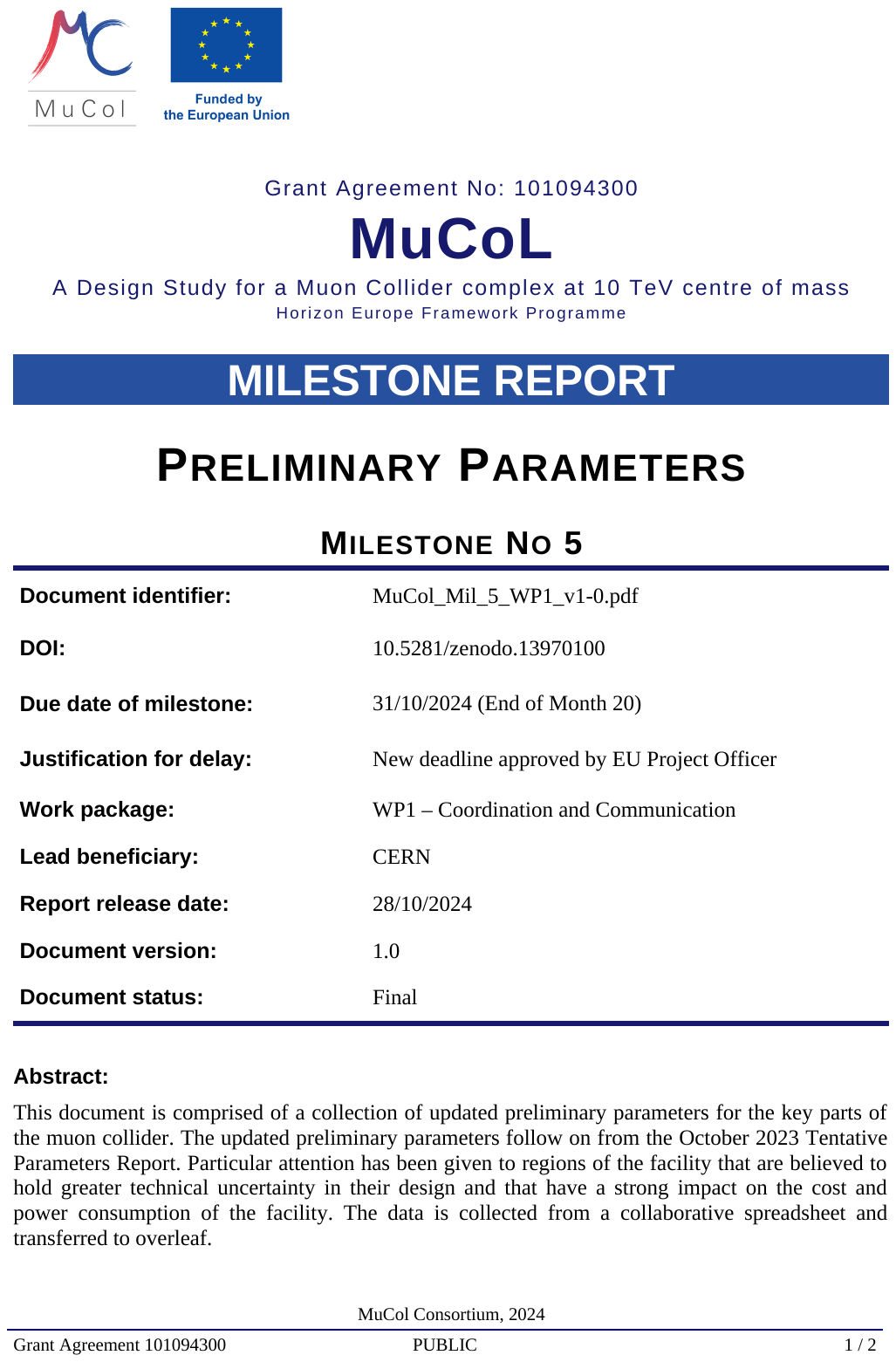}

\newpage

\vskip 10cm
\begin{center}
{\huge\bfseries
IMCC \& MuCol Authors }

\end{center}

\vspace{5mm}
{\small
\noindent
Carlotta~Accettura$^{1}$,
Simon~Adrian$^{2}$,
Rohit~Agarwal$^{3}$,
Claudia~Ahdida$^{1}$,
Chiara~Aim\'e$^{4}$,
Avni~Aksoy$^{1, 5}$,
Gian~Luigi~Alberghi$^{6}$,
Siobhan~Alden$^{7}$,
Luca~Alfonso$^{8}$,
Nicola~Amapane$^{9, 10}$,
David~Amorim$^{1}$
% ...
} 

\vspace{3mm}

\begin{flushleft}

{\em\footnotesize
$^{1}$ CH - CERN  \\
$^{2}$ DE - UROS, University of Rostock  \\
$^{3}$ US - LBL, Lawrence Berkely National Laboratory  \\
$^{4}$ IT - INFN - Pavia,  Istituto Nazionale di Fisica Nucleare Sezione di Pavia  \\
$^{5}$ TR - Ankara University   \\
$^{6}$ IT - INFN - Bologna, Instituto Nazionale Di Fisica Nucleare - Sezione di Bologna  \\
$^{7}$ UK - RHUL, Royal Holloway and Bedford New College  \\
$^{8}$ IT - INFN - Genova, Istituto Nazionale di Fisica Nucleare Sezione di Genova  \\
$^{9}$ IT - UNITO, Universit\`{a} di Torino  \\
$^{10}$ IT - INFN - Torino, Istituto Nazionale di Fisica Nucleare Sezione di Torino  \\
% ...
}
\end{flushleft}


\tableofcontents
\listoftables
\clearpage
\pagenumbering{arabic}
\setcounter{page}{1}

\section{Introduction}
\label{intro:sec}
This document contains updated parameters for the MuCol study.
This is the second iteration of the parameters, and is developed from the tentative parameters report of 2023~\cite{parameters2023}, with the goal of working towards the final consolidated parameters in 2025.\\
This preliminary collection of parameters includes high-level goals, such as the target beam parameters at different key interfaces of the collider complex.
It also contains many design and schematic-based parameters that have been developed bottom-up by the teams that work on the different parts of the complex and different technologies.
These parameters are already the fruit of the R\&D of each team, or the goals that the team considers realistic based on their expertise and studies carried out so far.
This allows for identification of further development needs to be addressed in future iterations of parameters.

\subsection{Muon Collider Design}
\label{intro:sec:design}
The design effort focuses on a high energy stage at 10 TeV with a luminosity of \SI[per-mode=reciprocal]{2.1E35}{\per\centi\meter\squared\per\second}.
This will demonstrate feasibility of a high energy stage matching approximately the physics reach of a \SI{100}{\tera\electronvolt} energy FCC-hh design.\\
This muon collider can be reached through one of two paths: either through \textit{energy staging }to build a \SI{3}{\tera\electronvolt} collider prior to the full \SI{10}{\tera\electronvolt}, or through \textit{luminosity staging} to begin with the full \SI{10}{\tera\electronvolt} collider, but with lower initial luminosity increased by a subsequent upgrade.
This could potentially be via initially a \SI{5}{\giga\electronvolt} \SI{2}{\mega\watt} proton beam on target (Option 1), then upgrade to a \SI{10}{\giga\electronvolt}, \SI{4}{\mega\watt} beam (Option 2).

\subsection{Structure of the Document}
\label{intro:sec:structure}
\begin{figure}[!h]
    \centering
    \includegraphics[width=1\textwidth]{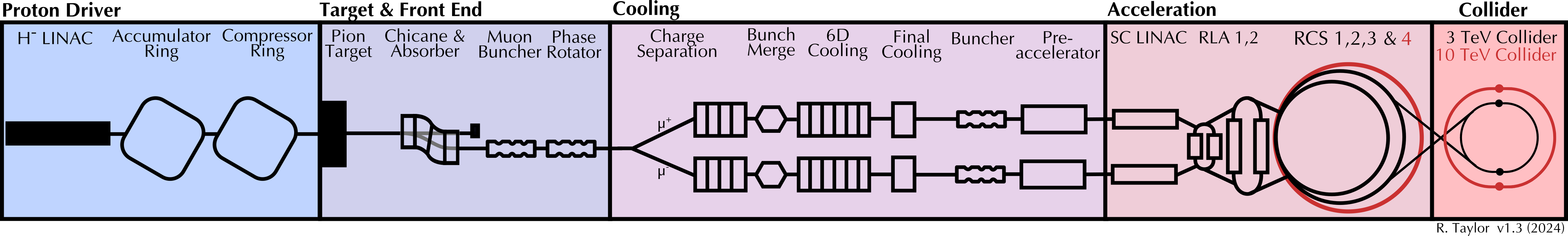}
    \caption{Simplified overview of the proton driver and muon collider accelerator complex.}
    \label{intro:fig:schematic}
\end{figure}

The overall parameters are listed in Section \ref{top:sec} followed by parameters for each subsystem split by section.
Figure \ref{intro:fig:schematic} demonstrates the present complex subsystems, starting with the proton driver (blue) in Section \ref{sec:proton}, passing through to the front end (purple) in Section \ref{target:sec}, the muon beam cooling (pink) in Section \ref{cool:sec}, acceleration (light red) in Section \ref{low:sec} and \ref{high:sec} and finally the collider ring (red) in Section \ref{col:sec}.
Then the Detector and Machine-Detector Interface (MDI) designs are described in Section \ref{mdi:sec} and \ref{detector:sec} respectively.
Details of underlying technologies are given in subsequent sections, including magnets (Section \ref{mag:sec}) and RF (Section \ref{rf:sec}).
Collective effects throughout the complex are described in Section \ref{imp:sec}, and the radiation shielding and protection considerations throughout the complex are described in Section \ref{rad:sec}.
Finally, site-specific parameters are defined, particularly for the demonstrators (Section \ref{demo:sec}) and the Rapid Cycling Synchrotrons (RCS) for CERN and Fermilab (Section \ref{site:sec}).
This document prioritises new and original parameters for each system.
Many of the initial baselines values for each system consider the results of the MAP study~\cite{MAP}.

\subsection{Differences from Tentative Parameter Report 2023}
\label{intro:sec:differences}
The tentative parameter set broadly assumed a new greenfield site.
In reality, the facility would likely be built at an existing laboratory with significant reuse of existing infrastructure where possible.
Studies to explore this opportunity are ongoing and the beginning of these results are indicated in this report within Section \ref{site:sec}.

Studies of a muon ionisation cooling demonstrator design have begun, considering both the mechanical integration of a single cooling cell, and also the optics overview of a cooling demonstrator facility.
Parameters defining the beginning of these designs are included within Section \ref{demo:sec}.

For an overview of the schedule considerations, please refer to the Tentative Parameters report~\cite{parameters2023}. 
\section{Top-Level Parameters}
\label{top:sec}
The top-level parameters for the Muon Collider are shown in Table \ref{top:tab:highlevel}.
These are the ideal design specifications that each subsystem aims to achieve.
The parameters are unchanged as of last year, with an additional acknowledgement of the asymmetry in the produced $\mu^+$, $\mu^-$ charges.
Machine transmission is the ideal transmission counted after the target and capture section.\\
Table \ref{top:tab:transmission} provides an overview of the whole complex, with the key numbers including the length and outgoing beam energy for each sub-system.
An estimate of transmission is provided, based on the current efficiency of each simulated system.

The total muon production efficiency after the front-end and cooling is \SI{4}{\percent}, which is less than the transmission budget of \SI{10}{\percent}.
This discrepancy motivates the need for an increase to a \SI{4}{\mega\watt} proton beam power to meet the intended number of muons per bunch at the collider.
\begin{table}[!htb]
    \centering
    \begin{tabular}{l|c|cc}
        Center of mass energy& Unit& \SI{3}{\tera\electronvolt}&\SI{10}{\tera\electronvolt}\\ \hline
        Luminosity for target parameters & \SI[per-mode=reciprocal]{E34}{\per\centi\meter\squared\per\second}& 2 &20 \\
        Transverse emittance& \si{\micro\meter} & \multicolumn{2}{c}{25}\\
        Proton beam power& \si{\mega\watt} & \multicolumn{2}{c}{2 - 4$^*$}\\
        Number of $\mu^+$ muons per bunch& \num{E12} & 2.2&1.8\\
        Number of $\mu^-$ muons per bunch& \num{E12} & 2.2&1.8\\
        Target integrated luminosity& \si[per-mode=reciprocal]{\per\atto\barn} & 1&10\\
        Luminosity lifetime& Turns&  1039&1158\\
        Collider peak field& \si{\tesla} &  11&16\\
        Repetition rate& \si{\hertz} &  \multicolumn{2}{c}{5}\\
        Beam power& \si{\mega\watt} & 5.3&14.4\\
        Longitudinal emittance& \si{\electronvolt\second} &  \multicolumn{2}{c}{0.025}\\
        IP bunch length& \si{\milli\meter} & 5&1.5\\
    \end{tabular}
    \caption[Top-level parameters for \SI{3}{\tera\electronvolt} and \SI{10}{\tera\electronvolt} muon colliders]{Preliminary parameters for a muon collider at two different energies. $^*$Currently muon transmission is not yet fully sufficient and further system design is required. As a back-up option a higher-power target can also mitigate this but affects e.g. beam loading. It should be noted that more mu+ are produced at the target, but currently we assume that this beam is collimated to have the same charge and hence the same beam loading everywhere.}
    \label{top:tab:highlevel}
\end{table}

\begin{table}
    \centering
    \begin{tabular}{cc|ccc}
         Parameter&  unit&  after Final Cooling&  Inj. at 3 TeV&  Inj. at 10 TeV\\ \hline
         Muon beam energy & GeV & 0.25 & 1500 & 5000 \\
         Number of muons & $10^{12}$ & 4 & 2.2 & 1.8\\
         Transverse norm. emittance&  um&  22.5&  25&  25\\
         Longitudinal emittance&  eVs&  0.0225&  0.025&  0.025\\
         RMS bunch length&  mm&  375&  5&  1.5\\
         RMS rel. momentum spread&  \% &  9&  0.1&  0.1\\
         Av. grad (0.2GeV-1.5TeV)&  MV/m&  &  2.4&  \\
         Av. grad (1.5-5TeV)&  MV/m&  &  &  1.1\\
    \end{tabular}
    \caption[Key target beam parameters along the collider.]{Key target beam parameters along the collider. Gradients given are average values required to achieve the transmission target. The two cooling options under development achieve transverse emittances of 24 (29.5)$ \mu m$ and longitudinal values of of $0.0124 (0.0289) \;\rm eVs$. The better performing one uses one solenoid with a field 25\% above the current target, however we expect that reducing the field will only increase the transverse emittance, remaining below 30 $\mu m$.}
    \label{tab:my_label}
\end{table}

\begin{table}[!ht]
    \centering
    \begin{tabular}{l|ccccl}
    Subsystem & Energy&   Length& Achieved &Achieved& Target\\
 & & & Transm.& $\mu^-$/bunch&  $\mu^-$/bunch\\
 & \si{\giga\electronvolt} &   \si{\meter}& \si{\percent} &$10^{12}$& $10^{12}$\\ \hline
    Proton Driver& 5 ($p^+$)&  1500  & -- &500 ($p^+$)& \\
    Front End& 0.17&   150& 9 &45.0& \\
    Charge Sep.& 0.17&   12& 95 &42.8& \\
    Rectilinear A& 0.14&   363& 50 &21.4& \\
    Bunch Merge& 0.12&   134& 78 &16.7& \\
    Rectilinear B& 0.14&   424& 32 &5.3& \\
    Final Cooling & 0.005&    100& 60 &3.2& \\
    Pre-Acc.& 0.25& 140 & 86 &2.8& 4.0 \\
 Low-Energy Acc.& 5 & --& \textit{90}$^*$ &2.5&  \\
    RLA2& 62.5&   $\circ$2430& 90 &2.3& \\
    RCS1 & 314&   $\circ$5990& 90 &2.1& \\
    RCS2 & 750&   $\circ$5990& 90 &1.9& \\
    RCS3 & 1500&   $\circ$10700& 90 &1.7& \\
 3 TeV Collider& 1500& $\circ$4500& -- &1.7&  2.2\\
    RCS4 & 5000&   $\circ$35000& 90 &1.5& \\
    10 TeV Collider& 5000&   $\circ$10000& -- &1.5& 1.8\\
    \end{tabular}
    \caption[Lengths, energies and transmission of each subsystem]{Preliminary beam parameters at the end of each section of the acceleration chain for the 2 MW target. Lengths are approximate and $\circ$ indicates that the length refers to the circumference.
    Muon numbers in the muon cooling systems refer to the yields from Option 1 as per Table \ref{target:tab:yield}. For $\mu^+$ the charge at the Front End is \num{60E12} but we assume that this is reduced by collimation to provide the same bunch charge and beamloading in both beams. Currently, the achieved muon transmission is lower than the target value in the cooling and somewhat higher than the target value in the muon accelerator part. Further improvement is expected. A 4 MW target would provide almost twice as many muons at the beginning. $^*$ For the initial muon acceleration no design exists at this moment, the target value is given.}    
    \label{top:tab:transmission}
\end{table}




%
 \clearpage
\section{Proton Driver}
\label{sec:proton}
This section is devoted to the Proton Complex parameters choice. The parameters are preliminary and based on previous studies as the MAP~\cite{MAP} and the Design for a Neutrino Factory at CERN~\cite{Accumulator1,Accumulator2,Compressor} as well as simulations and studies carried out during the years of the project.

The proton driver of a future Muon Collider is required to deliver a proton-beam of at least 2 MW at a repetition rate of 5 Hz to the pion-production target. The proton-beam energy must be in the multi-GeV range in order to maximize the pion yield. In addition, a particular time structure consisting of a single very short bunch, of rms pulse length on the order of 2 ns, is needed to allow the muon beam to be captured efficiently in the cooling section. The proton bunch parameters are intimately connected and constrained by beam loading and longitudinal acceptance in the downstream muon accelerator systems and by the acceptance (in time, energy, and power) of the target and pion capture system. The proton beam parameters necessary to produce the desired number of muons in the final storage rings of the Muon Collider are listed in Tables~\ref{proton:tab:H-Linac} and~\ref{proton:tab:compressor}.
Option 1 considers a \SI{5}{\giga\electronvolt} proton beam with a power of \SI{2}{\mega\watt}, and Option 2 considers a higher energy and higher power proton beam of \SI{10}{\giga\electronvolt} and \SI{4}{\mega\watt}.
These two options are equivalent to the luminosity scaling options.

\begin{table}[h!]
    \centering
    \begin{tabular}{l|ccc}
        Parameters & Unit & Option 1 & Option 2 \\ \hline
        Final energy & GeV & 5 & 10\\
        Repetition rate$^{1}$ & Hz &\multicolumn{2}{c}{5}\\
        Max. source pulse length$^{2}$ & ms & 3.4 & 5.0\\
        Max. source pulse current$^{2}$ & mA & \multicolumn{2}{c}{80.0}\\ 
        Source norm. emittance & mm.mrad & \multicolumn{2}{c}{0.25}\\
        Power & MW & 2 & 4\\
        RF frequency & MHz & \multicolumn{2}{c}{352 and 704} \\
    \end{tabular}
    \caption[H- LINAC parameters]{H- LINAC parameters for both options considering 1)  LINAC single use for muon production, 2) Chopping will later reduce the average current.}
    \label{proton:tab:H-Linac}
\end{table}

\begin{figure}[!h]
  \begin{center}
    \includegraphics[width=0.8\textwidth]{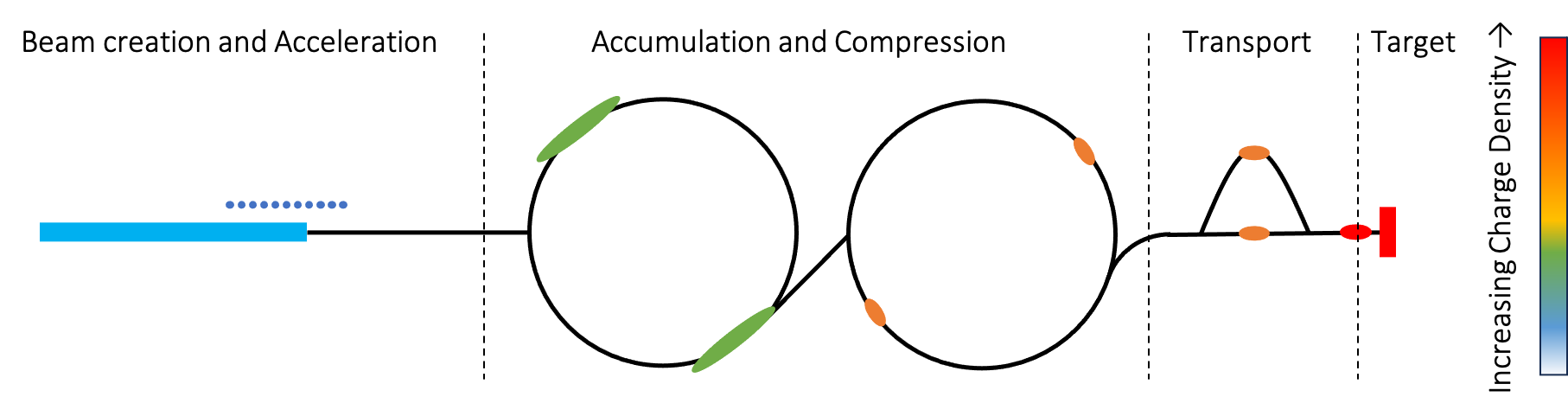}
  \end{center}
  \caption{Schematic of the baseline design for the Proton Complex.}
\label{fig:Baseline}
\end{figure}

The baseline proton driver design for the MuCol project is composed of a final energy multi-GeV H$^-$ LINAC followed by two rings, a compressor and an accumulator as shown in Fig.~\ref{fig:Baseline}, both operating at fixed energy.
It is likely that the two lattices will be part of the same ring or tunnel.
Depending on the final energy a single or a double bunch will be used with the latter needing a recombination line before reaching the target.

\subsection{High power Linac}
\label{sec:Section111}

The linear accelerator is the first stage of any hadron accelerator complex. The LINAC generates the initial transverse and longitudinal beam emittances and energy spread, defining the beam quality for the next stages of acceleration, accumulation and compression.
For a project like the Muon Collider, where the repetition rate is low, a high-energy high-power LINAC can be a versatile machine that can serve many other purposes including (and not restricted to) neutrino factories and nuclear science experiments.


The main parameters for a Muon Collider LINAC based injector are listed in Tab.~\ref{proton:tab:H-Linac} consisting of two options that will drive the final power of the facility. For the preliminary parameters of the proton complex we will assume that the LINAC final energy of 5 GeV or 10 GeV is also the final beam energy, removing the need of further acceleration in an intermediary storage ring. 

\begin{figure}[!h]
  \begin{center}
    \includegraphics[width=0.6\textwidth]{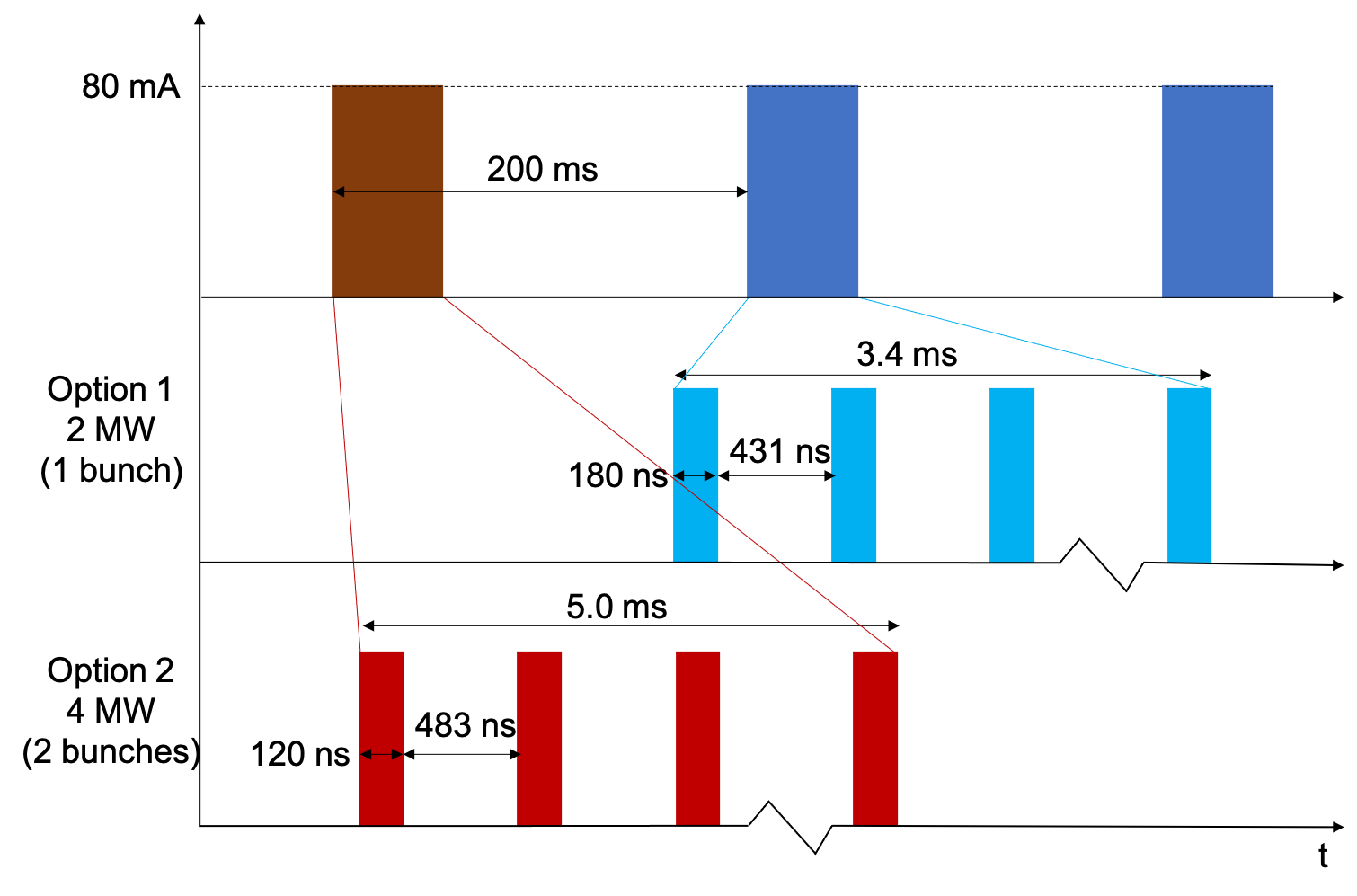}
  \end{center}
  \caption{Possible Chopping scheme for the LINAC considering a single bunch option as the baseline design.}
\label{fig:Chopping}
\end{figure}

Additional components required for the proton driver includes a $H^{-}$ source for charge-exchange injection, and a low-energy chopper.
Linac4 at CERN could be used as the first part of a CERN based proton drive and its development and design will be used as the reference for the initial parameters of the protons.
The current ongoing development of Linac4 beam studies and $H^{-}$ source will be beneficial to this project, as well as any advances in source design and R\&D in development\cite{Linac4_source}.
A 80 mA and 1 ms pulse is needed from the source to achieve the final target chargehowever chopping has not yet been taken into account.
Fig.~\ref{fig:Chopping} presents a chopping scheme for each of the options where the source current of 80 mA is assumed and a pulse length from the source of either 3.4 of 4.7 ms is needed.
The pulse length is long for a $H^{-}$ source, but an option of two sources working in parallel would have parameters close to current working sources.
   
A study on the losses in the high energy section of a full energy LINAC for both options was carried out~\cite{IPAC24}. The main contributor is black body radiation losses which indicates that the warm sections outside cryomodules and the transfer line to the accumulator ring will have to be cooled to temperatures below 200 K in order to maintain the losses within acceptable levels.

\subsection{Accumulator}

In order to test the stability and possible accumulation schemes for the beam coming from the LINAC we used two lattices developed for the Neutrino Factory at CERN~\cite{Accumulator1,Accumulator2} and based on the parameters listed in Tab.~\ref{proton:tab:accumulator}. For both final energies no instabilities are seen throughout the turns needed for accumulation. The total tune spread due to space charge forces is 0.15 for the 5 GeV (2 MW) option and less than 0.05 for the 10 GeV (4 MW) option.

\begin{table}[htb]
    \centering
    \begin{tabular}{l|ccc}
        Parameters & Unit & Option 1 & Option 2 \\ \hline
        Energy & GeV & 5 & 10\\
        Circumference & m & 180 & 300\\
        Final rms bunch length & ns & 180 & 120\\ 
        Geo.~rms.~emittance & $\pi$.mm.mrad & \multicolumn{2}{c}{5.0}\\
        Number of bunches &  --& 1 & 2\\
        Number of turns &  --& 5600 & 5900\\
    \end{tabular}
    \caption{Accumulator ring parameters}
    \label{proton:tab:accumulator}
\end{table}

\subsection{Compressor}

\begin{figure}[!h]
  \begin{center}
    \includegraphics[width=0.7\textwidth]{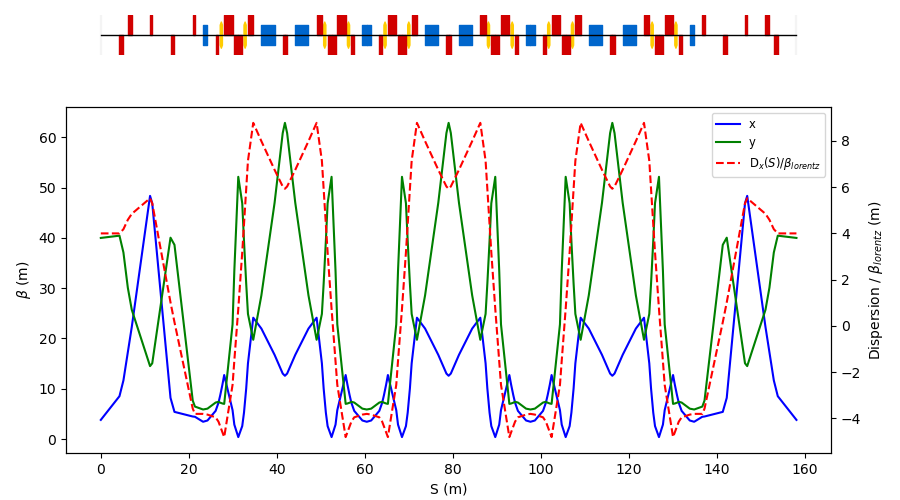}
  \end{center}
\caption{1 period of the compressor lattice with negative bending magnets for the 10 GeV case.\cite{Compressor}.}
\label{fig:compressor_lattice}
\end{figure}

A 10 GeV compressor lattice was developed, show in Fig.~\ref{fig:compressor_lattice} where the \SI{5}{\giga\electronvolt} option keeps the old lattice as developed for the  Neutrino Factory at CERN~\cite{Compressor}. Both lattices contain negative bends to minimize the momentum compaction factor while controlling the dispersion function along the ring. For both final energy options studies of the compression were carried out, including space-charge effects. The final tentative parameters for each option are listed in Tab.~\ref{proton:tab:compressor}.

\begin{table}[htb]
    \centering
    \begin{tabular}{lc|cc}
        Parameters & Unit & Option 1 & Option 2 \\ \hline
        Energy & GeV & 5 & 10\\
        Circumference & m & 300 & 600\\
        Protons on target & \num{E14} & \multicolumn{2}{c}{5.0} \\
        Final rms bunch length & ns & \multicolumn{2}{c}{2.0}\\ 
        Geo. rms. emittance & $\pi$.mm.mrad & \multicolumn{2}{c}{5.0}\\
        RF voltage & MV & 1 & 4 \\
        RF harmonic &  --& 1 & 2\\ 
        Number of turns & --& 60 & 70\\\end{tabular}
    \caption{Compressor ring parameters}
    \label{proton:tab:compressor}
\end{table}

\begin{figure}[!h]
    \centering
    \includegraphics[trim={0 0 0 1cm}, clip, width=0.4\textwidth]{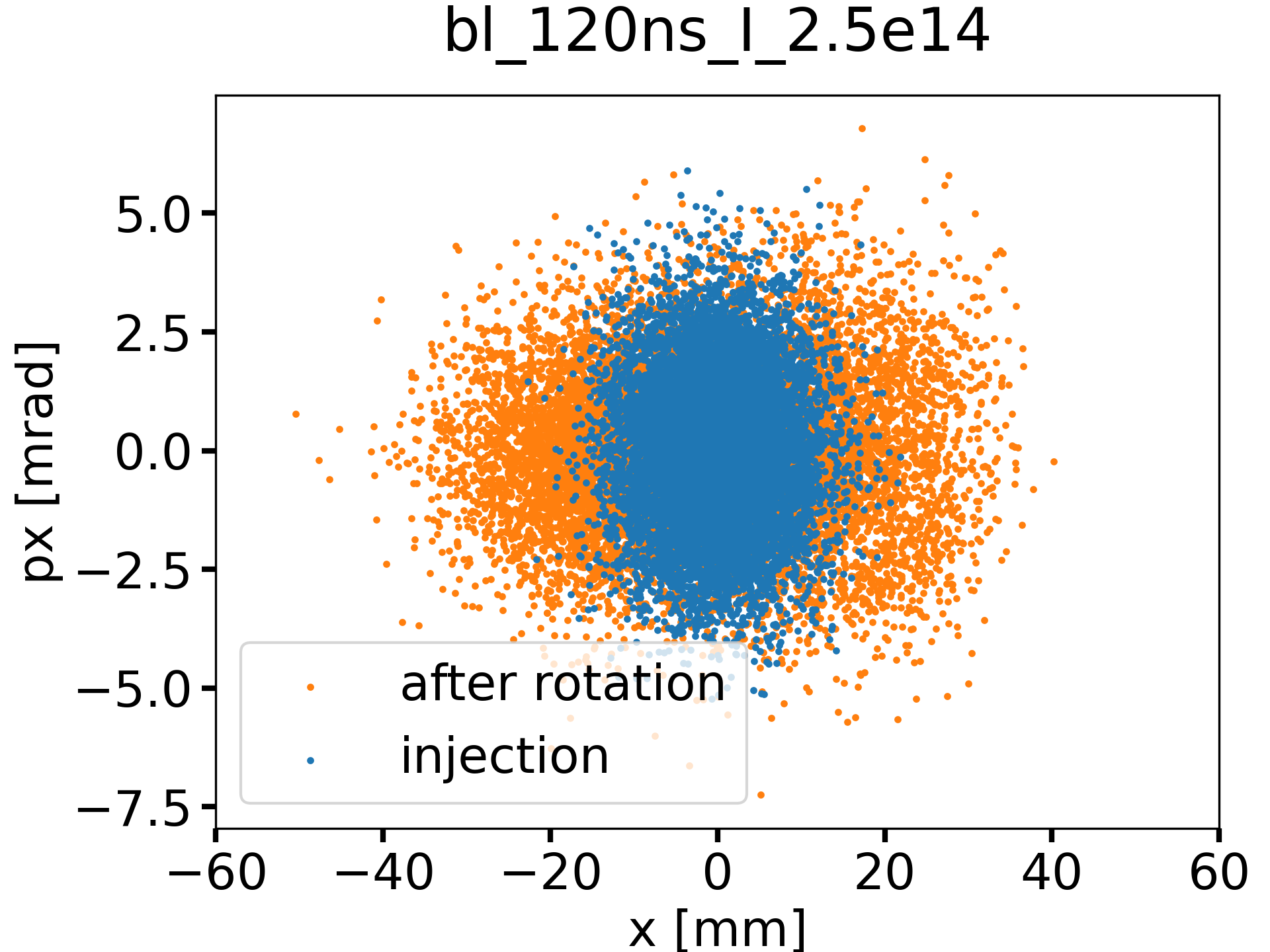}
    \includegraphics[trim={0 0 0 1cm}, clip, width=0.4\textwidth]{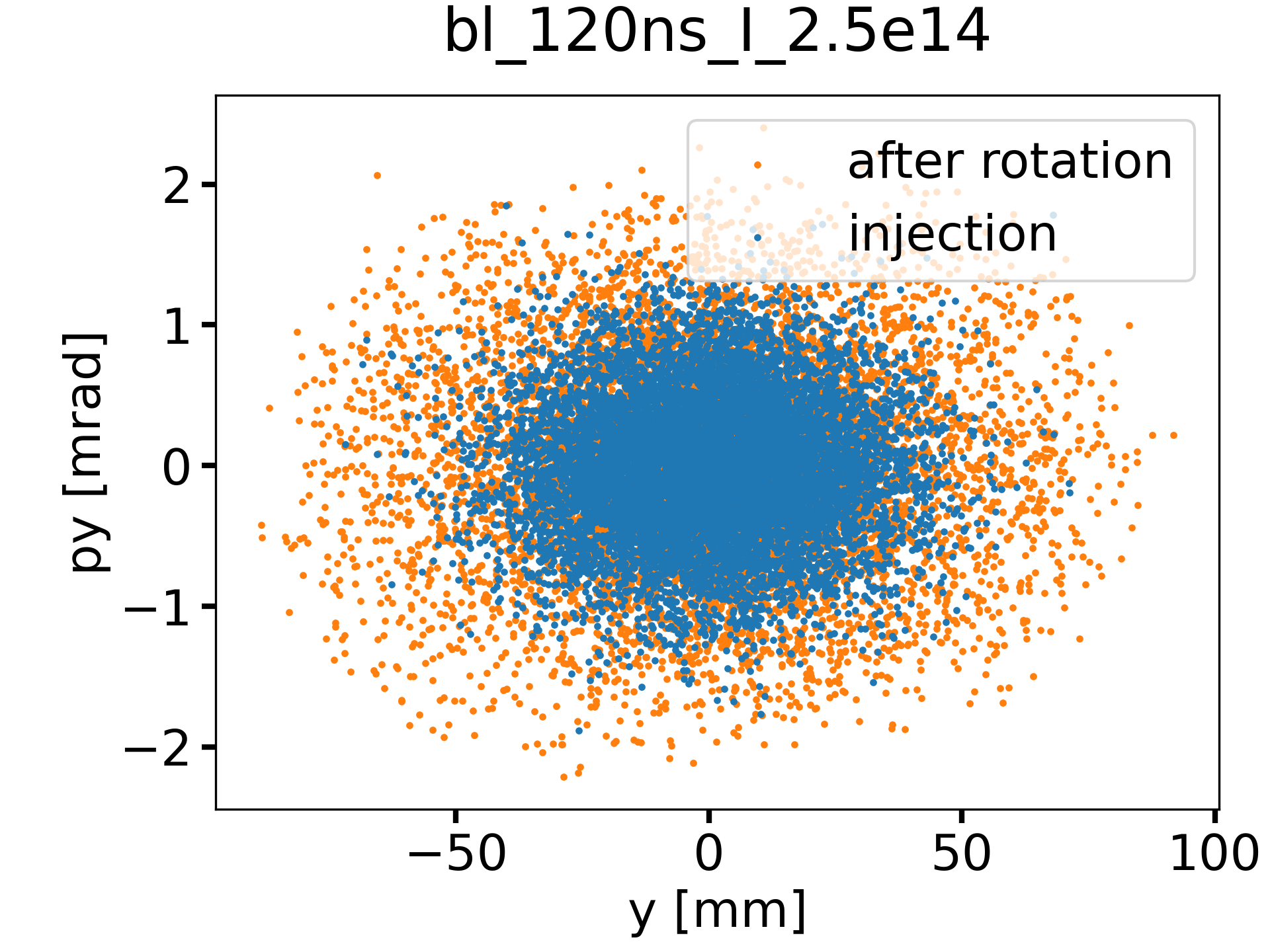}

    \includegraphics[trim={0 0 0 1cm}, clip, width=0.4\linewidth]{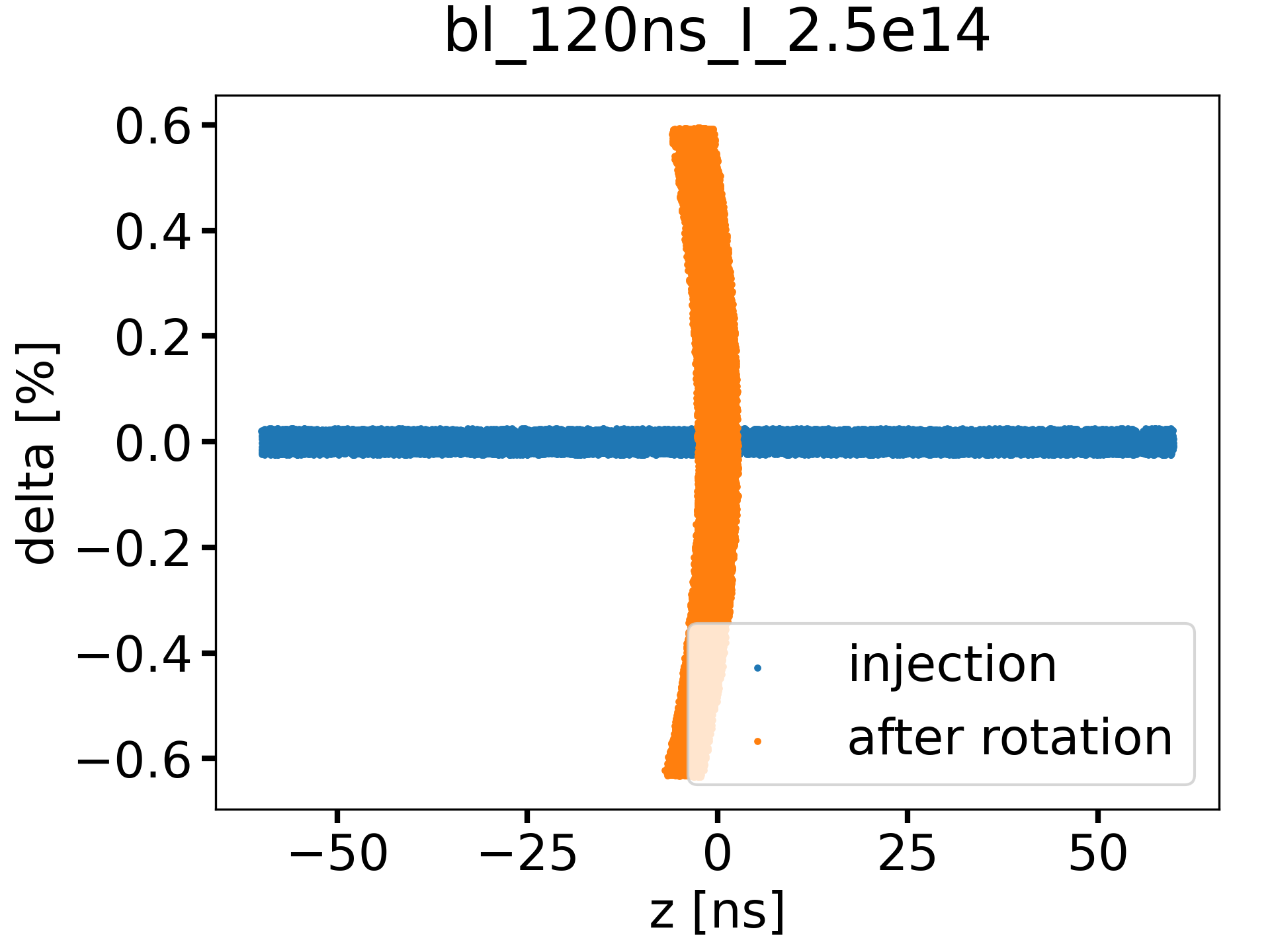}
    \includegraphics[trim={0 0 0 1cm}, clip, width=0.4\linewidth]{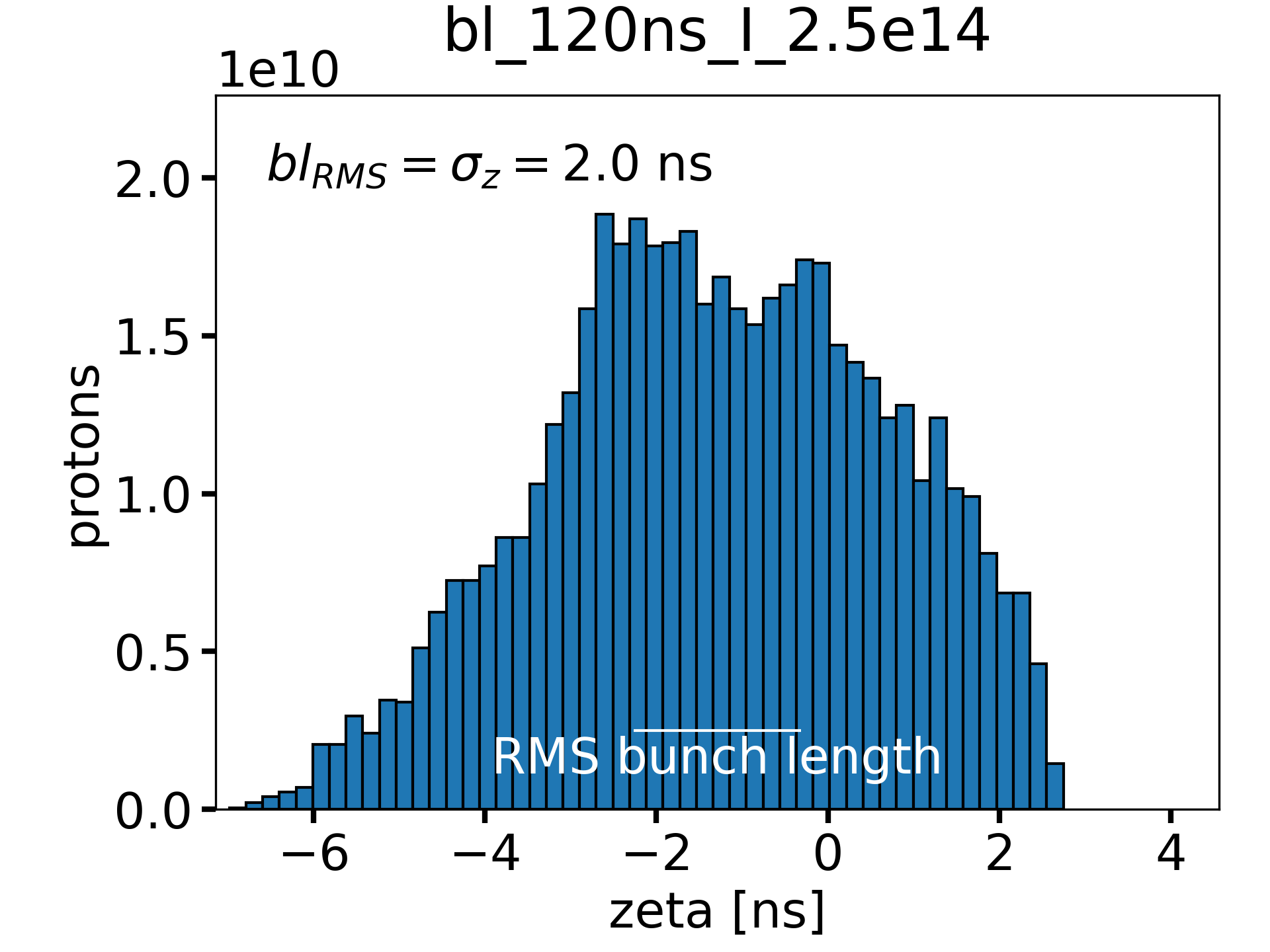}
    \caption{The figure shows the 6D phase space of a bunch at injection and after rotation in the proposed 10\~GeV compressor lattice. The longitudinal profile along the bunch is also plotted after rotation.}
    \label{fig:Com10GeV}
\end{figure}

For both final energies a full rotation simulation was performed and the final results are displayed in Fig.~\ref{fig:Com10GeV} where the sub 2~ns rms bunch length is achieved.

\vfill

\subsection{Target delivery system}
\begin{figure}[ht]
  \begin{center}
    \includegraphics[width=0.6\textwidth]{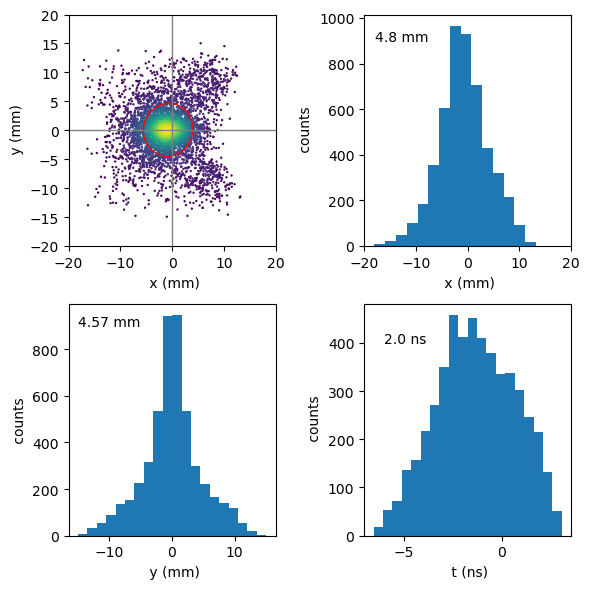}
  \end{center}
\caption{Beam Profile and bunch length on the target surface for the 10 GeV case. The values marked on the histogram are the rms trasnverse beam sizes and bunch length.}
\label{fig:target_beam}
\end{figure}

After rotation, the short bunch (or bunches) from the compressor have to be transported to the target.
The simple transfer line was designed and simulations of the beam transport for both options were carried out, not yet taking into account the extraction line from the compressor ring. No significant degradation of the bunch length was observed, however losses were observed due to the halo from the compressor. The maximum quadrupole gradients, including the triplet needed for the final focusing on the target surface, are between 30 -- 40 T/m depending on the beam final energy. Figure~\ref{fig:target_beam} shows an example of the result for a transport simulation of the compressed beam to the target surface.

\subsection{Next Steps}

The LINAC simulations including the chopper and accumulator ring injection studies have yet to be finalized.
A study on other possible alternative compressor lattices with slower rotation and flexible slip factor, like the JPARC RCS and Main ring, is needed. A slower rotation  makes the RF requirements less stringent and not having negative bends also make for weaker dipole fields, making the lattice easier to tune and cheaper to manufacture.
Significant beam loss was observed at the end of rotation that has to be investigated.
The study of the extraction line from the accumulation, transport and recombination should be initiated for the double bunch option.

Losses in the Proton Complex will need to be studied in detail since they cause activation of elements and make maintenance complex. The budget on the LINAC is taken from the main working machine as 1 W/m however for the rings and transport line downstream there was no in depth study so far. Injection and extraction regions are of high importance as well as halo formation of formation around the beam due to space charge effects.










 \clearpage
\section{Target \& Front-End}
\label{target:sec}

The front end of the muon collider comprises several sub-systems:
\begin{enumerate}
\item The target and solenoid (transverse) capture system, initially with field 15--20 T tapering to 1--2 T, in Section \ref{target:sec:target};
\item Extraction line for the spent proton beam in Section \ref{front:sec:extraction};
\item Solenoid chicane and proton absorber, in Section \ref{front:sec:chicane};
\item Longitudinal drift;
\item Sequence of RF cavities for bunching, in Section \ref{front:sec:rfbunch};
\item Further sequence of RF cavities for rotating in energy-phase space, also in Section \ref{front:sec:rfbunch}.
\end{enumerate}

These parameters are consistent with a \SI{2}{\mega\watt}, \SI{5}{\giga\electronvolt} proton beam (Option 1), with the \SI{4}{\mega\watt}, \SI{10}{\giga\electronvolt} (Option 2) design currently under active study.


\subsection{Target and solenoid}
\label{target:sec:target}
Pions are produced by sending protons onto a graphite target immersed in a strong magnetic field.
Solenoid parameters are listed in Table \ref{mag:tab:developments} of subsection \ref{mag:sec:target}.
In the MAP design, resistive magnets (RC1--RC5) were considered, however IMCC is developing a full HTS-based alternative.
The target and target systems are under design, thus no details regarding the expected operation temperature, mechanical response and life-time are listed here. 

Moreover, small discrepancies exist in the components dimensions reported in this chapter, particularly between pion/muon yield studies and design and engineering calculations.

Information on the radiation load on the target solenoids is within Subsec. \ref{rad:sec:target_sol}.

\subsubsection{Production target and beam parameters}
\label{target:sec:params}
The deep inelastic interactions of the proton beam with the target produces kaons and pions, which eventually decay into muons.
To capture the produced particles and keep the emittance under control, the production target and the subsequent line has to be kept in a strong solenoidal magnetic field, which confines the charged particles along helicoidal trajectories.

The baseline case considers a graphite target as the most suitable option.
This material allows operation at high temperatures and has a high thermal-shock resistance.
Therefore the majority of studies performed to optimize the pion-yield and estimate the radiation load on the front-end magnets have taken this target as baseline.
Nevertheless, it is worth acknowledging the ongoing parallel studies for a fluidised tungsten powder target and for a liquid lead target, which will be detailed in the future.

An overview of the proton driver parameters being used in the studies of the front-end target systems is shown in Table~\ref{target:tab:protonbeam}.
Different ranges of these parameters have been considered in order to optimise both the physics and engineering design.

\begin{table}[!ht]
    \centering
    \begin{tabular}{l|ccc}
     Parameters& Unit& Baseline&  Range\\ \hline
     Beam power& MW& 2&  1.5-3.0\\
     Beam energy& GeV& 5&  2-10\\
     Pulse frequency& Hz& 5&  5-50\\
     Pulse intensity& p+ $10^{14}$& 5&  3.7-7.5\\
     Bunches per pulse& & 1& 1-2\\
     Pulse length& ns& 2& 1-2\\
     Beam size& mm& 5& 1-1.5\\
     Impinging angle& \textdegree & 0& 0-10\\
    \end{tabular}
    \caption{Assumed beam from proton driver via carbon target used in studies}
    \label{target:tab:protonbeam}
\end{table}

\subsubsection{Production target engineering parameters}
As depicted in Table \ref{target:tab:target_baseline}, the target system is divided into production target, target vessel, target shielding, target shielding vessel, proton beam window and muon beam window. The auxiliary services for cooling of the target and shielding are equally part of the target system and are listed in Table \ref{target:tab:target_baseline-services}.  
For both tables, the main dimensions, key material considerations and important design and integration features are summarized.

\begin{table}[!ht]
    \centering
    \resizebox{\textwidth}{!}{
    \begin{tabular}{l|ccc}
     & Material& Box dimensions DxL  
[mm]& Integration \\ \hline \hline
     Production Target& Isostatic Graphite& D30 x L800& Rod supported with transverse CFC \\
 & & &supports attached to cylindrical frame\\ \hline
     Target Vessel& Titanium Grade 5& D346 x L920& Located in the bore of the Target shielding vessel.\\ \hline
     Proton Beam Window& Beryllium& D220 x L0.25& \\ \hline
     Muon Beam Window& Titanium Grade 5& 
D240 x L1& Welded on the vessel\\ \hline
     Target Shielding & Tungsten & D x L2000& Inside Shielding Vessel. Multiple pie-like \\
 & & &blocks stacked together with guiding rods.\\ \hline
     Target Shielding Vessel & Stainless-Steel& D1218 x L2065& Supported by transversal beam \\
 & & &across the cryostat of the solenoid.\\ 
    \end{tabular}
    }
    \caption{Baseline engineering parameters of the carbon target system}
    \label{target:tab:target_baseline}
\end{table}

\begin{table}[!ht]
    \centering
    \resizebox{\textwidth}{!}{
    \begin{tabular}{l|ccccc}
     Cooling& Coolant& Type& Mass Flow& Pressure& Integration \\
 Unit& & & \si{\kilo\gram\per\second}& \si{\bar} &\\ \hline \hline
     Target& Helium& Static / Natural & -& 1& Surrounding target rod \\
 & & convection& & &enclosed by windows and target vessel.\\ \hline
 Target Vessel& Helium& Forced convection& 0.5& 10&Inside double wall target vessel. \\
 & & & & &Routing upstream via the  Solenoid bore.\\ \hline
 Target Shielding& Helium& Forced convection& 0.33& 2&Inside Target shielding Vessel. \\
 & & & & &Routing upstream via the  Solenoid bore.\\ 
    \end{tabular}
    }
    \caption{Baseline engineering parameters of the carbon target auxiliary systems for 2MW.}
    \label{target:tab:target_baseline-services}
\end{table}

The baseline for a 2~MW-class target consists of a solid graphite target. The graphite rod is housed within a double-walled vessel filled with a static helium atmosphere. This helium confinement facilitates the initial stage of heat removal from the graphite rod through natural convection, while raising the sublimation temperature of the graphite when compared to a vacuum environment and providing a non-erosive heat transfer medium. Forced convection cooling is then applied through the vessel's double wall using a 10 bar helium flow.

Titanium is a suitable candidate for the target vessel  due to its low density (reduced interaction with produced pions and muons) and good thermal-shock resistance. However, it is required to use beryllium in proton and muon windows to guarantee a peak power density of approximately $\SI{800}{W/cm^3}$ and yearly DPA around 0.5. On the contrary, adopting titanium would increase these values by an order of magnitude.

The target vessel is surrounded by a helium-cooled, heavy tungsten shield, which reduces power deposition and radiation damage to the solenoid materials to acceptable levels.
For details on the radiation shielding, see Section \ref{rad:sec}.
The target proximity shielding is housed inside a large stainless-steel vessel, extending from just upstream of the target to around 2 meters downstream. The large size and weight, combined with the need to efficiently extract heat from each tungsten block, resulted in proximity shielding composed of multiple pie-shaped tungsten segments, perforated in specific locations to either guide helium flow or allow for the insertion of longitudinal rods to hold the assembly together. The shielding vessel also hosts a water layer to moderate the neutrons.

Both the cooling and instrumentation routing for the target systems are handled via the upstream side of the assembly.

Downstream of the target and its cooled shielding assembly, the shielding is made of tungsten and has an aperture following the parabolic shape defined in the MAP studies.

\subsubsection{Muons and pions yield}
\label{target:sec:yield}
To assess the most suitable conditions to operate the proton driver and to design the target, several FLUKA simulations were conducted, calculating the muon and the pion yield in each setting.
For this purpose, it was assumed that all the muon and pions going in the chicane can be captured if their momentum is below \SI[per-mode=symbol]{500}{\mega\electronvolt\per\clight}.
 The obtained yields are summarized as a function of beam energy in Table~\ref{target:tab:yield}, assuming a transverse beam sigma of 5~mm and a graphite target rod with a radius of 15~mm. 






\begin{table}[!ht]
    \centering
    \begin{tabular}{l|ccccclll}
     Yield [$10^{-2} $\si{\giga\electronvolt} $/p^+$]& 3& 4& 5& 6&  7& 8& 9&10\\ \hline
     $\mu^+$ & 2.8& 2.6& 2.4& 2.3&  2.2& 2.1& 1.9&1.9\\
     $\mu^-$ & 1.8& 1.8& 1.8& 1.8&  1.7& 1.7& 1.7&1.7\\
     $\pi^+$ & 1.3& 1.2& 1.1& 1.1&  1& 0.98& 0.92&0.9\\
     $\pi^-$ & 0.84& 0.81& 0.84& 0.82&  0.83& 0.8& 0.8&0.81\\
    \end{tabular}
    \caption{Yield per unit energy proton beam [$10^{-2} $\si{\giga\electronvolt} $/p^+$]}
    \label{target:tab:yield}
\end{table}

\subsubsection{Target radial build}
\label{target:sec:radialbuild}
A preliminary target radial build has been defined and is shown in Table~\ref{target:tab:radialbuild}. This build takes into account a 700~mm inner-radius solenoid coil, the baseline target system dimensions as described in Table~\ref{target:tab:target_baseline}, and the required shielding configuration with a water and Boron-Carbide neutron-absorbing layers (Table~\ref{rad:tab:DPA_coils}). The discrepancy in thickness of tungsten shielding between Table~\ref{target:tab:radialbuild} and Table~\ref{rad:tab:DPA_coils} is explained by the need to integrate other components in the prior as part of the exercise to engineer the entire target-solenoid cryostat. 

\vfill

\begin{table}[!ht]
    \centering
    \begin{tabular}{l|cccc}
     Component& Material& $r_i$ [mm]& $r_e$ [mm]& ${\Delta}r$ [mm]\\ \hline
     Solenoid coils& HTS& 700& -& -\\
     Insulation& Insulation& 690& 700& 10
\\
     Vacuum& Vacuum& 670& 690& 20
\\
     Thermal shield& Copper \& Water& 651& 670& 19
\\
 Vacuum& Vacuum& 631& 651& 20
\\
 Inner supporting tube& Stainless-steel& 619& 631& 12
\\
 Vacuum& Vacuum& 609& 619& 10
\\
 Outer Target shielding& Tungsten& 599& 609& 10
\\
 Neutron absorber& Boron Carbide& 594& 599& 5
\\
 Target shielding and neutron moderator& Stainless-steel& 589& 594& 5
\\
 & Water& 569& 589& 20
\\
 & Stainless-steel& 564& 569& 5
\\
 & Tungsten& 179& 564& 385
\\
 & Stainless-steel& 174& 179& 5
\\
 Vacuum& Vacuum& 173& 174& 1
\\
 Target vessel& Titanium& 168& 173& 5
\\
 & Helium& 155& 168& 13
\\
 & Titanium& 150& 155& 5
\\
 & Helium& 15& 150& 135
\\
 Target& Graphite& 0& 15&15
\\
    \end{tabular}
    \caption{Target System radial build for a graphite target.}
    \label{target:tab:radialbuild}
\end{table} 
\subsection{Front-End}
\label{front:sec}
\subsubsection{Chicane and proton absorber}
\label{front:sec:chicane}
The target solenoid is followed by a solenoid chicane which is terminated by a thick beryllium cylinder.
The cylinder absorbs low energy remnant protons which would otherwise irradiate equipment downstream of the chicane.
The concept was initially introduced in \cite{Rogers:2013kna} and initial parameters were defined. 
Further discussion was made in \cite{Stratakis:2014ina}.
In particular, the former study assumed \SI{1.5}{\tesla} solenoid fields, while the MAP and latter study considered \SI{2}{\tesla} solenoid fields in this region.
The latter study also noted that a large proportion of undecayed pions were stopped in the proton absorber which negatively impacted the muon yield.

Table \ref{front:tab:chicane} shows the current design parameters for the chicane and the proton absorber.

\begin{table}[h]
    \centering
    \begin{tabular}{l|cc}
     Parameters& Unit& Value\\ \hline
     Chicane bend angle& degree& 15 \\
     Chicane radius of curvature& m& 22 \\
     Proton absorber material& -& Be \\
     Proton absorber thickness& m& 0.1 \\
 Chicane field& T& 1.5 \\
    \end{tabular}
    \caption{Chicane and proton absorber parameters}
    \label{front:tab:chicane}
\end{table}

\subsubsection{Spent proton beam extraction}
\label{front:sec:extraction}
A non-negligible fraction of the primary protons do not have an inelastic nuclear collision in the production target and escape from the graphite rod.
At these energies, the protons are not bent significantly by the chicane and would be lost on the chicane aperture.
In absence of a mitigation strategy, the energy carried by these particles would lead to a high power deposition density in the normal-conducting chicane solenoids. In addition, a high cumulative ionizing dose and displacement damage would be reached within a short operational time. It is therefore necessary to extract the spent protons from the front-end and steer them onto an external beam dump.

Earlier studies explored a possible solution of injecting the proton beam at different angles into the front-end, with an extraction channel envisaged in a gap between the superconducting magnets upstream of the chicane. This concept proved to be unfeasible due to geometrical aspects and the increase of the radiation load to the superconducting coils. As an alternative solution, the spent proton beam could be extracted in the middle of the chicane, by using solenoids with different diameters in order to create a gap for the high-energy protons. Shower simulation studies showed that such an extraction channel in the chicane needs to have a transverse size of a few tens of centimeters, which is challenging for the magnet design. In addition, an internal radiation shielding would be needed to protect the coils from particles, which are still lost in the chicane. The chicane design studies are presently still ongoing. 



\begin{table}[h]
    \centering
    \begin{tabular}{l|cc}
     Parameters& Unit& \\ \hline
     Num. micro bunches& & 21\\
     Longit. emittance& mm& 46\\
     Transv. emittance& um& 17000\\
     Positive muon yield& 1/GeV per p+& 0.024\\
    Negative muon yield& 1/Gev per p+& 0.018\\
    \end{tabular}
    \caption{Outgoing muon beam}
    \label{front:tab:muonbeam}
\end{table}

\subsection{Buncher \& Phase Rotator}
\label{front:sec:rfbunch}
The buncher is comprised of a sequence of RF cavities.
The cavity frequency is chosen to match the distance between nominal RF bunches, so that it varies along the length of the buncher.
The phase is purely bunching.

In the phase rotator, cavities are dephased so that the low energy tail of the beam sees an accelerating gradient and the high energy front of the beam sees a decelerating gradient.

Cavities are placed in a two-cavity LINAC with \SI{0.25}{\meter} separation between adjacent cavity pairs.
Each cavity in the pair is independently phased.
Transversely, the beam is contained in a \SI{2}{\tesla} field.

 \clearpage
                                                                                                                                                                                                                                                                                                                                                                                 \section{Cooling}
\label{cool:sec}

The cooling channel is defined from the end of the RF capture system to the beginning of acceleration.
Five sub-systems are part of the cooling apparatus:
\begin{enumerate}
\item Charge separation, which splits the positive and negative muon species into separate beamlines;
\item Rectilinear cooling (A and B lattices) which cools the beam in 6D phase space;
\item Bunch merge after the A lattice merges the microbunches produced by the front end into a single bunch;
\item Final cooling, which produces the final low emittance beam;
\item Re-acceleration, which accelerates the low energy beam up to 200 MeV/c.
\end{enumerate}
For this first iteration, parameters are listed in Table \ref{cool:tab:summary} for the principal subsystems: rectilinear cooling and final cooling.
Additional details are available for the charge separation \cite{Yoshikawa:2013eba} and bunch merge \cite{Bao:2016fmz} subsystems.
Potential performance for re-acceleration is estimated.

\begin{table}[!ht]
    \centering
    \begin{tabular}{l|cccccc}
         Num. bunches&  Actual $\varepsilon_{\rm{T}}$&  Target $\varepsilon_{\rm{T}}$&  Actual $\varepsilon_{\rm{L}}$&  Target $\varepsilon_{\rm{L}}$&  Mean $p_z$&  Transm.\\
 Unit& um& um& mm (eVs)& mm (eVs)& MeV/c&\%\\ \hline
     End of charge separation&      -& 17000&            -&           46&  288&  95.0 \\
\hline
    6D Cooling end of \textbf{Stage 8}&    260&   300&         1.86&          1.5&  200&  14.9\\
         End of Final Cooling&   29.5&  22.5&  82 (0.0289)&           64&   50&  4.0\\ 
        End of Reacceleration&      -&  22.5&            -&  64 (0.0225)&  339&  3.8 \\
        \multicolumn{2}{l}{}\\
         Num. bunches&  Actual $\varepsilon_{\rm{T}}$&  Target $\varepsilon_{\rm{T}}$&  Actual $\varepsilon_{\rm{L}}$&  Target $\varepsilon_{\rm{L}}$&  Mean $p_z$&  Transm.\\
 Unit& um& um& mm (eVs)& mm (eVs)& MeV/c&\%\\ \hline
   6D Cooling end of \textbf{Stage 10}&    140&    140&         1.56&          1.56& 200&  10.5 \\
End of high field Final Cooling&    24&  22.5&         35.3&            64&36.4&  6.3 \\
        End of Reacceleration&       -&  22.5&            -&   64 (0.0225)& 339&  5.9 \\ 

    \end{tabular}
    \caption[Cooling system emittance overview]{Beam parameters entering and leaving the cooling system for short-rectilinear (top) and long-rectilinear (bottom) options. The target emittances are listed. They are 10 \% more demanding than the nominal emittances in the RCS and collider, allowing for some emittance growth at some point in the acceleration chain.}
    \label{cool:tab:summary}
\end{table}

\subsection{Rectilinear Cooling}
\label{cool:sec:6d}
The rectilinear cooling section consists of a number of solenoid magnets with dipole field superimposed.
In the MAP design the dipole field was achieved by means of introducing a tilt in the solenoids but separate dipoles are proposed for this IMCC design. The rectilinear cooling lattice described below is stored in the MuonCollider-WG4 github group, \verb|rectilinear| repository as release (branch) \verb|2024-09-27_release| and described in \cite{Zhu:2024vfe}.

The solenoid field is approximately sinusoidal with a period given by the cell length $L$ so that $B_z(z, r=0) = B_{peak} \sin(2 \pi z/L)$.
Cells in the Rectilinear B lattices are increasingly non-sinusoidal, with a component $B_z(z, r=0) = B_{peak} \sin(4 \pi z/L)$ that gets stronger further down the B lattice.
The peak $B_z$ listed in Table \ref{cool:tab:6d_cell} is the peak field on the axis of the solenoid.
Fields may be higher in the conductor volume.

RF cavities are modelled as perfect cylindrical pillbox cavities operating in TM010 mode.
Several RF cavities are included within each cell.
A thin conductive window electromagnetically seals the RF cavities so that the pillbox model is an adequate approximation to the real cavity field and the cavities can be assumed to be independently phased. 
The RF gradient listed in Table \ref{cool:tab:6d_rf} is the peak gradient.

Updates for the A and B stages of the rectilinear cooling system have been developed, comprising of 10 "B-type" stages, denoted S1 through S10 that yields improved performance over the MAP lattice listed above and has been designed using 352 MHz RF and harmonics.
The performance is summarised in Table  \ref{cool:tab:6d_emit}.

Hardware parameters are described in Table \ref{cool:tab:6d_cell}.
In this lattice, the dipoles were simulated as a magnet independent of the solenoids which were not tilted and the dipole field is listed.

\begin{table}[!h]
    \centering
    \begin{tabular}{l|ccccc}
         &  $\varepsilon_{\rm{T}}$ &  $\varepsilon_{\rm{L}}$ &  $\varepsilon_{\rm{6D}}$ & Stage & Cumulative\\
         &  mm&  mm&  mm$^3$& Transmission & Transmission \%\\ \hline
         Start&      16.96&  45.53&   13500& &  100\\ \hline
         A-Stage 1 &  5.17&  18.31&  492.60& 75.2 & 75.2\\
         A-Stage 2&   2.47&   7.11&   44.03& 84.4 & 63.5\\
         A-Stage 3&   1.56&   3.88&    9.59& 85.6 & 54.3\\
         A-Stage 4&   1.24&   1.74&    2.86& 91.3 & 49.6\\ \hline
         Bunch merge& 5.13&   9.99&   262.5& 78.0 & 38.7\\ \hline
         B-Stage 1&   2.89&   9.09&   76.07& 85.2 & 33.0\\
         B-Stage 2&   1.99&   6.58&   26.68& 89.4 & 29.4\\
         B-Stage 3&   1.27&   4.05&    6.73& 87.5 & 25.8\\
         B-Stage 4&   0.93&   3.16&    2.83& 89.8 & 23.2\\
         B-Stage 5&   0.70&   2.51&    1.32& 89.4 & 20.7\\
         B-Stage 6&   0.48&   2.29&    0.55& 88.4 & 18.2\\
         B-Stage 7&   0.39&   2.06&    0.31& 92.8 & 17.0 \\
         B-Stage 8&   0.26&   1.86&    0.13& 87.9 & 14.9\\
         B-Stage 9&   0.19&   1.72&    0.06& 85.2 & 12.7\\
         B-Stage 10&  0.14&   1.56&    0.03& 87.1 & 11.1\\ 
    \end{tabular}
    \caption[Rectilinear cooling performance]{Rectilinear cooling performance in terms of emittance reduction (transverse, longitudinal and 6D) and transmission per stage.}
    \label{cool:tab:6d_emit}
\end{table}

\begin{table}[!h]
    \centering
    \begin{tabular}{l|ccccccccc}
         &  Cell&  Stage&  Pipe&  Max. $B_z$ & Int.& $\beta_\perp$ & $D_x$ & On-Axis&Wedge\\
         & Length & Length & Radius & On-Axis & $B_y$ & & & Wedge Len.& Angle \\
         &  m&  m&  cm&  T& Tm& cm& mm& cm&deg\\ \hline
         A-Stage 1 &  1.8&  104.4&    28&   2.5&  0.102&   70&   -60& 14.5&  45\\
         A-Stage 2&   1.2&  106.8&    16&   3.7&  0.147&   45&   -57& 10.5&  60\\
         A-Stage 3&   0.8&  64.8&     10&   5.7&  0.154&   30&   -40&   15& 100\\
         A-Stage 4&   0.7&  86.8&      8&   7.2&  0.186&   23&   -30&  6.5&  70\\ \hline
         B-Stage 1&   2.3&  50.6&     23&   3.1&  0.106&   35& -51.8&   37& 110\\
         B-Stage 2&   1.8&  66.6&     19&   3.9&  0.138&   30& -52.4&   28& 120\\
         B-Stage 3&   1.4&  84.0&   12.5&   5.1&  0.144&   20& -40.6&   24& 115\\
         B-Stage 4&   1.2&  66.0&    9.5&   6.6&  0.163&   15& -35.1&   20& 110\\
         B-Stage 5&   0.8&  44.0&      6&   9.1&  0.116&   10& -17.7& 12.5& 120\\
         B-Stage 6&   0.7&  38.5&    4.5&  11.5&  0.087&    6& -10.6&   11& 130\\
         B-Stage 7&   0.7&  28.0&   3.75&    13&  0.088&    5&  -9.8&   10& 130\\
         B-Stage 8&  0.65& 46.15&   2.85&  15.8&  0.073&  3.8&    -7&    7& 140\\
         B-Stage 9&  0.65&  33.8&    2.3&  16.6&  0.069&    3&  -6.1&  7.5& 140\\
         B-Stage 10& 0.63& 29.61&    2.0&  17.2&  0.069&  2.7&  -5.7&  6.8& 140\\ \hline
    \end{tabular}
    \caption[Rectilinear cooling cell hardware]{Rectilinear cooling cell hardware in terms of cell geometry, solenoid fields, dipole fields and wedge geometry}
    \label{cool:tab:6d_cell}
\end{table}

\begin{table}[!h]
    \centering
    \begin{tabular}{l|ccccl}
         &  RF Frequency&  Num. RF&  RF Length&  Max. RF Gradient& RF phase\\
         &  MHz&  &  cm&  MV/m& deg\\ \hline
         A-Stage 1 &  352&  6&  19&   27.4& 18.5\\
         A-Stage 2&   352&  4&  19&   26.4& 23.2\\
         A-Stage 3&   704&  5&  9.5&  31.5& 23.7\\
         A-Stage 4&   704&  4&  9.5&  31.7& 25.7\\ \hline
         B-Stage 1&   352&  6&  25&   21.2& 29.9\\
         B-Stage 2&   352&  5&  22&   21.7& 27.2\\
         B-Stage 3&   352&  4&  19&   24.9& 29.8\\
         B-Stage 4&   352&  3&  22&   24.3& 31.3\\
         B-Stage 5&   704&  5&  9.5&  22.5& 24.3\\
         B-Stage 6&   704&  4&  9.5&  28.2& 22.1\\
         B-Stage 7&   704&  4&  9.5&  28.5& 18.4\\
         B-Stage 8&   704&  4&  9.5&  27.1& 14.5\\
         B-Stage 9&   704&  4&  9.5&  29.7& 11.9\\
         B-Stage 10&  704&  4&  9.5&  24.9& 12.2\\ 
    \end{tabular}
    \caption{Rectilinear cooling cell RF parameters. 0$^o$ phase is bunching mode.}
    \label{cool:tab:6d_rf}
\end{table} \clearpage
\subsection{Final cooling (short rectilinear)}
\label{cool:sec:final}
A \SI{75}{\meter} long final cooling system has been developed and optimised with RF-Track.
This system is made of 11 cells, each of which is composed of a high-field $\approx$\SI{40}{\tesla} solenoid which encompasses a liquid or gaseous hydrogen absorbers, and a long low-field solenoid which encompasses a series of RF cavities and their drift regions.
The RF cavities are split into acceleration, to restore the energy lost from the absorber, and rotation, to restore a more uniform momentum distribution of the beam. The beam conditions at the start of Cell 1 for this design were assumed based on previous rectilinear cooling cell designs from MAP; they are approximately compatible with the end of rectilinear B-8.

The hardware parameters for the final cooling cells, and the final cooling RF cavities are in Table \ref{cool:tab:fc_cell} and \ref{cool:tab:fc_rf} respectively.

The performance of this system is sufficient to reach \SI{29.5}{\micro\meter} in transverse emittance $\varepsilon_T$, which is \SI{4.5}{\micro\meter} away from the target emittance of \SI{25}{\micro\meter}.
The longitudinal emittance $\varepsilon_L$ increases from \SI{2.7}{\milli\meter} to \SI{82}{\milli\meter}, which is significantly larger than the target emittance of \SI{64}{\milli\meter}.
However the transmission does not presently meet the target, as only \SI{28.5}{\percent} of the beam remains, due to both decays and losses within the absorbers. 
An optimisation effort is ongoing to improve the capture of beam within the RF buckets and therefore have reasonable transmission throughout the cooling channel.
Matching coils between high and low field regions are required to prevent emittance blow-up due to mismatches.

\begin{table}[!h]
    \centering
    \begin{tabular}{l|cccc}
         Cell&  $\varepsilon_{\rm{T}}$ &  $\varepsilon_{\rm{L}}$ &  $\varepsilon_{\rm{6D}}$ & Cumulative\\
         no. &  \si{\micro\meter} &  \si{\milli\meter} &  \si{\micro\meter} & transmission \%\\ \hline
         Start&  300&  1.5&  & 100\\ \hline
         1&  275.2&  2.7&  586.1& 97.5\\
         2&  212.7&  5.9&  645.4& 94.1\\
         3&  170.4&  6.8&  582.8& 88.9\\
         4&  138&  12.4&  617.5& 81.9\\ 
         5&  102.5&  20.6&  600& 74.4\\ 
         6&  81.3&  25&  548.8& 61.1\\
         7&  59.5&  32.7&  486.9& 53.1\\
         8& 50.8& 43.6& 482.8&46.9\\
         9& 41.2& 48.4& 434.2&37\\
         10& 32.9& 66.1& 414.6&31.7\\
         11& 29.5& 82& 414.5&28.5\\
    \end{tabular}
    \caption[Short rectilinear final cooling performance]{Baseline final cooling performance in terms of emittance reduction (transverse, longitudinal and 6D) and cumulative transmission per stage.}
    \label{cool:tab:fc_emit}
\end{table}

\begin{table}[!h]
    \centering
    \begin{tabular}{l|ccccc}
         Cell  &  Solenoid&  Stage&  Max. $B_z$ & Low $B_z$& Absorber\\
         no. & length & length & on-axis & on-axis& length\\
         &  m&  m&  T& T& m\\ \hline
         1&  1.48&  1.48&  44.63& 4.63& 0.85\\
         2&  1.75&  4.57&  44.63& 4.63& 0.47\\
         3&  1.00&  6.61&  44.63& 4.63& 0.47\\
         4&  1.00&  7.75&  44.63& 4.63& 0.40\\
         5&  1.00&  5.09&  44.63& 4.63& 0.30\\
         6&  1.11&  6.86&  44.63& 4.63& 0.25\\
         7& 1.33& 7.06& 42.00& 2.00& 0.30\\
         8& 0.80& 6.70& 42.00& 2.00& 0.10\\
         9& 1.48& 8.37& 41.00& 1.00& 0.17\\
         10& 0.95& 6.76& 40.80& 0.80& 0.08\\
         11& 0.95& 7.60& 40.80& 0.80& 0.05\\
    \end{tabular}
    \caption[Short rectilinear final cooling cell hardware]{Baseline final cooling cell hardware in terms of cell geometry, solenoid fields and absorber geometry}
    \label{cool:tab:fc_cell}
\end{table}

\begin{table}[!h]
    \centering
    \begin{tabular}{l|ccccc|cccc}
        Cell&  RF&  Num.&  Tot. RF&  Max. RF& Rot RF & Initial& Final& Energy&Bunch\\
         no. & freq. & RF& len. & grad. & phase & KE & KE & spread & len. \\
         &  MHz&  &  cm&  MV/m& deg & MeV& MeV& MeV&mm\\ \hline
         1&  0.0&  0&  0&  0&  0& 73.8& 39.4& 4.4&141\\
         2&  111.1&  10&  2.5&  19.81&  -180& 53.7& 32.7& 2.8&241\\
         3&  56.9&  17&  4.25&  14.17&  90& 53.0& 32.5& 4.1&406\\
         4&  40.1&  17&  4.25&  11.9&  51& 49.0& 31.4& 3.9&348\\ 
         5&  34.9&  9&  2.25&  11.11&  -10& 35.6& 16.9& 5.7&781\\
         6&  30.6&  15&  3.75&  10.4&  -54& 28.3& 14.7& 2.7&1256\\
         7& 11.6& 19& 4.75& 6.823&  -82& 32.6& 13.3& 3.1&1319\\
         8& 16.2& 9& 2.25& 8.04&  67& 21.4& 14.0& 3.2&1692\\
         9& 13.4& 13& 3.25& 7.32&  67& 24.1& 12.4& 3.5&1962\\
         10& 8.2& 13& 3.25& 5.39&  -6& 16.5& 8.8& 2.8&2702\\
         11& 5.7& 15& 3.75& 4.48&  -96& 16.3& 11.2& 2.9&3013\\
    \end{tabular}
    \caption{Short rectilinear final cooling cell RF parameters. 0$^o$ phase is on-crest mode.}
    \label{cool:tab:fc_rf}
\end{table}

\clearpage

\subsection{Final cooling (long rectilinear)}
A high-field final cooling system has also been designed that would follow the end of rectilinear B-10. This system is more demanding in terms of magnet parameters (exceeding present development target) but yields good performance. The parameters are listed in Tables \ref{cool:tab:hf_fc_emit},  \ref{cool:tab:hf_fc_field}, \ref{cool:tab:hf_fc_rf} and  \ref{cool:tab:hf_fc_beam}.

The Final cooling (high-field) lattice described below is stored in the \verb|Final_cooling_updated| repo as release \verb|2024-10-03-prerelease|

\begin{table}[!h]
    \centering
    \begin{tabular}{l|cccc}
Stage & $\varepsilon_{\rm{T}}$ & $\varepsilon_{\rm{L}}$ &	$\varepsilon_{\rm{6D}}$ & Cumulative \\
& mm & mm & mm$^3$ & transmission \% \\
\hline
Start	& 0.14	& 1.5	& 0.030	& 100 \\
\hline
Stage 1	& 0.12	    & 2.0	& 0.030	& 99.60\\
Stage 2	& 0.099	& 3.8	& 0.038	& 96.60\\
Stage 3	& 0.082	& 5.0	& 0.034	& 87.80\\
Stage 4	& 0.060	& 7.1	& 0.026	& 81.60\\
Stage 5	& 0.046	& 9.7	& 0.022	& 72.20\\
Stage 6	& 0.034	& 17.9	& 0.021	& 63.90\\
Stage 7	& 0.024	& 35.3	& 0.022	& 60
    \end{tabular}
    \caption{Long rectilinear final cooling cell performance parameters}
    \label{cool:tab:hf_fc_emit}
\end{table}

\begin{table}[!h]
    \centering
    \begin{tabular}{l|cccc}
Stage&Stage length (m)&Peak on-axis Bz (T)&LH absorber length (m)\\
\hline
Stage 1&1.564&38.5&0.203\\
Stage 2&2.735&-45.2&0.188\\
Stage 3&2.984&28&0.0736\\
Stage 4&2.949&-43.4&0.0547\\
Stage 5&2.781&46.2&0.064\\
Stage 6&5.6&-40.7&0.0575\\
Stage 7&5.494&50&0.0654\\
    \end{tabular}
    \caption{Long rectilinear final cooling cell magnet lattice parameters}
    \label{cool:tab:hf_fc_field}
\end{table}

\begin{table}[!h]
    \centering
    \begin{tabular}{l|ccccc}
Stage&Frequency&Number of RF cells&Maximum gradient&Phase&RF cell length\\
 &MHz& & MV/m & $^\circ$ & m \\
\hline
stage 1&& 0 &&& \\
stage 2&133.09&3&15&14.26&0.25 \\
stage 3&109.84&2&11.1&44.94&0.25 \\
stage 4&69.6&3&5.64&12.24&0.25 \\
stage 5&54&5&7.4&41.97&0.25 \\
stage 6&23.6&9&5.5&21.3&0.25 \\
stage 7&11.2&9&5.25&46.56&0.25 \\
    \end{tabular}
    \caption{Long rectilinear final cooling cell RF parameters. 0$^o$ phase is bunching mode.}
    \label{cool:tab:hf_fc_rf}
\end{table}

\begin{table}[!h]
    \centering
    \begin{tabular}{l|ccc}
Stage&Final Pz&Final energy spread&Final $c\sigma_t$ \\
Units & \si{\mega\electronvolt\per\clight} & \si{\mega\electronvolt} & \si{\clight} \\
\hline
Start&95&3.35&0.04794 \\
\hline
Stage 1&77.1&4.218&0.07809\\
Stage 2&56.7&2.546&0.19776\\
Stage 3&53.9&2.117&0.3408\\
Stage 4&42.1&1.983&0.4467\\
Stage 5&42.68&2.681&0.3999\\
Stage 6&37.03&2.811&0.8124\\
Stage 7&36.42&2.694&1.4994\\
    \end{tabular}
    \caption{Long rectilinear
    inal cooling cell beam longitudinal parameters}
    \label{cool:tab:hf_fc_beam}
\end{table}

\pagebreak 

\subsection{Pre-accelerator}
No pre-accelerator design exists. Table \ref{cool:tab:preacc} gives estimations of design and performance based on induction LINAC technology.

\begin{table}[!h]
    \centering
    \begin{tabular}{cccccc}
           Injection Energy&Extraction Energy&  Pulse Length&  Transmission& Linac Length\\ 
   MeV&MeV& ns& \%&m\\ \hline
           5&250&  15&  86 & 140\\
    \end{tabular}
    \caption{Pre-Accelerator (Induction Linac) - see for example RADLAC-1}
    \label{cool:tab:preacc}
\end{table} \clearpage
\section{Low Energy Acceleration}
\label{low:sec}
The low energy acceleration chain brings the muon beams from \SI{250}{\mega\electronvolt} after the pre-accelerator to \SI{63}{\giga\electronvolt} for injection into the high energy acceleration chain described in Section \ref{high:sec}.\\
It is composed of a single-pass superconducting LINAC outlined in Table \ref{low:tab:linac}, followed by two recirculating linear accelerators (RLA), described in Table \ref{low:tab:RLA}.\\
RLA2 has an preliminary optics design. No optics design exists for LINAC and RLA1.
Both RLAs have an assumed racetrack geometry.
The transmission through RLA2 is 92.6\%. The target transmission for LINAC and RLA1 is 90\%, which corresponds to an effective average gradient of $4.1\;\rm MV/m$.

\begin{table}[!h]
    \centering
    \begin{tabular}{l|cc}
         &  CryoModule 1& CryoModule 2\\ \hline
         Initial energy [GeV]&  0.255& --\\
         Final energy [GeV]& -- & 1.25\\
         Frequency [MHz]&  325& 325\\
         RF gradient [MV/m]&  20& 20\\
         Passes&  1& 1\\
    \end{tabular}
    \caption{Parameters describing the single-pass LINAC that follows the final cooling section.}
    \label{low:tab:linac}
\end{table}

\begin{table}[!h]
    \centering
    \begin{tabular}{l|c:cl|c:c}
         &  \multicolumn{2}{c}{RLA1} & \ &\multicolumn{2}{c}{RLA2}\\ \hline
         Initial energy [GeV]&   \multicolumn{2}{c}{1.25} & \ &   \multicolumn{2}{c}{5}\\
         Final energy [GeV]&   \multicolumn{2}{c}{5} &\ &   \multicolumn{2}{c}{63}\\
         Energy gain per pass&   \multicolumn{2}{c}{0.85} & \ &   \multicolumn{2}{c}{13.5}\\
         Frequency [MHz]&   352&1056 &\ &   352&1056\\
         No.~SRF cavities&   36&4 &\ &   600&80\\
         RF length [m]&   61.2&3.4 &\ &   1020&68\\
         RF gradient [MV/m]&   15&25 &\ &   15&25\\
         Passes&   \multicolumn{2}{c}{4.5} &\ &   \multicolumn{2}{c}{4.5}\\
 Linac length [m]&  \multicolumn{2}{c}{--} &\ & \multicolumn{2}{c}{915}\\
 Arc lengths [m]&  \multicolumn{2}{c}{--} &\ & \multicolumn{2}{c}{$\approx$ 300}\\
    \end{tabular}
    \caption{Multi-pass recirculating LINACs}
    \label{low:tab:RLA}
\end{table}

 \clearpage
\section{High Energy Acceleration}
\label{high:sec}
\begin{figure}[ht!]
\centering
{\includegraphics[width=0.9\columnwidth]{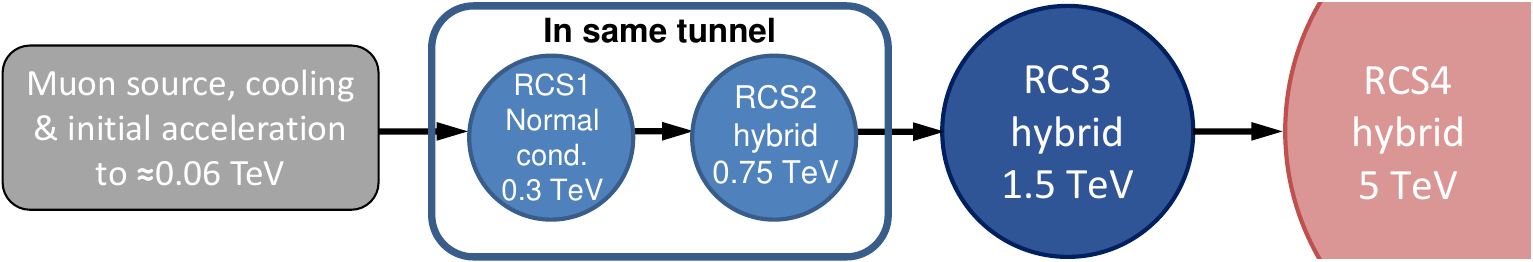}}
\caption{Schematic of the chain of rapid cycling-synchrotrons for the high-energy acceleration complex. From~\cite{batsch:ipac2023}.}
\label{high:fig:rcsoverview}
\vspace*{-1\baselineskip}
\end{figure}

As described in \cite{batsch:ipac2023}, an option for the chain of four rapid cycling synchrotrons~(RCS) foresees to accelerate two counter-rotating bunches at a repetition rate of 5\,Hz in stages of \SI{0.30}{\tera\electronvolt} (RCS1), \SI{0.75}{\tera\electronvolt} (RCS2) and \SI{1.5}{\tera\electronvolt} (RCS3) to inject into the \SI{3}{\tera\electronvolt} collider ring, or \SI{5}{\tera\electronvolt} (RCS4), to inject to the \SI{10}{\tera\electronvolt} collider ring.
This scenario is based on the US Muon Acceleration Program (MAP)~\cite{Berg:details,MAP} and applied for a general Greenfield site.
The high-energy stage of the accelerator chain with four RCS is illustrated in Fig.~\ref{high:fig:rcsoverview}.
Corresponding site-specific parameter designs can be found in Section~\ref{site:sec}.

The first two RCS share the same tunnel, meaning that they have the same circumference and layout~\cite{Berg:MCpsp}. The bending in the first RCS is provided by normal conducting magnets. The RCS2 to RCS4 are planned as hybrid RCSs where normal conducting magnets cycling from $-B_{\mathrm{nc}}$ to $+B_{\mathrm{nc}}$ are interleaved with strong fixed-field, superconducting magnets.
Within this section, NC magnets are referred to as \textit{pulsed}, and the SC magnets are referred to as \textit{steady}.
This is to reflect the alternative magnet technologies required for the hybrid RCS.
This combination allows for a large energy swing with a high average bending field to minimize the travel distance of the muons and thus their decay losses. 
The absolute value of magnetic field in the normal-conducting dipoles does not exceed \SI{\pm1.8}{T} at injection and extraction for all RCSs to avoid saturation of the magnet yoke.
For the hybrid RCS2 and RCS3, the magnetic field in the SC magnets is \SI{10}{T} to provide a compromise between the magnet filling factor and magnet costs. 
To protect the SC magnets from decay products, the inner aperture of the SC magnets is larger with \SI{10}{T}. Increasing the field to \SI{16}{T} implies higher technological and financial cost without a significant improvement of the machine performance. 
In the case of RCS4 however, the 
average magnetic field in the accelerator is assumed to be \SI{16}{T} as a higher magnetic field in the SC magnets helps to reduce the overall circumference and thus the muon decay and RF requirements.
This requirement may evolve with the optimization of the high-energy chain. 

The number of synchrotron oscillations per turn is extreme~\cite{batsch:ipac2023}, much larger than the conventional stability limit for stable synchrotron oscillations and phase focusing of $1/\pi$ in a synchrotron with one or few localized RF sections. 
To mitigate resulting beam losses, the RF system must be distributed over the entire RCSs. Tracking simulations on how the number of RF stations influences the longitudinal emittance have been performed. 
For the present design, the minimum number is around 32 RF stations for RCS1 and RCS4, and 24 stations for RCS2 and RCS3~\cite{batsch:ipac2023}. 

It is worth noting that the longitudinal dynamics used values of momentum compaction factor for an RCS lattice design based on FODO cells. 
With a more defined optics design, this number might change and with it the basic parameters of the longitudinal beam dynamics such as the synchrotron tune, bucket area and energy acceptance, which are all a function of the momentum compaction factor.

\subsection{Parameter tables}
 
Table \ref{high:tab:RCS_key} shows the general RCS parameters, and Table \ref{high:tab:RCS_add} specifies lattice parameters. 
The first parameters for the fourth RCS to accelerate to \SI{5}{TeV} are included but may evolve in the near future. 
We assume a survival rate of 90\,\% per ring and linear ramping only considering losses due to muon decay, even though these values are subject to further adjustments to optimize the RF and magnet powering parameters with respect to total costing, ramp shape, bunch matching, and the overall transmission of the entire chain.

\begin{table}[!h]
    \centering
    \begin{tabular}{lc|cccc} 
         Parameter&  Unit&  RCS1&  RCS2&  RCS3&  RCS4\\ \hline
         Hybrid RCS&  -& no & yes & yes & yes \\
         Repetition rate&  Hz&  5&  5&  5&  5\\
         Circumference& m& 5990& 5990& 10700& 35000\\
         Injection energy& GeV& 63& 314& 750& 1500\\
         Extraction energy& GeV& 314& 750& 1500& 5000\\
         Energy ratio&  &  5.0&  2.4&  2.0&  3.3\\
         Assumed survival rate& & 0.9& 0.9& 0.9&0.9\\
         Cumulative survival rate& &  0.9&  0.81&  0.729&  0.6561\\     
         Acceleration time&  ms&  0.34&  1.10&  2.37&  6.37\\
         Revolution period&  \textmu s&  20&  20&  36&  117\\
         Number of turns& -& 17& 55& 66& 55\\
         Required energy gain/turn& GeV& 14.8& 7.9& 11.4& 63.6\\
         Average accel.~gradient& MV/m& 2.44& 1.33& 1.06& 1.83\\ \hline
         Number of bunches& & 1& 1& 1& 1\\
         Inj.~bunch population& \num{E12}& 2.7& 2.4& 2.2& 2\\
         Ext.~bunch population& \num{E12}& 2.4& 2.2& 2& 1.8\\
         Beam current per bunch& mA& 21.67& 19.5& 9.88& 2.75\\
         Beam power&  MW&  640&  310&  225&  350\\
         Vert.~norm.~emittance& \textmu m& 25& 25& 25& 25\\
         Horiz.~norm.~emittance&  \textmu m&  25&  25&  25&  25\\
         Long.~norm.~emittance&  eVs&  0.025&  0.025&  0.025&  0.025\\
         Bunch length at injection & ps & 31& 30 & 23 & 13 \\
         Bunch length at ejection & ps & 20& 24 & 19 & 9\\ \hline
         Straight section length&  m & 2335 &  2335 &  3977 &  10367\\
         Length with pulsed dipole magnets& m & 3654& 2539 & 4366 & 20376\\
         Length with steady dipole magnets& m & -& 1115 & 2358 & 4257\\
         Max.~pulsed dipole field& T& 1.8& 1.8& 1.8&1.8\\
         Max.~steady dipole field& T& -& 10& 10&16\\
         Ramp rate& T/s& 4200& 3282& 1519&565\\
         Main RF frequency& GHz& 1.3& 1.3& 1.3&1.3\\
         Harmonic number&  &  25900&  25900&  46300&  151400\\
    \end{tabular}
    \caption{RCS acceleration chain key parameters}
    \label{high:tab:RCS_key}
\end{table}

\begin{table}[!h]
    \centering
    \begin{tabular}{lc|cccc} 
         Parameter&  Unit&  RCS1&  RCS2&  RCS3&  RCS4\\ \hline
         Fill ratio dipole& \%& 61& 61& 62.8& 70.4\\
         Cells per arc& &  5&  4&  6&  9\\
         Number of arcs& &  34&  26&  26&  26\\
         Cell length& m& 30.1& 53& 64.3& 133.6\\
         Relative path length difference& \num{E-06}& 0& 8.3& 2& 1.7\\
         Vertical aperture & mm & 40& 33.0 & 28.2 & 29.6 \\
         Transition gamma& & 46.2& 29.2& 36.9& 59.0\\
         Momentum compaction factor& \num{E-04}& 4.68& 11.74 & 7.35 & 2.87\\
    \end{tabular}
    \caption{RCS acceleration chain lattice parameters}
    \label{high:tab:RCS_add}
\end{table} \clearpage
\section{Collider}
\label{col:sec}

The present work concentrates on the design of a \SI{10}{\tera\electronvolt} center-of-mass collider.
The aim is to maximize the luminosity to the two possible experiments.
The basic assumptions are extrapolations from lower energy starting with a relative rms momentum spread of $\sigma_\delta = 1 \cdot 10^{-3}$.
Together with the longitudinal emittance, this fixes the rms bunch length $\sigma_z = 1.5$ \si{\milli\meter} and the $\beta^* = 1.5$ \si{\milli\meter} to the same value, such that the hour glass luminosity reduction factor $f_{hg} = 0.758$ starts to become significant.
Maximization of the luminosity requires to choose the shortest possible circumference $C$ compatible with feasibility of the magnets (average bending field assumed to be $\bar B \approx$ \SI{10.48}{\tesla} leading to $C \approx$ \SI{10}{\kilo\meter}).
Note that extrapolation of parameters to higher energies lead to very large chromatic effects further increasing with energy.

The main parameters are described in Table \ref{col:tab:param}, which contains a set of target parameters which meet the performance of Table \ref{top:tab:highlevel}.
The set of relaxed parameters considers a lattice with reduced beta oscillations and chromatic aberrations, to study imperfections and the effects of movers.\\
The radial build of arc dipoles is described in Table~\ref{col:tab:arcs}. The radial build assumes a radiation shielding thickness of 3~cm, which can be accepted from a cryogenics point of view if the operating temperature is 20~K. The estimated heat load and radiation damage in arc dipoles is summarized in Table~\ref{rad:tab:colliderarcdipoles}. 

\begin{table}[!h]
    \centering
    \begin{tabular}{l|c|c|c}
         &  & \multicolumn{2}{c}{version} \\
         Parameter&  Unit& relaxed & target \\ \hline
         Center of mass energy&  TeV &\multicolumn{2}{c}{10}  \\
         Geometric Luminosity\footnotemark & \SI[per-mode=reciprocal]{E34}{\per\centi\meter\squared\per\second} & 5.77 & 19.2 \\ 
         Beam energy& TeV & \multicolumn{2}{c}{5}\\
         Relativistic Lorentz factor& & \multicolumn{2}{c}{47322}\\
         Circumference&  km & \multicolumn{2}{c}{$\approx$ 10}\\
         Dist. of last magnet to IP& m& \multicolumn{2}{c}{6}\\
         Repetition rate& Hz& \multicolumn{2}{c}{5}\\
         Bunch intensity (one bunch per beam)& \num{E12} & \multicolumn{2}{c}{1.80}\\
         Injected beam power per beam&  MW& \multicolumn{2}{c}{7.2} \\
         Normalized transverse rms emittance&  \textmu m& \multicolumn{2}{c}{25} \\
         Longitudinal norm. rms emittance&  eVs& \multicolumn{2}{c}{0.025}\\
         Relative rms momentum spread& $10^{-3}$ & 0.3 & 1 \\
         RMS bunch length in space&  mm& 5 & 1.5 \\
         RMS bunch length in time domain&  ns& .017 & 0.005 \\
         Twiss betatron function at the IP&  mm& 5 & 1.5 \\
         Energy loss per turn\footnotemark & MeV & \multicolumn{2}{c}{$\approx$ 27.2}  \\
         Integrated RF gradient\footnotemark
          & MV & \multicolumn{2}{c}{30}
    \end{tabular}
    \caption{10 TeV collider main parameters}
    \label{col:tab:param}
\end{table}%

\setcounter{footnote}{\value{footnote}-2}
\footnotetext[\value{footnote}]{%
Luminosities for Gaussian beams with hour glass reduction factor and without beam-beam effect. Multiturn beam simulations with the correct lattice and tunes are needed in addition to first single pass simulations resulting in a modest luminosity increase.}%
\stepcounter{footnote}%
\footnotetext[\value{footnote}]{%
Assuming constant bending field of 15 T. The exact value will depend on the detailed lattice design and likely be lower.}%
\stepcounter{footnote}%
\footnotetext[\value{footnote}]{%
Assuming that only the synchrotron radiation losses have to be compensated. Some margin and no particular frequency requirements as long as the RF voltage does not vary too much over the bunch length of few 10s of ns.}

\begin{table}[!ht]
    \centering
    \begin{tabular}{l|c|cc}
         Parameter&  Unit&  Thickness& Outer radius\\ \hline
         Beam aperture&  mm&  23.49& 23.49\\
         Coating (copper)&  mm&  0.01& 23.5\\
         Radiation absorber (tungsten alloy)&  mm&  30& 53.5\\
         Shielding support and thermal insulation&  mm&  11& 64.5\\
         Cold bore&  mm&  3& 67.5\\
         Insulation (Kapton)&  mm&  0.5& 68\\
         Clearance to coils&  mm&  1& 69\\
    \end{tabular}
    \caption[Collider arcs, coil inner aperture.]{Collider arcs, coil inner aperture. For options using low temperature superconductor, i.e. at 3 TeV, the shielding thickness should be 40 mm and the other parameters changed accordingly.}
    \label{col:tab:arcs}
\end{table}
 \clearpage
\section{Machine-Detector Interface}
\label{mdi:sec}

The beam-induced background arising from muon decay poses a significant challenge for the physics performance of a multi-TeV muon collider.
The machine-detector interface relies on massive absorbers in close proximity to the interaction point (IP) to reduce the number of secondary particles reaching the detector.
This section describes the geometrical features of the shielding and quantifies the flux of secondary background particles.
In addition, the ionizing dose and displacement damage in different parts of the detector are presented. 

\subsection{Nozzle geometry and material composition}
\label{mdi:sec:nozzle}
The innermost part of the machine-detector interface consists of a nozzle-like shielding, which defines the inner detector envelope.
The nozzle extends from the last magnet ($L^*=6$~m) to almost the IP and must be made of a high-$Z$ and high density material to shield efficiently the electromagnetic showers induced by the decay electrons and positrons.
All studies carried out so far were based on the slightly modified nozzle geometry than the one developed within the Muon Accelerator Program (MAP) \cite{Mokhov2011,Mokhov2012}.
Although the MAP nozzle was optimized for a center-of-mass energy of 1.5~TeV, it has been used as a starting point for the first 10~TeV studies (see, for example, Refs.~\cite{Calzolari2022,Calzolari2023}). 

\begin{figure}[!h]
    \centering
    \includegraphics[width=0.6\linewidth]{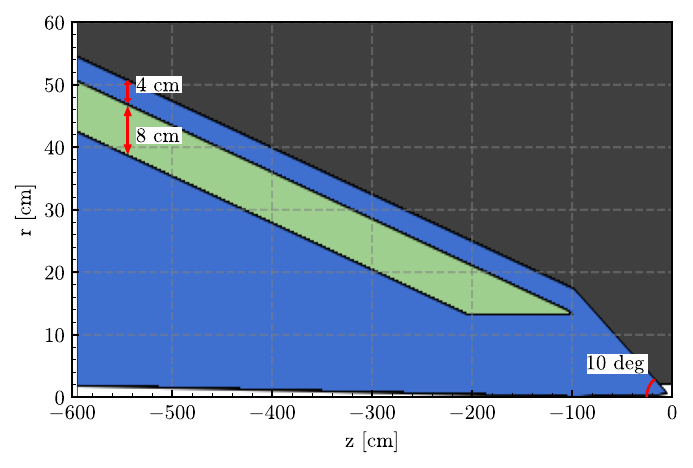}
    \caption{Left nozzle geometry dimensions. The blue layer is made of INERMET180 (registered trademark), a heavy tungsten alloy, while the green one is composed of borated polyethylene.}
    \label{mdi:fig:nozzle_geom}
\end{figure}

\begin{table}[!h]
    \centering
    \begin{tabular}{l|l}
         z [cm]& r [cm]\\ \hline
         \multicolumn{2}{l}{\textbf{Outer surface of nozzle}}\\
         595& 55\\
         100& 17.57\\
         6& 1\\ 
         \multicolumn{2}{l}{\textbf{Outer surface of the borated polyethylene layer}}\\
         595& 51\\
         100& 13.57\\
         \multicolumn{2}{l}{\textbf{Inner surface of the borated polyethylene layer}}\\
         595& 43\\
         204.49&13.47\\
         100&13.47\\
         \multicolumn{2}{l}{\textbf{Inner aperture of the nozzle}}\\
         595&1.78\\
         100&0.3\\
         15&0.6\\
         6&1\\
    \end{tabular}
    \caption{Nozzle Dimensions}
    \label{mdi:tab:nozzle_geom}
\end{table}

Figure~\ref{mdi:fig:nozzle_geom} illustrates the modified MAP nozzle geometry in the $z-r$ plane, where $z$ is the beam axis and $r$ is the radial coordinate.
The nozzle is assumed to have azimuthal symmetry around the $z$-axis.
The figure shows only the nozzle on the left side of the IP; the second nozzle has the same shape but is mirrored with respect to the interaction point.
The nozzle is assumed to consist mainly of INERMET180 (registered trademark), a tungsten-based alloy (blue color), with a layer of borated polyethylene on the outer surface (green color) to thermalize and absorb neutrons before they reach the detector.
Using a tungsten alloy (instead of pure tungsten) is required to allow the manufacture of such shielding elements, however such a choice reduces slightly the shielding effectiveness of the nozzle due to the lower material density.
The beam pipe connecting the two opposite nozzles is made of beryllium, with an internal radius of \SI{2.3}{\centi\meter} and a thickness of \SI{1}{\milli\meter}.

The nozzle tip is located at a distance of \SI{6}{\centi\meter} from the IP. The inner aperture of the nozzle features three different angles, with an aperture bottleneck at \SI{100}{\centi\meter} from the IP. In the region between \SI{100}{\centi\meter} and the first magnet at \SI{600}{\centi\meter}, the inner nozzle surface increases and is defined by the required beam clearance to avoid direct halo losses on the aperture.
The outer surface of the nozzle follows a conical shape, with two different angles.
Near the interaction point, the inclination amounts to 10 degrees, which determines the angular acceptance of the detector.
All the space outside the nozzle and the central beam pipe can be occupied by the detector.
The present setup is of conceptual nature, without yet considering engineering aspects or a possible support structure for the nozzle.

Table \ref{mdi:tab:nozzle_geom} summarizes the coordinates of the inner aperture and outer surface of the nozzle, respectively.
Table \ref{mdi:tab:nozzle_elements} provides the material components of the nozzle.

\begin{table}[!h]
    \centering
    \begin{tabular}{cccc} \hline
         Component&  Density [g/cm3]&  Element& Atomic Fraction (mass fraction if negative)\\ \hline
         EM Shower Absorber&  18&  W& -0.95\\
         &  &  Ni& -0.035\\
         &  &  Cu& -0.015\\ \hline
         Neutron Absorber&  0.918&  H& 0.5\\
         &  &  C& 0.25\\
         &  &  B& 0.25\\ \hline
    \end{tabular}
    \caption{Material composition of nozzle}
    \label{mdi:tab:nozzle_elements}
\end{table}

\subsection{Beam-induced background}
\label{mdi:sec:bib}
The number of background particles entering the detector per bunch crossing depends on the nozzle geometry, the nozzle material composition and the interaction region layout. Table~\ref{tab:MDI_bib_particles} summarizes the number of secondary electrons, positrons, photons and neutrons reaching the detector in a 10~TeV muon collider. The numbers were obtained with FLUKA Monte Carlo simulations, considering the nozzle introduced in the previous section. The bunch intensity was assumed to be 1.8$\times$10$^{12}$ muons. Only secondary particles with energies above a given threshold value were considered (see Table~\ref{tab:MDI_particle_thresholds}).

\begin{table}[ht]
\centering
\caption[Number of secondary particles entering the detector volume]{Number of secondary particles (muon decay) entering the detector volume (10~TeV). Only particles above the threshold values in Table~\ref{tab:MDI_particle_thresholds} were included. The multiplicities include only the contribution of one beam and correspond to one bunch crossing.}
\label{tab:MDI_bib_particles}
\begin{tabular}{lc}
\textbf{Particle type} & \textbf{Particles entering detector} \\ \hline
Photons             & \SI{1.0e8}{}       \\
Neutrons            & \SI{1.1e8}{}       \\
Electron/positrons  & \SI{1.2e6}{}       \\
Muons               & \SI{1.1e4}{}       \\
Charged hadrons     & \SI{4.0e4}{}       \\
\end{tabular}
\
\end{table}

\begin{table}[h]
\begin{center}
\caption{Particle production and transport thresholds assumed in the background simulations.}
\label{tab:MDI_particle_thresholds}
\begin{tabular}{lc}
\textbf{Particle type} & \textbf{Threshold} \\
\hline
Electrons, positrons and photons & 100 keV \\
Hadrons and muons & 100 keV \\
Neutrons & 0.01 meV \\
\end{tabular}
\end{center}
\end{table}

The number of background particles presented in this section includes only the contribution from muon decay, which is expected to be the dominant source of beam-induced background. Other background sources can include muon halo losses on the aperture and incoherent electron-positron pair production.

\subsection{Ionizing dose and displacement damage in detector}
\label{mdi:sec:dose}
To evaluate the cumulative radiation damage in detector equipment, two quantities have been considered: the total ionizing dose and the 1~MeV neutron-equivalent fluence in Silicon. The former is a measure for the radiation damage in organic materials and compounds, while the latter is related to the displacement damage. 
The studies assumed a CLIC-like detector and the nozzle described above. 
The results for the vertex detector, the inner tracker, as well as the electromagnetic calorimeter are presented in Table~\ref{mdi:tab:raddamagedetector} and correspond to one year of operation, assuming 1.2$\times$10$^7$ seconds of operation (139~days). The studies considered only muon decay, while neglecting the contribution of collision products and beam halo losses. The results were computed for IR lattice version 0.8.

\begin{table}[!h]
    \centering
        \caption[Ionizing dose and fluence in a CLIC-like detector]{Maximum values of the ionizing dose and the 1~MeV neutron-equivalent fluence (Si) in a CLIC-like detector. All values are per year of operation (10~TeV) and include only the contribution of muon decay.}
    \label{mdi:tab:raddamagedetector}
    \begin{tabular}{cccc} \hline
         &   Radius &Dose & 1 MeV fluence (Si) \\
 Unit& cm& kGy&$10^{14}$ n/\si{\centi\meter\squared}\\ \hline
         Vertex detector&   3&1000& 10\\
         Inner tracker&   12&70& 15\\
         ECAL&   150&2& 2\\ \hline
    \end{tabular}
\end{table} \clearpage
\section{Detectors}
\label{detector:sec}

The design of the detector for $\sqrt{s}=10$ TeV follows the concept already developed for $\sqrt{s}=3$ TeV with modifications to account for the higher energy. Two distinct detector concepts are presented, MAIA (Muon Accelerator Instrumented Apparatus) and MUSIC (MUon System for Interesting Collisions), to fully exploit the two interaction points of the collider.
Both designs share a similar structure, a cylinder $11.4$ m long with a diameter of $12.8$ m. The main detector components are:
\begin{itemize}
    \item Tracking system
    \item Electromagnetic calorimeter (ECAL)
    \item Hadron calorimeter (HCAL)
    \item A superconducting solenoid 
    \item A muon sub-detector 
\end{itemize}
The origin of the space coordinates is the beam interaction point. The z-axis has direction parallel to the beam pipe, the y-axis is parallel to gravity acceleration and the x-axis is defined as perpendicular to the y and z axes.

Table~\ref{tab:detector_general} summarises the detector parameters sub-system by sub-system for the two concepts. While the tracking system has a similar structure, the MAIA detector has the solenoid just after the tracker, before the ECAL while MUSIC places the solenoid magnet between ECAL and HCAL.

\begin{table}[ht]
\centering
\resizebox{\textwidth}{!}{
\begin{tabular}{l|ccccccc}
\textbf{Detector}& \textbf{R\_min}& \textbf{R\_max}& \textbf{|Z|\_min}& \textbf{|Z|\_max}& \textbf{Ang. Acc.}& \textbf{X / X0} & \textbf{L / L0} \\
\textbf{ MAIA / MUSIC} & \si{\milli\meter}& \si{\milli\meter}& \si{\milli\meter}& \si{\milli\meter}& \si{\degree}& &\\
\hline
\textbf{Inner Trackers} & 30 / 28 & 1500 & 0 & 2300 & 10 -- 170 & 0.1 to 0.3 & 0.04 to 0.1 \\
\hline
\textbf{EM Cal}& & & & & & & \\
Barrel & 1857 / 1690 & 2124 / 1960 & 0 & 2307 / 2210 & 10 -- 170 & 42 / 38 & 1.9 / 1.7 \\
Endcap & 310 & 2124 / 1960 & 2307 & 2577 & & 40 / 33 & 1.8 / 1.4 \\
\hline
\textbf{HAD Cal}& & & & & & & \\
Barrel & 2125 / 2902 & 4112 / 4756 & 0 & 2575 / 2609 & 10 -- 170 & 100 / 89 & 10.9 / 9.5 \\
Endcap & 307 & 4112 / 4756 & 2575 & 4562 / 4434 & & 114 / 116 & 12.3 / 12.5 \\
\hline
\textbf{Muon Systems} & & & & & & & \\
Barrel & 4150 / 4806 & 7150 / 6800 & 0 & 4565 / 4444 & 10 -- 170 & -- & -- \\
Endcap & 446 & 7150 / 6800 & 4565 / 4444 & 6025 / 5903 & & -- & -- \\
\hline
\textbf{Solenoid} & 1500 / 2055 & 1857 / 2862 & --& 2307 / 2509 & -- & 6 / 18 & 1.4 / 2.7 \\
\hline
\textbf{Nozzle} & 10 & 55 & --& 5950 & 0.2 -- 10 & -- & -- \\
\end{tabular}
}
\caption[Detector parameters for MAIA and MUSIC concepts]{Detector parameters for MAIA and MUSIC concepts. Values that are left empty ("--") are not relevant for the specific detector. X/X0 and L/L0 are for a particle traveling from the IP.}
\label{tab:detector_general}
\end{table}

\subsection{Tracking System}
\label{sec:trackingsistem}
The tracking detector is composed of the vertex and tracker sub-detectors, both of them structured in barrels and end-caps. The barrels are cylindrical surfaces with variable lengths and radii, whose axes coincide with the beam pipe and cover the central part of the detector.
The endcaps are annuli centered on the z axis, with variable distance from the interaction point and radii which cover the forward part of the detector.
The major characteristics of this sub-system are described in Table~\ref{det:tab:tracker}.\\
The vertex detector is close to the interaction point in order to allow a good resolution on track impact parameter. The building blocks of the barrel detection layers are rectangular staves of sensors, arranged to form a cylinder, while the endcaps are constituted by trapezoidal modules of sensors, arranged as "petals" to form a disk.
The MAIA detector has 5 layers, with the first two structured as a double layer, while MUSIC has 5 distinct layers. The length of the MUSIC barrel is 26 cm, which is double that of MAIA.

The barrel layers have silicon pixels of size $25 \times 25$ \textmu m$^2$, and thickness $50$ \textmu m. 
The eight endcaps layers, four for each side of the interaction point are composed of silicon pixels of size $25 \times 25$ \textmu m$^2$ and thickness 50\,\textmu m and 16 modules. 

The inner and outer trackers are based on the same technology for MAIA and MUSIC, single layer of silicon sensors of 100 \textmu m thickness.

Strips on the barrels are oriented with the long side parallel to the beam axis while the end-caps are composed of radial modules composed by rectangular pads. 

\begin{table}[ht]
\centering
\resizebox{\textwidth}{!}{
\begin{tabular}{l|cccccccc}
\textbf{Sub-Detector} & \textbf{Technology} & \textbf{\# Layers} & \textbf{"Cell"} & \textbf{Sensor} & \textbf{Hit Time} & \textbf{Signal Time} & \textbf{Max} & \textbf{Max Fluence} \\
 \textbf{MAIA/MUSIC} & & \textbf{/Rings} & \textbf{Size} & \textbf{Thickness} & \textbf{Resolution} & \textbf{Window} & \textbf{Dose} & \textbf{1 MeV (Si)}\\
 Units & & & \si{\micro\meter\squared}& \si{\micro\meter}& \si{\pico\second}& \si{\nano\second}& \si{\kilo\gray}& 10$^{14}$ n \si[per-mode=reciprocal]{\per\centi\meter\squared} \\ \hline
Vertex Barrel & Pixels & 4*/5 & 25 x 25 & 50 & 30 & [-0.18, 15.0] & 1000 & 10 \\
Vertex Endcap & Pixels & 4 & 25 x 25 & 50 & 30 & [-0.18, 15.0] & 1000 & 10 \\
Inner Barrel & Macro-Pixels & 3 & 50 x 1000 & 100 & 60 & [-0.36, 15.0] & 70 & 15 \\
Inner Endcap & Macro-Pixels & 7 & 50 x 1000 & 100 & 60 & [-0.36, 15.0] & 70 & 15 \\
Outer Barrel & Macro-Pixels & 3 & 50 x 10000 & 100 & 60 & [-0.36, 15.0] & < 70 & -- \\
Outer Endcap & Macro-Pixels & 4 & 50 x 10000 & 100 & 60 & [-0.36, 15.0] & < 70 & -- \\
\end{tabular}
}
\caption[Specifications for MAIA and MUSIC Tracker Sub-Detectors]{Specifications for MAIA and MUSIC Tracker Sub-Detectors. * The first layer is a double-layer with a 2mm gap.}
\label{det:tab:tracker}
\end{table}

\subsection{Calorimeter System}
The calorimeter system is composed of the electromagnetic and hadron sub-detectors. A summary of the main characteristics are in Table~\ref{det:tab:calo}.

\begin{table}[ht]
\centering
\resizebox{\textwidth}{!}{
\begin{tabular}{l|cccccccc}
\textbf{Sub-Detector}& \textbf{Technology} & \textbf{Cell} & \textbf{\# Longitudinal } & \textbf{Time} & \textbf{Integration} & \textbf{Signal Time} & \textbf{Max} & \textbf{Max Fluence} \\
\textbf{MAIA / MUSIC}& \textbf{} & \textbf{Size} & \textbf{Slices} & \textbf{Resolution} & \textbf{Time} & \textbf{Window} & \textbf{Dose} & \textbf{1 MeV (Si)}\\
 Units & & \si{\milli\meter\squared} & & \si{\pico\second} & \si{\nano\second} & \si{\nano\second} & \si{\kilo\gray} & 10$^{14}$ n \si[per-mode=reciprocal]{\per\centi\meter\squared} \\\hline
\textbf{EM Cal - Barrel} & W+Si / Crystal & 5 x 5 & 50 / 6 & /50 & /25 & [-0.25, 10] & 2 & 2 \\
\textbf{EM Cal - Endcap} & W+Si / Crystal & 5 x 5 & 50 / 6 & /50 & /25 & [-0.25, 10] & 2 & 2 \\
\textbf{HAD Cal - Barrel} & Iron + Scint. & 30 x 30 & 75 / 70 &  --&  --& [-0.25, 10] &  --&  --\\
\textbf{HAD Cal - Endcap} & Iron + Scint. & 30 x 30 & 75 / 70 &  --&  --& [-0.25, 10] &  --&  --\\
\end{tabular}
}
\caption{Electromagnetic and Hadronic Calorimeters parameters for MAIA and MUSIC.}
\label{det:tab:calo}
\end{table}

The MAIA ECAL configuration is inspired by CLIC. It consists of a dodecagonal barrel and two endcap systems. It is composed of 40 interlaced layer of Tungsten as absorber material $2.2$ mm thick and Si sensor as active material with $5 \times 5$ mm$^{2}$ silicon detector cells. It is located outside of the superconducting solenoid.

The MUSIC ECAL, has the same shape of MAIA, but is positioned immediately after the tracking system and within the superconducting solenoid. It is a semi-homogeneous calorimeter based on Lead Fluoride ($PbF_2$) crystals read out by surface mounted ultraviolet extended Silicon Photomultipliers. It represents a modern design approach that aims to combine the intrinsic high-energy resolution of homogeneous calorimeters with the longitudinal segmentation typically found in sampling calorimeters.

MAIA and MUSIC currently share the same technology for HCAL. It consists of a dodecagonal barrel and two endcap systems, structured in 60 interlaced layers of iron absorber 20 mm thick and plastic scintillating tiles with cell size $30 \times 30 $ mm$^2$. It allows the reconstruction of hadronic jets and helps in particle identification, to separate hadrons from leptons and photons. 

The characteristics of the superconducting solenoid are reported in table~\ref{det:tab:magnet}
\begin{table}[ht]
\centering
\begin{tabular}{cccc}
\textbf{B [T]} & \textbf{Thickness [mm]} & \textbf{max |z| [m]} & \textbf{Bore Radius [m]} \\
\hline
5 & 356 / 393 & 2307 / 2509 & 1680 / 2459 \\
\end{tabular}
\caption{Magnetic field of both detector concepts, thickness of the coil for MUSIC, and dimensions.}
\label{det:tab:magnet}
\end{table}

\subsection{Muon System}
The current configuration of the two detector concepts does not include a magnetic field outside the calorimetric system, so the role of the muon detector must be reconsidered. In particular, for high-energy muons, new methods based on machine learning, which combine tracking detector and calorimeter information, could be employed. In this case, the muon detector would primarily serve to identify that the particle is a muon.

 \clearpage
\section{Magnets}
\label{mag:sec}

Here we provide a summary of the magnet parameters for the study so far.

\subsection{Magnet Needs and Challenges}
\label{mag:sec:needs}
The short muon lifetime (2.2 \textmu s at rest) and production of bright muon beams results in a unique set of demands for magnet technologies, including large-bore high-field solenoids, dipoles and quadrupoles, compact ultra-high-field solenoids, and very fast-ramping dipoles.
Activities within the scope of the IMCC has led to the most advanced set of main magnet conceptual designs and performance parameters.
These parameters are an evolution of previous studies, in particular the U.S. Muon Accelerator Program (MAP) \cite{palmer2015}, extending the performance space by considering recent advances in magnet technology.

First an overview of the magnet options is provided, then four key sub-sections of the accelerator complex are addressed, with corresponding demands in terms of magnet performance.
\begin{enumerate}
    \item Front-end: Target solenoids in Section \ref{mag:sec:front}.
    \item Cooling: HTS 6D solenoids and high-T final cooling solenoids in Section \ref{mag:sec:cool}.
    \item Acceleration: Rapid-cycling and hybrid-cycling dipoles in Section \ref{mag:sec:acc}.
    \item Collider: Dipoles in Section \ref{mag:sec:col}.
\end{enumerate}

\subsection{Magnet Studies and Technology Options}
\label{mag:sec:tech}
The main performance targets and target ranges (i.e., not yet to specification) of the most challenging magnets of the muon collider are shown Table \ref{mag:tab:developments}.
Though these targets are bound to adapt as the study proceeds, they already provide a good basis to feedback on beam optics and accelerator performance, and to identify outstanding issues to be addressed by future work and dedicated R\&D. 
The whole accelerator complex functions in steady state, apart from the fast ramped magnets in the rapid cycling synchrotrons.

\begin{table}[!h]
    \centering
    \resizebox{\textwidth}{!}{
    \begin{tabular}{l|cccccccc}
         Complex&  Magnet &No.&  Aper.&  Length&  Field&Grad.&  Ramp rate&  Temp.\\
 Unit& & & [mm]& [m]& [T]&[T/m]& [T/s]&[K]\\ \hline
         Target, capture&  Solenoid Coils &23&  1380&   $\approx$ 0.4 -- 0.8&  2 -- 20 &&  SS&  20\\
         6D cooling&  Solenoid Coils & $\approx$ 6000&  90-1500&  0.08 -- 0.5&  2 -- 17 &&  SS&  4.2-20\\
 Final cooling&  Solenoid Coils &14& 50& 0.5& >40 && SS&4.2\\
 RCS&  NC dipole & $\approx$ 1500& 30x100& 5& $\pm$ 1.8 && 4200&300\\
 &  SC dipole & $\approx$ 2500& 30x100& 1.5& 10 && SS&4.2-20\\
         Collider arc&  Dipoles& $\approx$ 1050&  140&  5&  14* &&  SS&  \\
         &  CF& $\approx$ 628&  140&  5 -- 10&  4 -- 8&$\pm$100--$\pm$150*&  SS&  4.2-20\\
         IR &  quadrupoles&  $\approx$ 20&  100 - 280&  5 -- 10&  &$\pm$110 -- $\pm$330**&  SS&  4.2-20\\
    \end{tabular}
    }
    \caption[Summary of main magnet development targets]{Summary of main magnet development targets. For the collider magnet values marked with a * slightly higher values are assumed in the lattice design but no important changes are expected adjusting to the specified performances. The values marked with ** correspond to the lattice design but might be too high for the magnets; the lattice design will be updated accordingly. Specific configurations still need to be evaluated and this is a work in progress. CF stands for combined-function magnets.}
    \label{mag:tab:developments}
\end{table}

\subsection{Front End (muon production and capture)}
\label{mag:sec:front}
The details for the current target and front-end parameters are shown in Section \ref{target:sec:target} and \ref{front:sec}.\\
The target for muon production is inserted in a steady-state, high field solenoid which has outer dimension in the range of \num{150} to \SI{250}{\milli\meter}, depending on technology.
It captures the pions and guides them into a decay and capture channel, also embedded in solenoid magnets.
To maximize capture efficiency, the magnetic field profile along the axis of the channel needs to have a specific shape, with peak field of \SI{20}{\tesla} on the target, and an adiabatic decay to approximately \SI{1.5}{\tesla} at the exit of the channel, over a total length of approximately \SI{18}{\meter}.

Besides the high field values, another challenge derives from the radiation environment due to the interaction of the multi-MW proton beam with the target. The radiation requirements for this system are explained further in Section \ref{rad:sec}. A large bore dimension implies high stored magnetic energy, which in turn affects electromagnetic forces, magnet protection, and cost as we will discuss later. 

\subsubsection{Target solenoid}
\label{mag:sec:target}
Following recent advances in HTS magnets for fusion \cite{MITcable} \cite{MITpress} we have proposed a configuration based on an HTS cable operated at \SI{20}{\kelvin} \cite{botturaAPS2023} \cite{portoneMT28}.
The analysis performed so far shows that it is possible to eliminate the resistive insert and reduce the magnet bore to \SI{1380}{\milli\meter}, almost half of that of the US-MAP LTS coil, still producing the desired field profile for muon capture efficiency.
Operation at temperature higher than liquid helium reduces the need to shield the radiation heat, maintaining good overall energy efficiency.
The proposed system has a stored energy of $\approx$ \SI{1}{\giga\joule}, a coil mass of $\approx$ \SI{100}{\tonne} and wall-plug power consumption of $\approx$ \SI{1}{\mega\watt}, i.e.~a considerable reduction with respect to the hybrid solution proposed earlier.

\subsection{Cooling}
\label{mag:sec:cool}
The overview of the cooling system parameters are in Section \ref{cool:sec}, which factors in our evolving understanding of acceptable solenoid parameter limits. We are presently performing analysis and optimization on this latest configuration.

To the first order, the final emittance of the muon beam is inversely proportional to the strength of the final cooling solenoids.
The design study from MAP was based on a \SI{30}{\tesla} final cooling solenoid, and demonstrated that an emittance roughly a factor of two greater than the transverse emittance goal can be achieved \cite{Stratakis2015}.
Other studies \cite{palmer2011muon} show that fields in the range of \SI{50}{\tesla} improve the final emittance requirements and offer further gains in beam brightness.
To improve upon these results, we are considering an HTS final cooling solenoid with the potential to reach an excess of \SI{40}{\tesla}.

\subsubsection{6D Cooling solenoids}
\label{mag:sec:6d_cool}

\begin{table}[!h]
\begin{center}
\begin{tabularx}{\linewidth}{lXX|XXXXXX}
\hline\hline
Cell & $E_\text{Mag}$ & $e_\text{Mag}$ & Coil & \textbf{$J_E$} & $B_\text{peak}$   & $\sigma_\text{Hoop}$ (Max.)  &$\sigma_\text{Radial}$ (Min.) &$\sigma_\text{Radial}$ (Max.)  \\

& (MJ) & (MJ/m$^3$) & & (A/mm$^2$) & (T)  & (MPa) & (MPa) & (MPa) \\
\hline\hline
A1  & 5.4   & 21     & A1-1   & 57.6  & 5.2   & 42   & -8  & 0 \\
A2  & 22.1  & 106.1  & A2-1   & 149.5 & 11.6  & 194 & -48 & 0 \\
A3  & 5.0   & 49.5   & A3-1   & 131.5 & 10.1  & 121 & -25 & 0 \\
A4  & 8.0   & 92.3   & A4-1   & 193.2 & 13.8  & 225 & -51 & 1 \\
B1  & 9.1   & 49.8   & B1-1   & 96.9  & 7.7   & 104 & -24 & 0 \\
B2  & 15.6  & 64.2   & B2-1   & 102.1 & 9.2   & 131 & -32 & 0 \\
B3  & 36.9  & 105.9  & B3-1   & 127.9 & 12.9  & 208 & -57 & 0 \\
B4  & 75.6  & 149.9  & B4-1   & 88.5  & 16.1  & 260 & -1  & 29 \\
B5  & 17.3  & 88.9   & B5-1   & 179.6 & 14.7  & 295 & -2  & 17 \\
B5  &       &        & B5-2   & 154.0 & 14.7  & 212 & -57 & 1 \\
B6  & 8.3   & 96.6   & B6-1   & 214.4 & 15.3  & 339 & -5  & 18 \\
B6  &       &        & B6-2   & 211.5 & 12.0  & 214 & -6  & 6 \\
B6  &       &        & B6-3   & 212.7 & 12.4  & 162 & -46 & 0 \\
B7  & 8.2   & 87.7   & B7-1   & 183.3 & 14.7  & 264 &  0  & 25 \\
B7  &       &        & B7-2   & 153.9 & 11.1  & 175 & -4  & 10 \\
B7  &       &        & B7-3   & 210.3 & 13.2  & 180 & -45 & 1 \\
B8  & 8.8   & 92.1   & B8-1   & 193.7 & 16.5  & 270 & -6  & 38 \\
B8  &       &        & B8-2   & 202.1 & 15.4  & 270 & -6  & 29 \\
B8  &       &        & B8-3   & 212.8 & 13.2  & 187 & -50 & 0 \\
B9  & 7.5   & 76.5   & B9-1   & 256.4 & 17.2  & 281 &  0  & 37 \\
B9  &       &        & B9-2   & 88.4  & 10.0  & 95  & -2  & 12 \\
B9  &       &        & B9-3   & 204.9 & 13.2  & 184 & -46 & 0 \\
B10 & 5.0   & 68.6   & B10-1  & 326.8 & 19.2  & 378 &  0  & 49 \\
B10 &       &        & B10-2  & 146.1 & 11.1  & 105 & -4  & 13 \\
B10 &       &        & B10-3  & 207.8 & 12.5  & 158 & -43 & 1 \\
\end{tabularx}
\end{center}
\caption[Solenoid types in the latest 6D cooling optics]{Table of various parameters for 14 cell types and 26 unique solenoid types in the latest 6D cooling optics \cite{zhu2024performance}. Values correspond to solenoids operating in their respective cells within a lattice. Note that if the solenoid is operating stand-alone or in a single cell, some parameters take on higher or lower values.}
\label{tab:rectilinear_magnet}
\end{table}


In the current configuration, a total of 3054 solenoids are spread over a 0.85 km distance. There are 14 unique cell types, and 26 unique solenoid types.
During the beam dynamics studies, we integrated a magnet design guide to constrain allowable magnet geometries and current densities based on key solenoid parameters (stresses $\sigma$, stored magnetic energy $e_m$, critical current density $J_c$).
To assess limits on these properties, we use HTS (ReBCO) from Fujikura FESC-SCH tape as a reference \cite{Fujikura}.
The parameters and limits implemented (considering a single solenoid) are: hoop stress, $\sigma_{\theta}< 300$ MPa; radial tensile stress, $\sigma_{r}< 20$ MPa; and stored magnetic energy density, $e_m< 150$ MJ/m$^3$.
Additionally, we constrained the current density to not exceed the critical current density based on a large dataset of $J_c$ measurements \cite{fujita2019flux}, taking HTS operating at 20 K with 2.5 K margin.
We report in Tab. \ref{tab:rectilinear_magnet} main parameters of each cooling cell type and unique solenoid type.

Observing Tab. \ref{tab:rectilinear_magnet}, we find some solenoids exceed allowed design limits, primarily in terms of large hoop stresses (B6-1, B10-1) and tensile radial stresses (B4-1, B7-1, B8-1, B8-2, B9-1, B10-1).
The most problematic solenoid is B10-1, with a hoop stress of 378 MPa, tensile radial stress of 49 MPa, and peak field on the coil of 19.2 T which corresponds to it exceeding its $J_c$ by 114\%.
Importantly though, most of the solenoids are within or near the allowed design limits demonstrating the success of the iteration of design parameters with beam optics to produce an initial set of solenoids.


\subsubsection{Final Cooling solenoid}
\label{mag:sec:fc_cool}
A total of 17 final cooling cells were part of the scheme devised by US-MAP to achieve minimum beam emittance, with bore field up to \SI{30}{\tesla}.
To improve upon the results obtained by US-MAP we are considering for the final cooling a solenoid design with the potential to reach and exceed \SI{40}{\tesla}, a clear bore of \SI{50}{\milli\meter}, a magnet length of \SI{500}{\milli\meter}, and sufficiently compact in size as required for an accelerator magnet (considerations of mass, footprint, and cost) \cite{Bordini2024}.
The operating current density targeted is high, \SI{650}{\ampere\per\milli\meter\squared}, to reduce the coil size, as well as the forces and stored energy. The coil size is exceptionally small, with a \SI{90}{\milli\meter} outer radius.

The mechanics of the final cooling solenoid is designed to achieve a maximum hoop stress of \SI{650}{\mega\pascal} and no tensile stress in any condition experienced by the coil. To this aim, the wound and soldered pancakes are loaded in radial direction by a stiff external ring that introduces a radial pre-compression of \SI{200}{\mega\pascal}, at room temperature. The radial pre-compression is chosen to nearly balance the outward electro-magnetic stress at \SI{40}{\tesla}.

For the transverse resistance, our goal is to achieve quench protection through a low transverse resistance (possibly with means to actively trigger quench), while at the same time allowing full ramp in less than 6 hours, as well as field stability at flat-top better than 10 ppm/s.


\subsection{Acceleration}
\label{mag:sec:acc}
An overview of the high-energy accelerator parameters is in Section \ref{high:sec}.
In the present baseline, the NC dipoles in the first RCS need to sweep from \num{0.36} to \SI{1.8}{\tesla} within \SI{0.35}{\milli\second} (i.e. a rate of \SI{4}{\kilo\tesla\per\second}).
In the last HCS the NC dipoles swing from \SI{-1.8}{\tesla} to \SI{1.8}{\tesla}, in \SI{6.37}{\milli\second} (i.e. a rate of about \SI{560}{\tesla\per\second}).

Design concepts of NC fast ramped magnets were developed by US-MAP, for peak operating field of \SI{1.5}{\tesla} \cite{berg2016pulsed}.
SC dipoles for HCS were not yet studied in detail, besides setting target values for bore field and magnet length.
Beyond magnet engineering, the primary challenge of an accelerator ring of the required dimension is that the stored energy is of the order of several tens of \si{\mega\joule}.
Powering at a high-pulse rate with good energy recovery efficiency between pulses will require mastery in the management of peak power in the range of tens of \si{\giga\watt}.
Resonant circuits combined with energy storage systems seem to be the only viable solution.
A high energy storage density and high quality factor are mandatory to limit foot-print, energy consumption, capital and operating cost.

\subsubsection{Synchrotrons (RCS and HCS) dipoles}
\label{mag:sec:acc_dip}
A lower bound for the stored energy is the magnetic energy in the beam aperture, a nominal 30 mm (gap) x 100 mm (width). To limit saturation, affecting losses and field quality, we have taken an upper design field limit of 1.8 T for the resistive magnets. This corresponds to a magnetic energy of 3.9 kJ/m in the beam aperture, while the energy stored in the magnet will be forcibly higher. The analysis of several resistive magnet configurations, of different iron cross section and materials, coil design and current density, shows that the lowest magnet stored energy is in the range of 5.4 kJ/m, a factor 1.4 higher than the magnetic energy in the beam aperture, quoted above \cite{Breschi2024}. A second issue is the magnitude of the resistive, eddy current and hysteresis loss. This is the power drawn from the grid, and dissipated. A suitable target, though not yet settled, is in the range of 500 J/m per pulse. Among all configurations analyzed, we have found that the best compromise of stored energy, loss and field quality is obtained with a “H” and “Hourglass” shaped iron core \cite{Breschi2024}. These configurations will be retained for further magnetic analysis, including 3D and end effects. 


Finally, the design of the steady state superconducting magnets of the HCS’s is in development. Rejecting the cos-theta coil geometry due to its inefficiency for a rectangular aperture, our conceptual design focuses on flat racetrack coils which appear feasible to achieve a target field of around 10\,T. With HTS, this could operate at temperatures significantly above liquid helium (10\,to 20\,K), offering gains in efficiency. We are currently progressing with detailed magnetic and mechanical simulations of potential configurations which can satisfy the field quality requirements while minimizing cost and engineering complexity.

\subsection{Collider}
\label{mag:sec:col}
An overview of the collider parameters is in Section \ref{col:sec}, including the radiation shielding required for the head load due to muon decay.
To allow for a compact collider ring and maintain sufficient space for shielding, the ring and Interaction Region (IR) dipole and quadrupole magnets thus need to be high-field and large aperture.

It is assumed that the main arc magnets have combined functions (e.g. dipole/quadrupole and dipole/sextupole) and generate a steady-state magnetic field up to \SI{16}{\tesla} in a \SI{160}{\milli\meter} aperture.
The most recent optics requires dipole fields in the range of \SI{10}{\tesla} and gradients of \SI{300}{\tesla\per\meter}.
These field demands, combined with the aperture constraints, are presently only an initial evaluation, but they exceed practical limits of what is possible, and will require iteration.
For the IR quadrupole magnets the assumption from the optics studies is of a peak field of \SI{20}{\tesla}, also associated with large apertures, up to \SI{200}{\milli\meter}.

\subsubsection{Collider dipoles}
\label{mag:sec:col_dip}
\begin{figure}[h!]
    \centering
    \begin{minipage}{0.45\textwidth}
        \centering
        \includegraphics[width=\textwidth]{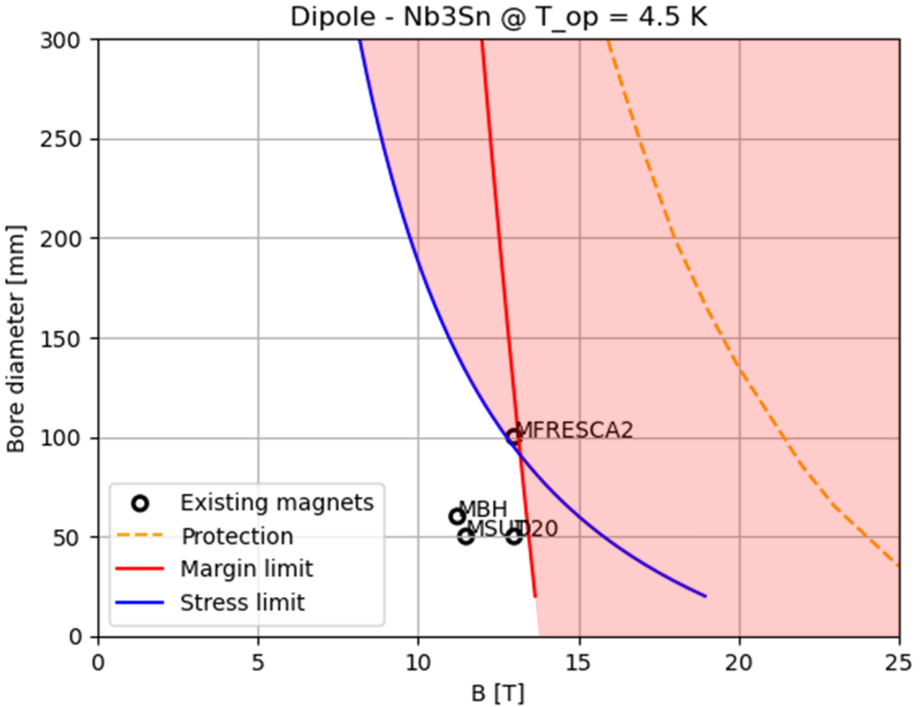}
        Dipole Nb$_3$Sn
    \end{minipage}
    \begin{minipage}{0.45\textwidth}
        \centering
        \includegraphics[width=\textwidth]{Chapters/N-Magnets/figures/AB_dipole_NbSn.png}
        Dipole ReBCO
    \end{minipage}

    \vspace{0.5cm} 

    \begin{minipage}{0.45\textwidth}
        \centering
        \includegraphics[width=\textwidth]{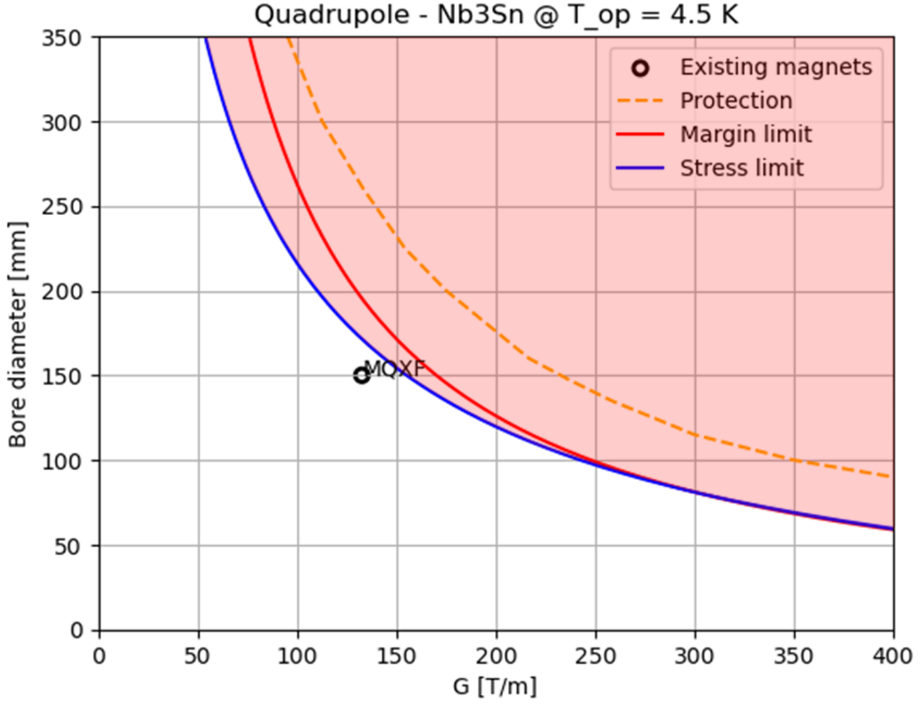}
        Quadrupole Nb$_3$Sn
    \end{minipage}
    \begin{minipage}{0.45\textwidth}
        \centering
        \includegraphics[width=\textwidth]{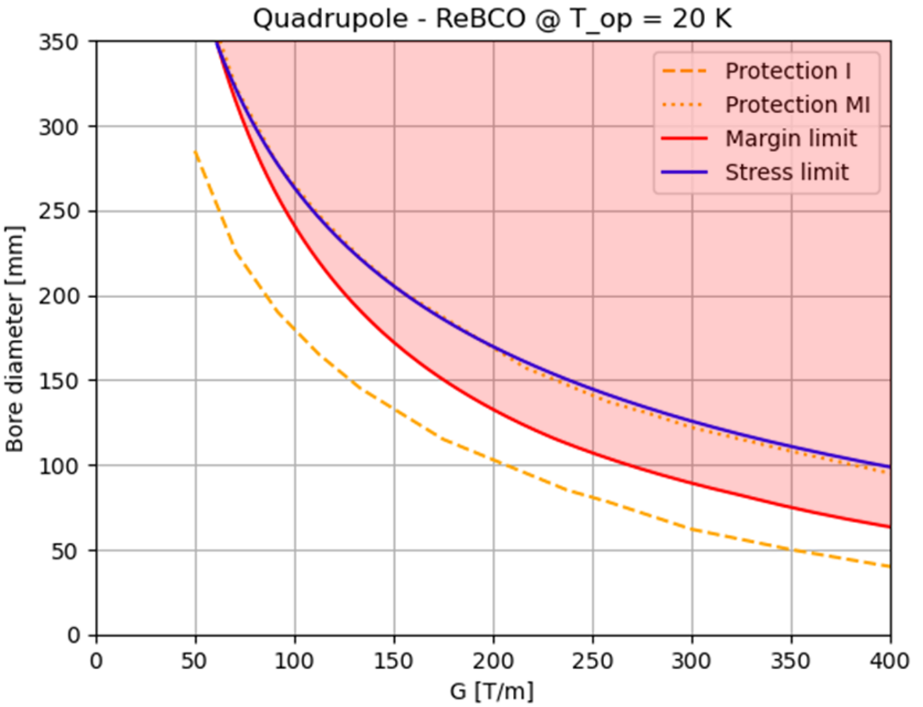}
        Quadrupole ReBCO
    \end{minipage}

    \caption{A-B plots for dipole (top) and quadrupole (bottom) collider magnets made of either Nb$_3$Sn (left) at \SI{4.5}{\kelvin} or ReBCO (right) at \SI{20}{\kelvin}. For ReBCO, the white region is the allowed area, assuming the magnet is metal-insulated~\cite{ColliderAB}.}
    \label{mag:fig:ColliderAB}
\end{figure}

Using analytical evaluations of operating margin, peak stress, hot-spot temperature, and magnet cost under the assumption of a sector coil geometry \cite{Novelli2023} we have produced design charts of maximum magnet aperture (A) vs. bore field (B), which is a form convenient for iterating with the beam optics. Such A-B charts are shown in Figure \ref{mag:fig:ColliderAB} for a choice of superconductor and operating point of Nb3Sn at 4.5 K and ReBCO at 20 K. 

For a 10 TeV collider, Nb-Ti at 1.9 K does not appear as a good solution because of low operating margin (recall the large energy deposition), as well as considerations of cryoplant efficiency and energy consumption. Similarly, Nb$_3$Sn at 4.5 K falls short of the required field performance for the arc magnets, being limited by peak stress and operating margin. It can provide feasible solutions only up to 14\,T, which can be considered for a 3 TeV MuC ($\sim$ 11 T, 150 mm aperture). Our initial evaluation of ReBCO shows that also in this case the available design space does not match the required performance. For ReBCO, however, operating margin is not an issue, and operation in the range of 10 K to 20 K could be envisaged. The main limitations come rather from the cost of the superconductor, and from quench protection. Cost considerations drive the current density in an all-HTS coil towards high values, in the range of 800 to 1000 A/mm2, where standard detect-and-dump protection strategies are not sufficiently fast. It is hence clear that alternative protection schemes need to be devised to benefit from the large current carrying capacity and margin of present REBCO conductors. Provided that the cost per m of REBCO tape can be reduced by a factor three to four, relaxing the need for very high values of current density, we have found that a suitable design range for the arc magnets can be defined using two points, from a nominal aperture of 140 mm at reduced bore field of 14 T, up to nominal bore field of 16 T but reduced aperture of 100 mm. The whole range can be achieved with REBCO at 4.5 K and 20 K, while the low field range can be reached also with Nb3Sn at 4.5 K, thus providing at least two technology options.

The semi-analytic tool \cite{Novelli2023} discussed in this section has been very successful to provide quick feedback and iteration with beam dynamics, energy deposition and cryogenics team during this design stage. Currently, this tool is being adapted to consider quadrupole performance limits (for the design of the interaction regions) and combined function magnets, as required because of neutrino flux mitigation. Upcoming work will focus on more detailed magnetic and mechanical designs of ARC dipoles and IR quadrupoles. \clearpage
\section{RF}
\label{rf:sec}
The RF parameters which should be considered in the design are listed in Table \ref{rf:tab:RFparameters}.
\begin{table}[h]
\begin{center}
\caption{RF frequencies and gradients to be used in the beam dynamics studies.}
\label{rf:tab:RFparameters}
\begin{tabular}{lcccccc}
\hline\hline
\textbf{Proton driver}\\
\hline
Linac\\
\hline
RF frequencies & MHz & 352 & 704\\
\hline
\textbf{Muon cooling complex}\\
\hline
6D-cooling channels\\
\hline
RF frequencies & MHz & 352 & 704 & 1056\\
Maximum accelerating field in cavity (conservative) & MV/m & 22 & 30 & 30 \\
Maximum accelerating field in cavity (optimistic) & MV/m & 35 & 50 & 50 \\
\hline
\textbf{Acceleration complex}\\
\hline
Linacs\\
\hline
RF frequencies & MHz & 352 & 704 & 1056\\
Maximum accelerating field in cavity (conservative) & MV/m & 20& 25& 30\\
Maximum accelerating field in cavity (optimistic) & MV/m & 30& 38& 45\\
\hline
RCSs\\
\hline
RF frequency & MHz & & 704 & 1056, 1300 \\
Maximum accelerating field in cavity (conservative) & MV/m & & 25 &  30 \\
Maximum accelerating field in cavity (optimistic) & MV/m & &  38 & 45 \\
\hline

\hline\hline
\end{tabular}
\end{center}
\end{table}

In the other sub-systems of the muon cooling complex: capture, bunch merge, final cooling, etc many different RF frequencies are necessary. It is recommended to keep these RF frequencies as high as reasonable possible from the beam dynamics point of view, since the size of the achievable gradient scales approximately as $\sqrt{(f_{RF})}$.

\subsection{RF systems for rectilinear cooling}
\label{rf:sec:cool}

The preliminary RF cavity design for each stage of the rectilinear cooling channel was developed based on the shape presented in \cite{Barbagallo:2024zak} following the beam dynamics specification in Table~\ref{cool:tab:6d_rf}. 
The other geometrical parameters characterizing the cavity shape are chosen to maximize the shunt impedance ($R/Q\cdot Q_{0}$) and reduce surface losses ($P_{\mathrm{diss}}$) on the windows and cavity walls.
The peak surface electric field ($E_{\mathrm{peak}}$) is also minimized to avoid RF breakdown risk. The RF cavity frequency ($f_{0}$), the cavity length ($L_{\mathrm{cav}}$), and the nominal RF gradient along the cavity axis ($E_{\mathrm{nom}}$) for the studied RF cavities are reported in Table~\ref{cool:tab:6d_rf}. Table~\ref{rf:tab:cool_cavity} summarizes the relevant RF figures of merit computed for the operating frequencies of the studied cavities. Most of the power is dissipated in the cavity walls. 

\begin{table}[!h]
    \centering
    \begin{tabular}{c|cccccccc}
          & $Q_0$& $t_f$& $DF$& R/Q&  ${P_{\mathrm{diss}}}$ & $\frac{{P_{\mathrm{diss,Be}}}}{{P_{\mathrm{diss}}}}$ & $E_{\mathrm{peak,Cu}}$&$E_{\mathrm{peak,Be}}$\\
         & $10^4$& \textmu s& $10^{-4}$& $\Omega$ & MW/cavity & - & MV/m & MV/m\\ \hline
         Stage A1& 3.06& 31.203& 1.17& 171.73& 4.25&0.377& 11.72&27.383\\
         Stage A2& 3.14& 32.087& 1.21& 149.68& 4.34&0.085& 23.249&26.511\\
         Stage A3& 2.20& 11.248& 0.43& 160.36& 2.06&0.201& 20.802&31.507\\
         Stage A4&  2.22&  11.345&  0.43&  150.21&  2.21& 0.085& 27.873&31.829\\
         Stage B1&  3.91&  39.954&  1.51&  183.70&  2.78& 0.23& 12.392&21.651\\
         Stage B2&  3.56&  36.323&  1.37&  170.47&  3.24& 0.164& 16.376&23.361\\
         Stage B3&  3.15&  32.148&  1.21&  141.27&  4.07& 0.031& 26.175&24.429\\
         Stage B4&  3.59&  36.71&  1.38&  154.02&  3.92& 0.009& 27.732&22.823\\
         Stage B5&  2.23&  11.366&  0.43&  140.85&  1.18& 0.026& 24.116&22.027\\
         Stage B6&  2.22&  11.36&  0.43&  137.40&  2.34& 0.007& 37.092&29.544\\
         Stage B7&  2.22&  11.354&  0.43&  136.87&  2.11& \num{3.08E-03} & 36.487&26.892\\
         Stage B8&  2.22&  11.347&  0.43&  137.39&  1.92& \num{8.32E-04} & 35.95&23.73\\
         Stage B9&  2.22&  11.344&  0.43&  138.11&  2.14& \num{3.16E-04} & 38.528&23.268\\
         Stage B10&  2.22&  11.342&  0.43&  139.06&  1.51& \num{1.56E-04} & 32.522&18.341\\
    \end{tabular}
    \caption{RF figures of merit for the RF cavities in the rectilinear cooling channel}
    \label{rf:tab:cool_cavity}
\end{table}

 The filling time $t_{\mathrm{f}}$, which is the time required to fill the cavity to the nominal voltage~$V_{\mathrm{nom}} = E_{\mathrm{nom}} L_{\mathrm{cav}}$, is given by:

\begin{equation}
 t_{\mathrm{f}} \approx \frac{2Q_{\mathrm{L}}}{\omega_{0}} \ln\left(\frac{2\beta_{\mathrm{c}}}{\beta_{\mathrm{c}} - 1}\right),
\label{eq:filling_time}
\end{equation}

\noindent
where $Q_{\mathrm{L}}=Q_{0}/(1+\beta_{\mathrm{c}})$, with $Q_{0}$ being the intrinsic quality factor, $\beta_{\mathrm{c}}$ the coupling factor, and $\omega_{0}$ is the angular frequency of the cavity's operating mode.
The beam duty factor ($DF$) can be calculated as the ratio between the average power and the peak dissipated power: 

\begin{equation}
     DF = \frac{P_{\mathrm{ave}}}{P_{\mathrm{diss}}} = \frac{\int_{0}^{\infty}P(t)\mathrm{d}t \cdot f_{\mathrm{b}}}{V_{\mathrm{acc}}^{2}/(R/Q \cdot Q_{0})},
\label{eq:duty_factor} 
\end{equation}
\noindent
where $P(t)$ is the time-dependent power calculated from the cavity voltage profile and $V_{\mathrm{acc}} = TTF\cdot V_{\mathrm{nom}}$ the accelerating cavity voltage, with $TTF$ being the Transit-Time factor, given by:

\begin{equation}
TTF = \frac{\int_{z_{\mathrm{min}}}^{z_{\mathrm{max}}}{E}_ze^{j k z} \, dz}{\int_{z_{\mathrm{min}}}^{z_{\mathrm{max}}}{E}_z \, dz},
\label{eq:transit_time_factor} 
\end{equation}

\noindent
where $k=\omega_{0}/(\beta c)$ is the wave number with $c$ being the speed of light in a vacuum and $\beta$ the relativistic velocity factor. The geometric shunt impedance, $R/Q$, is calculated, considering the TTF as:

\begin{equation} \left(\frac{R}{Q}\right)=\frac{| V_{z}(0,0)|^{2}}{\omega_{0}U_{0}}TTF^2,
\label{Geometric shunt impedance}
\end{equation}

\noindent
where $U=\omega_{0}$ is the energy stored in the cavity. 

Table~\ref{rf:tab:cool_power} reports the power requirements for each stage of the cooling channel. The peak input RF power is given by:

\begin{equation} 
P_{\mathrm{g}}=P_{\mathrm{diss}}\beta_{\mathrm{c}}.
\label{Peak input RF power}
\end{equation}

The duty factor of the RF power source ($DF_{\mathrm{g}}$) is given as the ratio between the average power of the generator and the peak input RF power. 

\begin{equation}
     DF_{\mathrm{g}} = \frac{P_{\mathrm{ave,g}}}{P_{\mathrm{g}}}  = \frac{P_{\mathrm{g}} t_{\mathrm{f}} \cdot f_{\mathrm{b}}}{P_{\mathrm{g}}},
\label{eq:duty_factor_generator} 
\end{equation}

The total plug power for the RF systems was calculated considering the generator ($\eta_{\mathrm{G}}$) and modulator ($\eta_{\mathrm{M}}$) efficiencies reported in Table~\ref{rf:tab:cool_const} as:

\begin{equation}
     P_{\mathrm{g,ave,tot}} = \frac{N_{\mathrm{cav}}P_{\mathrm{ave,g}}}{\eta_{\mathrm{G}}\eta_{\mathrm{M}}} ,
\label{eq:total_plug_power} 
\end{equation}
\noindent
where $N_{\mathrm{cav}}$ is the total number of cavities for each stage.

\begin{table}[!h]
    \centering
    \begin{tabular}{ccc|c}
         Parameters&  Symbol&  Unit&  Value\\ \hline
         Coupling factor&  $\beta_{\text{c}}$&  -&  1.2\\
         Bunch repetition frequency&  $f_\text{b}$&  Hz&  5\\
         Generator efficiency&  $\eta_\text{G}$&  -&  0.7\\
         Modulator efficiency&  $\eta_\text{M}$&  -&  0.9\\
    \end{tabular}
    \caption{RF parameters for the rectilinear cooling channel}
    \label{rf:tab:cool_const}
\end{table}

 For the RF frequency, cavity length and nominal RF gradient of the rectilinear cooling RF system, please refer to Table \ref{cool:tab:6d_rf}.
Table \ref{rf:tab:cool_dyn}
 displays in addition the RF cavity window radius, window thickness and the relativistic beta of the muon beam at each stage. 
\begin{table}[!h]
    \centering
    \begin{tabular}{l|ccc}
         & Window&  Window&Relativistic\\
          & radius &  thickness &$\beta$ \\
      & mm&  $\mu$m&- \\ \hline
         Stage A1 & 240&  120&0.923\\
         Stage A2 & 160&  70&0.894\\
         Stage A3 & 100&  45&0.894\\
         Stage A4 &  80&   40&0.901\\
         Stage B1 &  210&   100&0.882\\
         Stage B2 &  190&   80&0.879\\
         Stage B3 &  125&   50&0.882\\
         Stage B4 &  95&   45&0.889\\
         Stage B5 &  60&   30&0.889\\
         Stage B6 &  45&   20&0.888\\
         Stage B7 &  37&   20&0.887\\
         Stage B8 &  27&   20&0.884\\
         Stage B9 &  23&   10&0.881\\
         Stage B10&  21&   10&0.884\\
    \end{tabular}
    \caption{Beam dynamics specifications for the RF cavities in the rectilinear cooling channel}
    \label{rf:tab:cool_dyn}
\end{table}

\begin{table}[!h]
    \centering
    \begin{tabular}{l|cccccc}
          & $P_g$& $DF_g$& $N_{cav}$& $P_{g,tot}$& $P_{g,av}$&$P_{g,av,tot}$\\
         & MW/cavity& $10^{-4}$& -& MW& kW&kW\\ \hline
         Stage A1& 5.094& 1.560& 348& 1772.7& 277.09&439.83\\
         Stage A2& 5.21& 1.610& 356& 1854.9& 297.87&472.82\\
         Stage A3& 2.468& 0.567& 405& 999.4& 56.70&90.00\\
         Stage A4&  2.655&  0.573&  496&  1317.1&  75.41& 119.70\\
         Stage B1&  3.336&  2.001&  144&  480.4&  96.11& 152.56\\
         Stage B2&  3.883&  1.819&  170&  660.1&  120.09& 190.61\\
         Stage B3&  4.882&  1.611&  216&  1054.5&  169.92& 269.72\\
         Stage B4&  4.701&  1.843&  183&  860.3&  158.54& 251.65\\
         Stage B5&  1.419&  0.573&  275&  390.1&  22.37& 35.51\\
         Stage B6&  2.809&  0.572&  220&  617.9&  35.35& 56.11\\
         Stage B7&  2.531&  0.572&  204&  516.4&  29.55& 46.90\\
         Stage B8&  2.304&  0.572&  276&  635.8&  36.39& 57.77\\
         Stage B9&  2.573&  0.571&  212&  545.4&  31.15& 49.45\\
 Stage B10& 1.806& 0.572& 196& 354.1& 20.26&32.16\\
    \end{tabular}
    \caption{RF power requirements in the rectilinear cooling channel}
    \label{rf:tab:cool_power}
\end{table}

\pagebreak

\subsection{RF systems for low-energy acceleration}
\label{rf:sec:low}
In the low-energy acceleration, only the design of RLA2 is being considered for the computation of RF parameters. 
The baseline cavity geometry is chosen to be the LEP2 cavity. A summary of the assumed parameters can be found in Table \ref{rf:tab:LEP2_cavity_parameters}. 
For the calculation of the losses in the power generation, the parameters of the ILC-powering system were used (Table \ref{rf:tab:ILC_power_consumption}). The resulting powering parameters for the RLA2 cavities can be found in table \ref{rf:tab:low}.

\begin{table}[!h]
    \centering
    \caption{Parameters of the LEP2 cavity from \cite{rf:LEP2_cavity}} 
    \label{rf:tab:LEP2_cavity_parameters}
    \begin{tabular}{lcc|c|c}
Parameter & Symbol & Unit & Value & Value \\ 
& & & linearizer & accelerator\\
\hline
Fundamental mode RF frequency	& $f_\text{RF}$	& MHz & 352 & 1056 \\
Accelerating gradient & $G_\text{acc}$ & MV/m & 15 & 25 \\
Geometric shunt impedance & $R/Q$ & [$\Omega$] & 247.25 & 360.72\\
Active length & $l_\text{active}$ & m & 1.686 & 0.845\\
Total length & $l_\text{total}$ & m & 1.851 & 1.011\\
Number of cells & - & - & 4 & 6\\
$E_\text{peak} / E_\text{acc}$ & - & - & 2.4 &  2.4\\
$B_\text{peak} / E_\text{acc}$ & - & mT/(MV/m) & 3.9 & 3.9\\
Iris aperture (inner/end cell) & - & mm & 286/241 & 94/80\\
Cavity quality factor & $Q_0$ & - & $\geq 1\times10^{10}$ & $\geq 1\times10^{10}$\\
Cell-to-cell coupling & $k_\text{cc}$ & \% & 1.51 & 1.62\\
    \end{tabular}
\end{table}
\begin{table}[!h]
    \centering
    \begin{tabular}{lc|cc}
         Parameter &  Unit &  RLA2 acc& RLA2 lin\\ \hline
         Synchronous phase &  \textdegree &  95& 275\\
         Frequency&  MHz&  352& 1056\\
         Number of bunches/species &  - &  \multicolumn{2}{c}{1} \\
         Combined beam current ($\mu^+$, $\mu^-$) &  mA&  \multicolumn{2}{c}{134} \\ \hline
         Total RF voltage &  GV&  15.2& 1.69\\
         Total number of cavities &  - &  600& 80\\
         Total number of cryomodules & -  & 200&16\\
         Total RF section length & m& 1110.6&80.8\\
         External Q-factor & \num{E6}& 0.38&0.21\\
         Cavity detuning for beam loading comp. & kHz& 0.04&0.21\\ \hline
         Beam acceleration time & \textmu s& \multicolumn{2}{c}{35.5}\\
         Cavity filling time & \textmu s& 344&65\\
         RF pulse length & ms& 0.38&0.1\\
         RF duty factor & \%& 0.19&0.05\\
         Peak cavity power & kW& 3425&2965\\
         Average RF power & MW& 5.16&0.16\\
    \end{tabular}
    \caption[RF parameters for the low-energy acceleration chain.]{RF parameters for the low-energy acceleration chain. For the synchronous phase, \SI{90}{\degree} is defined as being on-crest}
    \label{rf:tab:low}
\end{table}

\begin{table}[!h]
    \centering
    \caption{ILC RF-power parameters \cite{rf:ILC-tdr} in the Distributed Klystron Scheme (DKS)}
    \label{rf:tab:ILC_power_consumption}
    \begin{tabular}{lc|c}
Parameter & Unit & Value \\ 
\hline
Max. klystron power & MW & 10 \\
Klystron efficiency & \% & 65 \\
Wall plug RF power efficiency & \% & $\sim$48\\
Klystron repetition rate &  Hz & 5 \\
Klystron frequency & MHz & 1300\\
RF pulse length & ms & 1.65 \\
RF duty factor & \% & 0.83 \\
\end{tabular}
\end{table}

\vfill
\pagebreak

\subsection{RF systems for high-energy acceleration}
\label{rf:sec:high}
A first approximation of the power requirements for the RCS chain has been performed using the ILC cavities, cryomodules, and powering infrastructures \cite{rf:ILC-tdr} as a baseline, the results of which can be found in Table~\ref{rf:tab:high}.

\begin{table}[!h]
    \centering
    \begin{tabular}{lc|ccccc}
         &  &  RCS1&  RCS2&  RCS3&  RCS4&All\\ \hline
         Synchronous phase &  \textdegree &  135&  135&  135&   135&- \\
         Number of bunches/species &  - &  1&  1&  1&   1&- \\
         Combined beam current ($\mu^+$, $\mu^-$) &  mA &  43.3&  39&  19.8&   5.49&- \\
         Total RF voltage &  GV&  20.9&  11.2&  16.1&   90&138.2\\
         Total number of cavities &  - &  683&  366&  524&   2933&4506\\
         Total number of cryomodules &  -  &  76&  41&  59&   326&502\\
         Total RF section length &  m &  962&  519&  746&   4125&6351\\ \hline
         Combined peak beam power ($\mu^+$, $\mu^-$) &  MW &  640&  310&  225&   350&- \\
         External Q-factor &  \num{E6} &  0.696&  0.775&  1.533&   5.522&- \\
         Cavity detuning for beam loading comp. & kHz & -1.32& -1.186& -0.6&  -0.166&- \\
         Beam acceleration time & ms & 0.34& 1.1& 2.37& 6.37&- \\
         Cavity filling time & ms & 0.171& 0.19& 0.375& 1.352&- \\
         RF pulse length & ms & 0.51& 1.29& 2.73& 7.77&- \\
         RF duty factor & \% & 0.19& 0.57& 1.22& 3.36&- \\
         Peak cavity power & kW & 1128& 1017& 516& 144&- \\ \hline
         Total peak RF power & MW & 1020 & 496 & 365 & 561 & - \\
         Total number of klystrons & - & 114& 53& 38&  57&262\\
         Cavities per klystron & - & 6& 7& 14&  52&- \\
         Average RF power & MW & 1.919& 2.84& 4.43& 18.92&28.1\\
         Average wall plug power for RF system & MW & 2.95& 4.38& 6.811& 29.1&43.25\\
         HOM power losses per cavity per bunch & kW & 25.85& 26.16& 16.24&  5.75&- \\
        Average HOM power per cavity& W & 366& 384& 287&  86&-\\  
    \end{tabular}
    \caption[RF parameters for the RCS chain]{RF parameters for the RCS chain. The average RF power includes losses from the cavity to the klystron, while the wall plug power also includes the klystron efficiency. For the synchronous phase, \SI{90}{\degree} is defined as being on-crest}
    \label{rf:tab:high}
\end{table}

The parameters of the ILC cavity can be found in Table \ref{rf:tab:TESLA_cavity_parameters}. To calculate the losses, parameters from the ILC DKS powering scheme are used (Table \ref{rf:tab:ILC_power_consumption}).
While these parameters are used for initial beam dynamics and power requirements studies, other frequencies and cavities are under investigation for muon acceleration. 
The requirements do not consider HOM power contributions, cryogenic losses and the impact of the detuning, which is necessary due to the orbit change during the acceleration. 
The calculated parameters assume a linear ramp of the magnet system.
In the accelerator, a harmonic magnet ramp is foreseen, which will require additional cavities. 

\begin{table}[!h]
    \centering
    \caption[~Parameters of the TESLA cavity]{Parameters of the TESLA cavity from \cite{rf:ILC-tdr} and \cite{rf:TESLA_wakefields}}
    \label{rf:tab:TESLA_cavity_parameters}
    \begin{tabular}{lcc|c}
Parameter & Symbol & Unit & Value \\ 
\hline
Fundamental mode RF frequency	& $f_\text{RF}$	& MHz	&1300 \\
Accelerating gradient & $G_\text{acc}$ & MV/m & 30 \\
Geometric shunt impedance & $R/Q$ & $\Omega$ & 518\\
Geometry factor & $G$ & $\Omega$ & 271\\
Active length & $l_\text{active}$ & m & 1.065\\
Total length & $l_\text{total}$ & m & 1.247 \\
Number of cells & - & - & 9\\
$E_\text{peak} / E_\text{acc}$ & - & - & 2.0 \\
$B_\text{peak} / E_\text{acc}$ & - & mT/(MV/m) & 4.26 \\
Iris aperture (inner/end cell) & - & mm & 70/78 \\
Cavity quality factor & $Q_0$ & - & $\geq 1\times10^{10}$ \\
Longitudinal loss factor ($\sigma_z = 1mm$) & $k_{||}$ & V/pC & 11.05\\
Cell-to-cell coupling & $k_\text{cc}$ & [\%] & 1.87 \\
    \end{tabular}
\end{table}

The change in the cavity detuning and external quality factor stems from the inclusion of transient beam loading effects in the calculation of the powering parameters. 
As a result, the power consumption of the system also changes. 

%
%
%
 \clearpage
\section{Impedance}
\label{imp:sec}
This section is devoted to beam intensity limitations that could be encountered in the different machines due to collective effects.

\subsection{Impedance model for the Rapid Cycling Synchrotrons}
\label{imp:sec:rcs}
Impedance models for the four RCS of the acceleration chain were developed.
The Rapid Cycling Synchrotrons (RCS) will be comprised of many RF cavities to provide the large acceleration voltage needed to reach the muon survival target, as developed in Section \ref{high:sec}.
It is assumed that the RCS 1, 2, 3 and 4 have respectively 700, 380, 540 and 3000 cavities.
Because of their number, the cavities are expected to be a large contributor to the RCS impedance model.
The models assume that superconducting TESLA cavities~\cite{rf:TESLA_cavity} are used for the RF system, and include the High-Order Modes (HOMs) generated by these cavities~\cite{impedance:bib:tesla_homs}.
The HOMs parameters for a single cavity are reported in Table~\ref{impedance:tab:HOMS_t}.

\begin{table}[!h]
    \centering
    \begin{tabular}{cccc}
         Frequency $f_{res}$ &  $\frac{R_s}{Q}$ &  Q factor & Shunt impedance $R_s$\\
         \si{\giga\hertz} & [\si{\kilo\ohm\per\meter}] & [\num{1E4}] & [\si{\mega\ohm\per\meter}]\\ \hline
         1.659&    0.10&    \num{31.4}    &   32.61\\
         1.705&    1.05&    \num{1.35}    &   14.16\\
         1.706&    1.21&    \num{1.34}    &   16.27\\
         1.728&    0.97&    \num{0.0413}  &   0.4\\
         1.729&    0.45&    \num{0.0381}  &   0.17\\
         1.736&    1.25&    \num{0.0516}  &   0.64\\
         1.737&    0.95&    \num{0.0574}  &   0.54\\
         1.761&    0.35&    \num{0.583}   &   2.04\\
         1.762&    0.28&    \num{0.621}   &   1.72\\
         1.788&    0.16&    \num{0.867}   &   1.43\\
         1.789&    0.18&    \num{0.890}   &   1.61\\
         1.798&    0.11&    \num{1.23}    &   1.29\\
         1.799&    0.10&    \num{1.21}    &   1.27\\
         1.865&    0.79&    \num{3.91}    &   30.87\\
         1.865&    0.83&    \num{4.12}    &   34.07\\
         1.874&    1.09&    \num{3.88}    &   42.32\\
         1.874&    1.07&    \num{4.39}    &   47.14\\
         1.88&     0.22&    \num{4.23}    &   9.38\\
         1.88&     0.24&    \num{5.15}    &   12.21\\
         2.561&    0.13&    \num{0.0620}  &   0.08\\
         2.561&    0.12&    \num{0.0527}  &   0.07\\
         2.577&    2.05&    \num{0.364}   &   7.46\\
    \end{tabular}
    \caption{HOMs from TESLA cavity, complete table, for a single cavity.}
    \label{impedance:tab:HOMS_t}
\end{table}

An alternate type of cavity, the Low Losses~\cite{bib:LL_cavities} type based on the TESLA one, was considered in previous studies and the HOMs of this cavity are reported in Table~\ref{impedance:tab:HOMS_llt}.

\begin{table}[!h]
    \centering
    \begin{tabular}{cccc}
         Frequency $f_{res}$ &  $\frac{R_s}{Q}$ &  Q factor & Shunt impedance $R_s$ \\
         \si{\giga\hertz} & [\si{\kilo\ohm\per\meter}]& [\num{1E4}]&[\si{\mega\ohm\per\meter}]\\ \hline
         1.717&  0.70&  \num{4.0} & 27.8\\
         1.738&  0.41&  \num{6.0} & 24.7\\
         1.882&  0.43&  \num{0.6} & 2.6\\
         1.912&  0.57&  \num{0.9} & 5.2\\
         1.927&  1.93&  \num{1.5} & 29\\
         1.94&   1.49&  \num{2.0} & 29.8\\
         2.451&  3.08&  \num{10} & 307.8\\
         2.457&  2.16&  \num{5.0} & 107.9\\
         3.057&  0.04&  \num{30} & 11.7\\
         3.06 &  0.03&  \num{80}&25\\
    \end{tabular}
    \caption{HOMs from Low Loss TESLA cavity, complete table, for a single cavity.}
    \label{impedance:tab:HOMS_llt}
\end{table}

\vfill
\pagebreak

However the HOMs of the Low Losses cavities generate stronger wakefields and are more detrimental to beam stability~\cite{bib:amorim_tesla_vs_low_loss}.
The RCS parameters relevant for the impedance and coherent stability simulations are reported in Table~\ref{impedance:tab:collective}.

\begin{table}[!h]
    \centering
    \begin{tabular}{l|ccc}
         Parameter&  Unit&  \multicolumn{2}{c}{All RCS rings}\\
         &  &  Horizontal& Vertical\\ \hline
         Average Twiss beta &  m&  50& 50\\
         Chromaticity Q' &  -&  +20& +20\\
         Detuning from octupoles &  \si[per-mode=reciprocal]{\per\meter} &  0& 0\\
         Transverse damper&  turns&  \multicolumn{2}{c}{20}\\
         Bunch intensity at injection&  muons/bunch&  \multicolumn{2}{c}{\num{2.7E12}}\\
    \end{tabular}
    \caption{RCS Collective Effects Parameters used in simulations.}
    \label{impedance:tab:collective}
\end{table}

A second important contributor to the impedance model of the RCS is the normal conducting magnets vacuum chamber.
Because of the high ramping rate, a large eddy current would appear if a fully metallic chamber was used~\cite{bib:kvikne_eddy_currents_imcc_2024}.
A ceramic chamber with a thin metallic coating on the inner surface would therefore be used~\cite{impedance:bib:rcs_vacuum_chamber}.
Its dimension and characteristics are reported in Table~\ref{impedance:tab:rcs2_nc}.

\begin{table}[!h]
    \centering
    \begin{tabular}{l|cc}
         Parameter&  Unit& Value\\ \hline
         Inner dimension width, height&  \si{\milli\meter}, \si{\milli\meter}& 30, 20\\
         Titanium coating thickness&  \si{\micro\meter}& 1 to 10\\
         Ceramic thickness&  \si{\milli\meter}& 5\\
         Outer dimension width, height&  \si{\milli\meter}, \si{\milli\meter}& 40, 30\\
    \end{tabular}
     \caption{RCS 2 normal conducting magnets vacuum chamber used in simulations.}
    \label{impedance:tab:rcs2_nc}
\end{table}

Transverse coherent stability simulations were performed to evaluate the impact of the RF cavities and vacuum chambers.
To mitigate the instabilities, a transverse damper system can be used to damp the transverse centroid motion of the bunches, and/or chromaticity can be introduced with sextupoles.
Parametric scans were performed to find if those are needed and, if necessary, the chromaticity $Q^{\prime}$ required.
The chromaticity was scanned from $Q^{\prime}=-20$ to $Q^{\prime}=+20$, and the transverse damper from a 4-turn to a 100-turn damping time, with an additional case without damper.

Tracking simulations were performed using Xsuite~\cite{impedance:bib:xsuite} and PyHEADTAIL~\cite{bib:pyheadtail}.
The bunch motion is simulated through the complete RCS chain.
Muon decay is not included in these simulations, therefore the bunch intensity remains constant through the chain, equal to the intensity of \num{2.7e12}~muons per bunch at injection in RCS 1.
Results showed that a positive chromaticity of $Q^{\prime} = +20$ is needed in the accelerators to stabilize the beams and leave enough margin for some initial transverse offset of the bunches, and a 20-turn transverse damper also helps stabilize the beams~\cite{impedance:bib:amorim_rcs_imcc_2024, bib:amorim_tesla_vs_low_loss}.

\subsection{Impedance model for the 10 TeV collider ring}
\label{imp:sec:coll}

In the \SI{10}{\tera\eV} collider ring, the main impedance source would be the resistive-wall contribution from the magnets' vacuum chamber.
To protect the superconducting magnet coils from muon decay induced heating and radiation damage, a tungsten shield is proposed to be the inserted in the magnet cold bore as detailed in Section~\ref{rad:sec} and described in Ref.~\cite{bib:shielding_requirements}.

Previous parametric studies performed with Xsuite and PyHEADTAIL showed that a minimum chamber radius of \SI{13}{\milli\metre}, together with a copper coating on the inner diameter are required to ensure coherent transverse beam stability.
The current dipole magnet radial build detailed in Section~\ref{mag:sec:col} foresees a \SI{23.5}{\milli\metre} inner radius, with a \SI{10}{\micro\metre} copper coating.
The vacuum chamber properties used for the impedance model computation are summarized in Table~\ref{impedance:tab:collider}.

\begin{table}[!h]
    \centering
    \begin{tabular}{cc|c}
         Parameter&  Unit& Value\\ \hline
         Chamber geometry&  & circular\\
         Chamber length&  \si{\meter} & 10000\\
         Copper coating thickness&  \si{\micro\meter}& 10\\
         Copper resistivity at \SI{300}{\kelvin} &  \si{\nano\ohm\meter}& 17.9\\
         Tungsten resistivity at \SI{300}{\kelvin}&  \si{\nano\ohm\meter}& 54.4\\
         Chamber radius (from magnet radial build)&  \si{\milli\meter}& 23\\
 Min. chamber radius required (50-turn damper)& mm&13\\
    \end{tabular}   
    \caption{\SI{10}{\tera\electronvolt} collider parameters for impedance model simulations.}
    \label{impedance:tab:collider}
\end{table}

A particularity of the collider ring is its isochronous operation (i.e. with $\eta\approx0$)~\cite{impedance:bib:ng_quasi_isochronous_buckets}, obtained with the flexible momentum compaction cells described in Section~\ref{col:sec}.
This is to avoid the large RF voltage that would be needed to bunch beams with very short length and large energy spread.
However this freezes the synchrotron motion of the particles within the bunch and can lead to beam breakup instabilities such as those encountered in Linacs~\cite{impedance:bib:kim_transverse_instability}.

Transverse coherent beam stability simulations were performed with Xsuite and PyHEADTAIL, including the effect of muon beam decay~\cite{impedance:bib:amorim_coll10tev_imcc_2024}.
The beam parameters used for these simulations are summarized in Table~\ref{impedance:tab:collider_param}.
With a chromaticity of $Q^{\prime} = 0$, the beam becomes unstable over its lifetime in the collider, leading to large transverse emittance growth~\cite{impedance:bib:amorim_coll10tev_imcc_2024}.
A slightly positive chromaticity of $Q^{\prime} = +2$ is needed to introduce a betatron frequency spread that helps stabilize the beam.


\begin{table}[!ht]
    \centering
    \begin{tabular}{cc|c}
         Parameter&  Unit& Value\\ \hline
         Circumference &   \si{\meter}& 10 000\\
         Beam energy &  \si{\tera\electronvolt} & 5\\
         Bunch intensity at injection &  muons/bunch & \num{1.80E+12}\\
         1$\sigma$ bunch length& \si{\milli\meter} & 1.5\\
         Longitudinal emittance $\epsilon_l = \sigma_z \sigma_E$ &  \si{\mega\electronvolt\meter} & 7.5\\
         Transverse normalized emittance&  \si{\micro\meter\radian} & 25\\
         Momentum compaction factor&  & 0\\
         Total RF voltage & \si{\mega\volt} & 0\\
    \end{tabular}
    \caption{\SI{10}{\tera\electronvolt} collider machine and beam parameters.}
    \label{impedance:tab:collider_param}
\end{table}

 \clearpage
\section{Radiation}
\label{rad:sec}
This Section presents radiation studies for the following systems:

\begin{itemize}
    \item Target solenoids considering proton impact on a Graphite target in Section \ref{rad:sec:target_sol}.
    \item Magnets in the arcs and interaction regions of the collider ring due to muon decay in Section \ref{rad:sec:collider_muon}.
    \item Neutrino-induced dose in soil for mono-directional muons in Section \ref{rad:sec:neutrino}.
\end{itemize} 
The latter can serve as dose kernel for computing surface dose levels under consideration of the beam optics and the collider placement.

\subsection{Radiation load on the target superconducting solenoids}
\label{rad:sec:target_sol}
Generic radiation load studies for the superconducting solenoid were performed by means of  FLUKA Monte Carlo simulations. A 5\,GeV proton beam with a beam sigma of 5\,mm and a beam power of 2\,MW was assumed to impinge on a graphite target rod (see Table~\ref{target:tab:radialbuild} for the target dimensions). The target was centered along the beam axis and therefore no dependence on the azimuthal angle can be expected. The simulation results for the coils are presented in Table \ref{rad:tab:DPA_coils}, showing the maximum displacement per atom (DPA) per year and the maximum yearly absorbed dose. The studies were carried out for different target shielding thicknesses and shielding compositions. The shielding inner radius in the area of the target vessel is fixed at \SI{17.8}{cm}. The gap between the shielding outer radius and the magnet coils is always kept at \SI{7.5}{cm}. The shielding outer radius can be read from the table by subtracting \SI{7.5}{cm} from the magnet coils' inner radius. The target shielding was either assumed to be made of pure tungsten or tungsten with an outer, neutron-absorbing layer made of water combined with boron-carbide.

\begin{table}[b!]
\caption[Radiation load on target solenoids]{Radiation load on the target superconducting magnet coils in terms of the maximum displacement per atom (DPA) and the maximum absorbed dose per year of operation for various shielding configurations.}
\label{rad:tab:DPA_coils}
\centering\resizebox{1.0\textwidth}{!}{%
\begin{tabular}{cccc}
\hline\hline
& \textbf{Tungsten + Water  + Boron-Carbide} & & \\
Inner radius of the magnet coils & Shielding thickness around the target & DPA/year [$10^{-3}$] & Dose [MGy/year] \\
\hline
\hline
\SI{60}{cm} & \textbf{(B)W \SI{31.2}{cm} + H\textsubscript{2}O  \SI{2}{cm} + B\textsubscript{4}C \SI{0.5}{cm} + W \SI{1}{cm}} & 1.70 ± 0.02&  10.0 ± 0.3\\
\SI{65}{cm} & W \SI{36.2}{cm} + H\textsubscript{2}O  \SI{2}{cm} + B\textsubscript{4}C \SI{0.5}{cm} + W \SI{1}{cm} &  0.90 ± 0.02&  5.6 ± 0.2\\
\SI{70}{cm} & W \SI{41.2}{cm} + H\textsubscript{2}O  \SI{2}{cm} + B\textsubscript{4}C \SI{0.5}{cm} + W \SI{1}{cm} & 0.49 ± 0.01&  3.1 ± 0.1\\
\SI{75}{cm} & W \SI{46.2}{cm} + H\textsubscript{2}O  \SI{2}{cm} + B\textsubscript{4}C \SI{0.5}{cm} + W \SI{1}{cm} & 0.29 ± 0.01& 1.9 ± 0.1\\
\SI{80}{cm} & W \SI{51.2}{cm} + H\textsubscript{2}O  \SI{2}{cm} + B\textsubscript{4}C \SI{0.5}{cm} + W \SI{1}{cm} &  0.16 ± 0.01&  1.0 ± 0.1\\
\SI{85}{cm} & W \SI{56.2}{cm} + H\textsubscript{2}O  \SI{2}{cm} + B\textsubscript{4}C \SI{0.5}{cm} + W \SI{1}{cm} & 0.09 ± 0.01&  0.6 ±  0.1\\
\hline\hline
\end{tabular}
}
\end{table}

The maximum DPA per year in the magnet coils is reduced by a factor of 1.8 with every \SI{5}{cm} of extra tungsten.
Exchanging \SI{2.5}{cm} of tungsten for a layer of \SI{2}{cm} of water followed by \SI{0.5}{cm} of boron-carbide enclosed by an external layer of \SI{1}{cm} of tungsten is equivalent to thickening pure tungsten shielding by \SI{5}{cm} in the context of reducing the yearly DPA in the magnet coils.
In this way, the DPA can be decreased without the necessity of increasing the total shielding radius.
In the presence of the external layer of tungsten enclosing the water-boron-carbide layer, the absorbed dose is not affected by the reduction in the total thickness of tungsten.
Previous studies have shown that if the tungsten was not placed after the special neutron-absorbing layer, the absorbed dose in the magnets would increase, most likely due to the  photon production in the neutron capture of hydrogen (sharp line at \SI{2.2}{MeV}).
Moreover, it was found that the reduction of the DPA per year thanks to the presence of water saturates for a \SI{3}{cm}-thick water layer.
Therefore, only the results for one water-boron-carbide layer configuration is presented in Table \ref{rad:tab:DPA_coils}.



\subsection{Muon decay in the collider ring}
\label{rad:sec:collider_muon}

The radiation-induced power load and radiation effects in collider equipment are dominated by the products of muon decay.
While decay neutrinos yield a negligible contribution to the radiation load on the machine, the decay electrons and positrons induce secondary particle showers, which dissipate their energy in the surrounding materials.
A continuous shielding is therefore needed, which dissipates the induced heat and protects the superconducting magnets against long-term radiation damage.
Shielding studies for muon colliders have been previously carried out within MAP~\cite{Mokhov2011PAC,Kashikhin2012IPAC,Mokhov2014IPAC}.
In particular, the shielding must: 
\begin{itemize}
    \item prevent magnet quenches,
    \item reduce the thermal load to the cryogenic system (by reducing the heat load to the cold mass of magnets),
    \item prevent magnet failures due to the ionizing dose in organic materials (e.g. insulation, spacers) and atomic displacements in the superconductor. 
\end{itemize}

The assumed beam parameters and operational scenarios for the radiation studies are summarized in Table~\ref{rad:tab:ring}.
The beam parameters (\SI{10}{\tera\electronvolt}) originate from Table~\ref{col:tab:param}, but are repeated here for completeness.

\begin{table}[h]
\begin{center}
\caption[Parameters for collider radiation]{Parameters for radiation studies (collider ring). The number of decays  consider the contribution of both beams.}
\label{rad:tab:ring}
\begin{tabular}{ll|cc}
 &Units& \textbf{\SI{3}{\tera\electronvolt}} & \textbf{\SI{10}{\tera\electronvolt}} \\
\hline
Particle energy  &\si{\tera\electronvolt}& 1.5& 5\\
Bunches/beam  && 1& 1\\
Muons per bunch  &\num{E12}& 2.2& 1.8\\
Circumference  &\si{\kilo\meter}& 4.5& 10\\\hline
Muon decay rate per unit length  &\SI[per-mode=reciprocal]{E9}{\per\meter\per\second}& 4.9&  1.8\\
Power ($e^{\pm}$)/meter  &\si{\kilo\watt\per\meter}& 0.411& 0.505\\\hline
Operational years  &years& \multicolumn{2}{c}{5-10}\\
Operational time per year (average)  &days& \multicolumn{2}{c}{139}\\
\end{tabular}
\end{center}
\end{table}

The power carried by decay electrons and positrons is on average \SI{35}{\percent} of the energy of decaying muons. 
With the presently assumed beam parameters, this amounts to about \SI{500}{\watt\per\meter}.
The remaining \SI{65}{\percent} of the energy released in decays is carried away by neutrinos.
Assuming five years of operation and an average operational time of \SI{1.2E7}{\second\per year}, the total number of decays in the collider ring reaches almost \SI[per-mode=reciprocal]{3E17}{\per\meter} for the \SI{3}{\tera\electronvolt} collider, and about \SI[per-mode=reciprocal]{1E17}{\per\meter} for the 10~TeV collider. In an alternative scenario, the collider might operate for ten years, which means that that the number of decays increases by a factor of two. The results presented in the following subsections are given for one year of operation and need to be scaled to the actual operational scenario (5 or 10 years).

\subsubsection{Power deposition, dose and DPA in arc magnets}
\label{rad:sec:magarcs}

In order to estimate the required shielding thickness for a \SI{10}{\tera\electronvolt} collider, generic shielding studies for the arc magnets were performed with FLUKA~\cite{Calzolari2022IPAC,Lechner2024IPAC}.
The studies considered only muon decay, whereas other source terms (e.g. beam halo losses) still have to be addressed in the future.
Table~\ref{rad:tab:colliderarcdipoles} summarizes the calculated power load and radiation damage in collider ring magnets as a function of the radial absorber thickness (\SI{10}{\tera\electronvolt} collider).
For simplicity, the FLUKA simulation model consisted of a generic string of \SI{16}{\tesla} dipoles, each six meters long; the drift regions between dipoles were assumed to be \SI{20}{\centi\meter} long.
As absorber material, we used tungsten due to its high atomic number and density (tungsten was also considered as shielding material by the MAP collaboration).
For engineering reasons, pure tungsten may be substituted by tungsten-based alloys without significantly affecting the shielding efficiency if the alloy has a similar material density.
As beam aperture, we considered 23.5~mm like in the 1D radial build summarized in Table~\ref{col:tab:param}.
A gap of 15.5~mm was assumed between the radiation absorber and inner coil aperture, which leaves space for the shielding support and thermal insulation (both not simulated), as well as cold bore and Kapton (both included in the simulation).
As shielding thicknesses, we considered 2~cm, 3~cm and 4~cm. The second case (3~cm) corresponds to the radial build in Table~\ref{col:tab:param}.

As can be seen in Table~\ref{rad:tab:colliderarcdipoles}, the power penetrating the tungsten absorber (mostly in the form of electromagnetic showers) amounts to 3.7\% in the case of a 2~cm shielding, and decreases to 0.8\% in the case of a 4~cm shielding.
Most of this power is deposited in the cold bore and cold mass of the superconducting magnets.
A small fraction of the power escapes from the magnets and is dissipated in the surrounding materials, in particular the tunnel wall and soil.
These power estimates do not consider the power carried by the decay neutrinos as they are not relevant for the radiation load to the machine. 

\begin{table}[!h]
\begin{center}
\caption[Power load and radiation damage in collider]{Power load and radiation damage in collider ring arc magnets (\SI{10}{\tera\electronvolt}) as a function of the radial tungsten absorber thickness. The power penetrating the shielding does not include neutrinos, since they are not relevant for the radiation load to the machine; the percentage values are given with respect to the power carried by decay electrons and positrons. The results include the contribution of both counter-rotating beams.}
\label{rad:tab:colliderarcdipoles}
\begin{tabular}{ll|ccc}
         &Unit&  \textbf{2~cm} & \textbf{3~cm} & \textbf{4~cm} \\
        \hline
        Beam aperture  &\si{\milli\meter}& 23.5& 23.5& 23.5\\
        Outer shielding radius  &\si{\milli\meter}& 43.5& 53.5& 63.5\\
        Inner coil aperture  &\si{\milli\meter}& 59& 69& 79\\
        \hline
        Absolute power penetrating tungsten absorber&\si{\watt\per\meter}& 18.5& 8& 4\\
        Fractional power penetrating tungsten absorber&\si{\percent}& 3.7& 1.6& 0.8\\
        Peak power density in coils&\si{\milli\watt\per\centi\meter\cubed}& 6.3& 2.1& 0.7\\
        Peak dose in Kapton insulation (1 year)&\si{\mega\gray}& 10.6& 3.3& 1.3\\
        Peak dose in coils (1 year)&\si{\mega\gray}& 8.5& 2.8& 1\\
 Peak DPA in coils (1 year)& \num{E5} DPA& 1.5& 1.2&1\\
\end{tabular}
\end{center}
\end{table}

For all considered shielding thicknesses, the power density in the coils remains below 10~mW/cm$^{3}$, which is expected to be significantly less than the quench level of HTS or Nb3Sn-based magnets.
The 5-year ionizing dose exceeds 50~MGy in the Kapton insulation in case of a 2~cm shielding, but is less than 20~MGy if the shielding thickness is 3~cm or more.
This is compatible with typical dose limits of Kapton tapes (usually of the order of 30~MGy).
Likewise, the cumulative DPA in the superconductor, which is mainly induced by secondary neutrons, remains below critical values for any of the considered shielding thicknesses.
In summary, the results show that the power leaking from the shielding and deposited in the cold mass is the most important factor for the shielding thickness. The studies presented in this section are representative for dipoles in the collider arcs, but a separate assessment for insertion region magnets is needed (see next section).

\subsubsection{Dose in IR magnets}
\label{rad:sec:magir}

In the interaction region, which accommodates the final focus magnets and a chicane for background reduction, more radiation is expected to arrive on the machine elements. This is a consequence of the long straight section between the chicane and the chromaticity correction section, which leads to a build-up of decay products. As a consequence, the radial shielding thickness generally needs to be larger than in the arcs in order to remain below critical dose levels. Moreover, the beam size in this section is substantially larger than the one in the arc sections, therefore increasing the aperture requirements.

In Table \ref{rad:tab:focus_geom}, the different IR magnets and the corresponding ionizing dose is reported. Thicker shielding elements are required for the first three dipoles than for the final focus quadrupoles.

\begin{table}[!h]
    \centering
        \caption[Final focusing magnets and dose]{Cumulative ionizing dose in final focus quadrupoles and chicane dipoles located in the insertion region (lattice version 0.8).}
    \label{rad:tab:focus_geom}
\begin{tabular}{l|cccc}
\textbf{Name} & \textbf{L [m]} & \textbf{Shield thickness [cm]} & \textbf{Coil aperture [cm]} & \textbf{Peak TID [MGy/y]} \\ \hline
IB2     & 6  & 6   & 16.0 & 1.3  \\
IB1     & 10 & 6   & 16.0 & 3.1  \\
IB3     & 6  & 6   & 16.0 & 4.9  \\
IQF2    & 6  & 4   & 14.0 & 7.7  \\
IQF2\_1 & 6  & 4   & 13.3 & 4.6  \\
IQD1    & 9  & 4   & 14.5 & 1.1  \\
IQD1\_1 & 9  & 4   & 14.5 & 3.7  \\
IQF1B   & 2  & 4   & 10.2 & 6.4  \\
IQF1A   & 3  & 4   & 8.6  & 3.6  \\
IQF1    & 3  & 4   & 7.0  & 3.5  \\
    \end{tabular}
\end{table}

In case of 5 years of operation, the dose would remain below 40~MGy in all magnets, which is considered acceptable. However, the dose would become too high for 10 years of operation, exceeding even 70~MGy for one of the final focus quadrupoles (IQF2). Therefore, in case of an extended operational period, even more stringent requirements on the shielding would be required.

\subsection{Neutrino radiation}
\label{rad:sec:neutrino}
The decay of muons in the collider ring produces very energetic neutrinos that have a non-negligible probability to interact far away from the collider in material near to the Earth’s surface producing secondary particle showers.
The goal is to ensure that this effect does not entail any noticeable addition to natural radiation and that the environmental impact of the muon collider is negligible, i.e. an effective dose of the order of 10~$\mu$Sv/year, similar, for instance, to the impact from the LHC.
For the environmental impact assessment, detailed studies of the expected neutrino and secondary-particle fluxes are being performed with FLUKA.
The latter can be folded with the realistic neutrino source term taking into account the collider lattice to predict the effective dose and to design suitable methods for mitigation and demonstration of compliance. 

\begin{figure}[!h]
    \centering
    \includegraphics[width=0.49\textwidth]{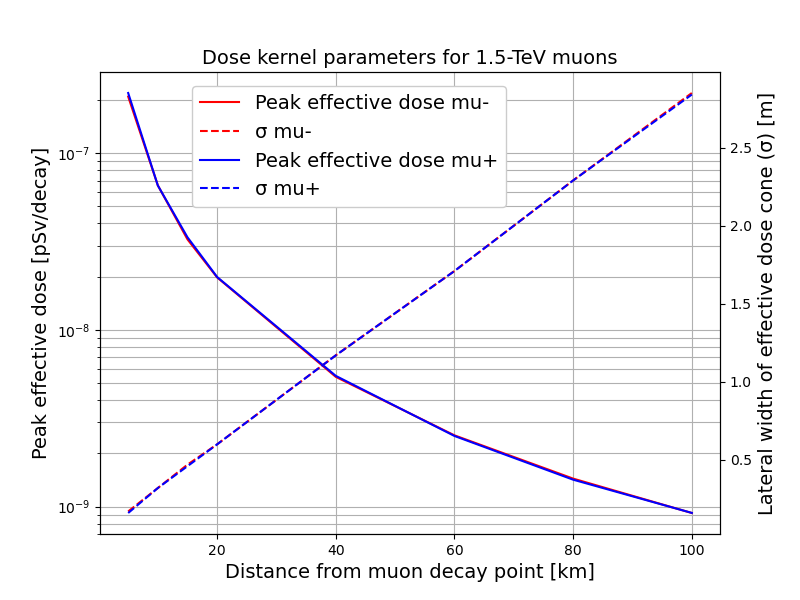}
    \includegraphics[width=0.49\textwidth]{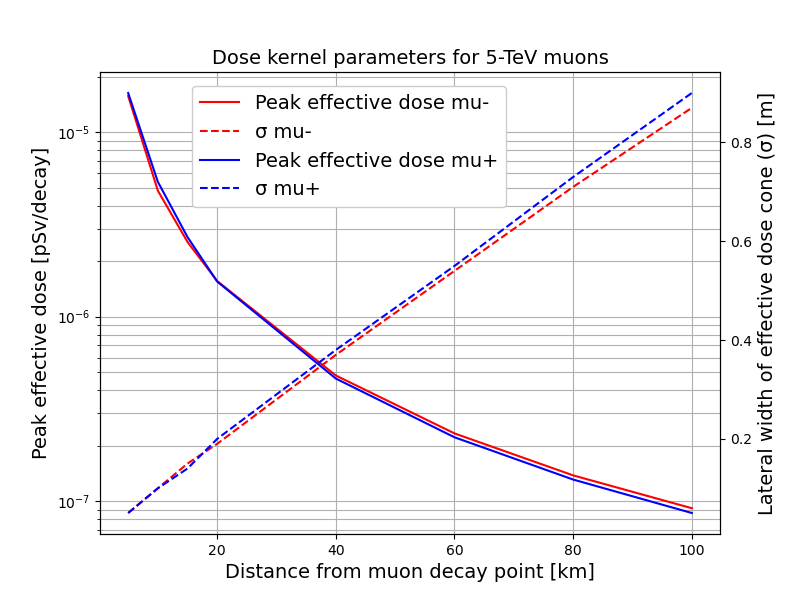}
    \caption{Effective dose kernel parameters within soil as a function of the baseline distance from the muon decay point, for muon energies of $1.5$~TeV (left) and $5$~TeV (right). }
    \label{fig:dosekernel_summary}
\end{figure}

FLUKA simulations were conducted to obtain the effective dose within soil resulting from the interaction of the neutrinos from the decay of $1.5$~TeV and $5$~TeV mono-directional muons.
The angular distribution and the energy of the neutrinos were sampled taking into account the respective distributions from the muon decay.
Moreover, the interactions were sampled such that the dose values corresponded to the values obtained when the neutrino-induced showers reach a plateau condition, i.e., after at least several meters of path through the material.
The latter is a very conservative worst-case scenario as it does not consider that neutrino-induced showers rapidly decrease in the transition between soil and air.


The results of the FLUKA simulations are shown in Figure~\ref{fig:dosekernel_summary} in terms of dose kernel parameters, i.e., peak and lateral width of the effective dose profile at different baseline distances from the muon decay position.
The values are reported in Tables~\ref{tab:effdose_kernel_1.5TeV}~and~\ref{tab:effdose_kernel_5TeV} for the 1.5~TeV and 5~TeV muon beams.

\begin{table}[!h]
    \centering
        \caption[Neutrino radiation at distances from muon decay (1.5 TeV beam)]{Effective dose kernel parameters of neutrino-induced radiation within soil at different baseline distances from the muon decay, for a muon beam energy of $1.5$~TeV. The peak dose per muon decay and the lateral width of the dose profile ($\sigma$) have been derived from Gaussian fits of the FLUKA results. 
        }
    \label{tab:effdose_kernel_1.5TeV}
    \begin{tabular}{rcccc}
         \hline\hline
         \multicolumn{1}{c}{}& \multicolumn{2}{c}{\textbf{$\mu^-$}} & \multicolumn{2}{c}{\textbf{$\mu^+$}} \\ 
         \textbf{Distance} & \textbf{Peak eff. dose [pSv/decay]} & \textbf{$\sigma$ [m]} & \textbf{Peak eff. dose [pSv/decay]} & \textbf{$\sigma$ [m]} \\ \hline 
       $5$~km  & $2.09 \cdot 10^{-7}$ & $0.17$ & $2.19 \cdot 10^{-7}$  & $0.16$ \\
       $10$~km  & $6.57 \cdot 10^{-8}$ & $0.32$ & $6.56 \cdot 10^{-8}$ & $0.32$ \\
       $15$~km  & $3.28 \cdot 10^{-8}$ & $0.47$ & $3.34 \cdot 10^{-8}$ & $0.46$ \\
       $20$~km  & $1.98 \cdot 10^{-8}$ & $0.60$ & $1.99 \cdot 10^{-8}$  & $0.60$ \\
       $40$~km  & $5.42 \cdot 10^{-9}$ & $1.17$ & $5.49 \cdot 10^{-9}$ & $1.17$ \\
       $60$~km  & $2.53 \cdot 10^{-9}$ & $1.71$ & $2.51 \cdot 10^{-9}$ & $1.71$ \\
       $80$~km  & $1.44 \cdot 10^{-9}$ & $2.29$ & $1.42 \cdot 10^{-9}$  & $2.29$ \\
       $100$~km  & $9.20 \cdot 10^{-10}$ & $2.85$ & $9.21 \cdot 10^{-10}$ & $2.84$ \\
       \hline\hline
    \end{tabular}
\end{table}

\begin{table}[!h]
    \centering
        \caption[Neutrino radiation at distances from muon decay (5 TeV beam)]{Effective dose kernel parameters of neutrino-induced radiation within soil at different baseline distances from the muon decay, for a muon beam energy of $5$~TeV. The peak dose per muon decay and the lateral width of the dose profile ($\sigma$) have been derived from Gaussian fits of the FLUKA results. 
        }
    \label{tab:effdose_kernel_5TeV}
    \begin{tabular}{rcccc}
         \hline\hline
         \multicolumn{1}{c}{}& \multicolumn{2}{c}{\textbf{$\mu^-$}} & \multicolumn{2}{c}{\textbf{$\mu^+$}} \\ 
         \textbf{Distance} & \textbf{Peak eff. dose [pSv/decay]} & \textbf{$\sigma$ [m]} & \textbf{Peak eff. dose [pSv/decay]} & \textbf{$\sigma$ [m]} \\ \hline 
       $5$~km  & $1.57 \cdot 10^{-5}$ & $0.05$ & $1.63 \cdot 10^{-5}$  & $0.05$ \\
       $10$~km  & $4.86 \cdot 10^{-6}$ & $0.10$ & $5.38 \cdot 10^{-6}$ & $0.10$ \\
       $15$~km  & $2.54 \cdot 10^{-6}$ & $0.15$ & $2.70 \cdot 10^{-6}$ & $0.14$ \\
       $20$~km  & $1.56 \cdot 10^{-6}$ & $0.19$ & $1.55 \cdot 10^{-6}$  & $0.20$ \\
       $40$~km  & $4.80 \cdot 10^{-7}$ & $0.37$ & $4.62 \cdot 10^{-7}$ & $0.38$ \\
       $60$~km  & $2.33 \cdot 10^{-7}$ & $0.54$ & $2.22 \cdot 10^{-7}$ & $0.55$ \\
       $80$~km  & $1.38 \cdot 10^{-7}$ & $0.71$ & $1.31 \cdot 10^{-7}$  & $0.73$ \\
       $100$~km  & $9.16 \cdot 10^{-8}$ & $0.87$ & $8.63 \cdot 10^{-8}$ & $0.90$ \\
       \hline\hline
    \end{tabular}
\end{table}

These numbers shall be used as the basis for the calculation of the effective dose which takes into account realistic lattice parameters, i.e. factoring the associated distribution of muon trajectories in the sections of interest of the accelerator in, as well as the reduction due to realistic geometries and exposure scenarios. Additionally, mitigation methods, such as optimization of the source term, location and orientation of the collider, should be considered.

The neutrino flux density arising from the collider ring arcs is expected to be reduced to a negligible level by deforming the muon beam trajectory, achieving a wide-enough angular spread of the neutrinos. 
Wobbling of the muon beam within the beam pipe would be sufficient for 1.5 TeV muon beam energy.
At 5 TeV muon beam energy, the beam line components in the arcs may have to be placed on movers to deform the ring periodically in small steps such that the muon beam direction would change over time. 
Table~\ref{tab:wobbling_dose_5TeV} presents the effective dose within soil similar to Table~\ref{tab:effdose_kernel_5TeV}, but taking into account the vertical deformation of the beam within $\pm$1 mrad by the movers. It results in a reduction factor of 80-90 of the saturated dose kernels within soil.

\begin{table}[!h]
    \centering
        \caption[Neutrino radiation at distances after vertical deformation by movers (5 TeV beam)]{Effective dose of neutrino-induced radiation within soil at different baseline distances from the muon decay after the vertical deformation by the movers is applied. The muon beam energy is $5$~TeV. The reduction factor is the ratio between the peak dose value of the corresponding kernel from Table \ref{tab:effdose_kernel_5TeV} and the dose value mitigated by the movers. 
        }
    \label{tab:wobbling_dose_5TeV}
    \begin{tabular}{rcccc}
         \hline\hline
         \multicolumn{1}{c}{}& \multicolumn{2}{c}{\textbf{$\mu^-$}} & \multicolumn{2}{c}{\textbf{$\mu^+$}} \\ 
         \textbf{Distance} & \textbf{Mitigated dose [pSv/decay]} & \textbf{Reduction factor} & \textbf{Mitigated dose} & \textbf{Reduction factor} \\ \hline 
       $5$~km  & $1.97 \cdot 10^{-7}$ & $80$ & $2.04 \cdot 10^{-7}$  & $80$ \\
       $10$~km  & $6.09 \cdot 10^{-8}$ & $80$ & $6.74 \cdot 10^{-8}$ & $80$ \\
       $15$~km  & $3.18 \cdot 10^{-8}$ & $80$ & $3.15 \cdot 10^{-8}$ & $85$ \\
       $20$~km  & $1.86 \cdot 10^{-8}$ & $84$ & $1.94 \cdot 10^{-8}$  & $80$ \\
       $40$~km  & $5.56 \cdot 10^{-9}$ & $86$ & $5.50 \cdot 10^{-9}$ & $84$ \\
       $60$~km  & $2.63 \cdot 10^{-9}$ & $89$ & $2.55 \cdot 10^{-9}$ & $87$ \\
       $80$~km  & $1.53 \cdot 10^{-9}$ & $90$ & $1.50\cdot 10^{-9}$  & $87$ \\
       $100$~km  & $9.99 \cdot 10^{-10}$ & $92$ & $9.73 \cdot 10^{-10}$ & $89$ \\
       \hline\hline
    \end{tabular}
\end{table}

\pagebreak

\begin{figure}[!h]
    \centering
    \includegraphics[width=\textwidth]{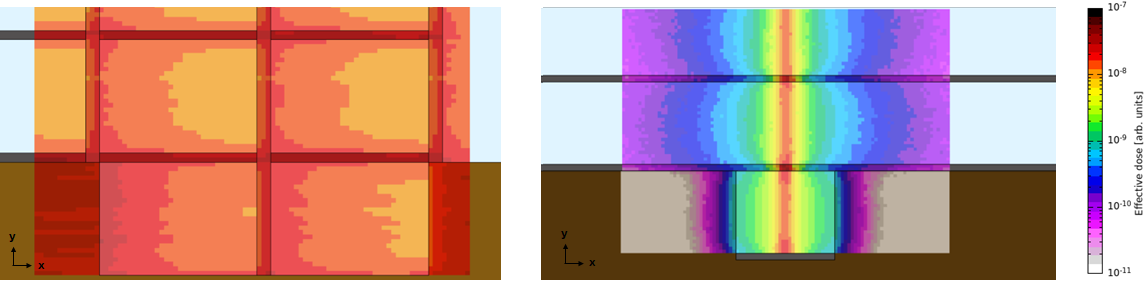}
    \caption{Side (left) and cross-sectional view (right) of the effective dose (in arb. units) for an underground building structure exposed to the neutrino flux from the decay of negative muons after the vertical deformation by the movers.}
    \label{fig:cellar_geo}
\end{figure}

Instead of considering the saturated effective dose within soil, which is unrealistic for an annual exposure, various more realistic, yet conservative scenarios are under investigation, including exposure in building structures below and above the ground. The most conservative of these is illustrated in Figure~\ref{fig:cellar_geo}, where two consecutive underground rooms are aligned along the neutrino flux path. Assuming a very conservative annual exposure scenario with a 100\% occupancy in the two underground rooms would lead to a effective dose for various relevant distances as given in Table~\ref{tab:cellar_wobbling_dose_5TeV}.   

\begin{table}[htbp]
    \centering
        \caption[Effective dose of neutrino-induced radiation for an underground building structure]{Effective dose of neutrino-induced radiation for an underground building structure at different baseline distances from the muon decay after the vertical deformation by the movers is applied. The muon beam energy is $5$~TeV.        }
    \label{tab:cellar_wobbling_dose_5TeV}
    \begin{tabular}{rcc}
         \hline\hline
         \multicolumn{1}{c}{}& \textbf{$\mu^-$} & \textbf{$\mu^+$} \\ 
         \textbf{Distance} & \textbf{Mitigated dose [pSv/decay]} &  \textbf{Mitigated dose[pSv/decay]} \\ \hline 
        $15$~km & $6.21 \cdot 10^{-9}$ & $6.49 \cdot 10^{-9}$ \\
        $20$~km & $4.57 \cdot 10^{-9}$ & $4.77 \cdot 10^{-9}$ \\
        $30$~km & $2.92 \cdot 10^{-9}$ & $3.06 \cdot 10^{-9}$ \\
        $60$~km & $1.21 \cdot 10^{-9}$ & $1.25 \cdot 10^{-9}$ \\
       \hline\hline
    \end{tabular}
\end{table} \clearpage
\section{Demonstrators}
\label{demo:sec}
Demonstrator parameters are to be reviewed pending the October 2024 demonstrator workshop.
\subsection{Cooling Cell Demonstrator}
\label{demo:sec:coolcell}

A cooling cell is composed of two or more solenoids (coils), an RF Structure made of one or multiple RF cells, and one or two absorbers made of low-Z materials. 

\begin{figure} [h]
    \centering
    \includegraphics[width=1\textwidth]{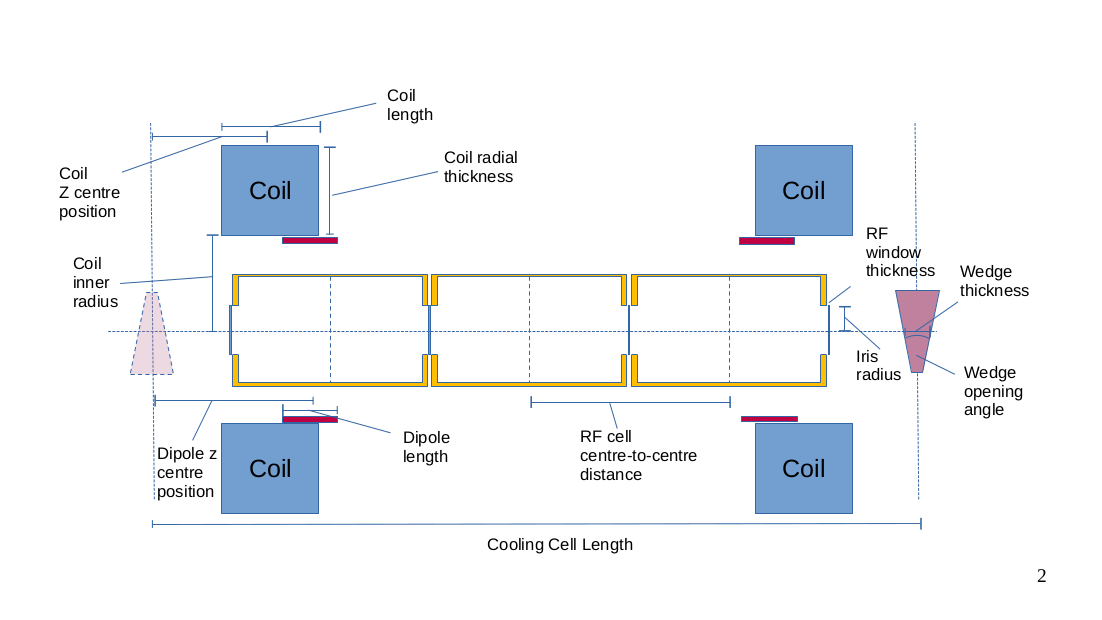}
    \caption{Cooling Cell Schematic}
    \label{coolcell:fig:schematic}
\end{figure}

The Collaboration adopts the terminology in Fig. \ref{coolcell:fig:schematic} to designate the elements of a cooling cell.

The cooling cells efficiency impacts the performances of the entire complex, since it determines the intensity of the two beams and their emittance at the entrance of the acceleration sections. This means that an efficient design relaxes the requirements of all the downstream chain, and therefore both the overall cost and the feasibility. For this reason designing and testing a real scale cooling cell has the highest priority for the IMCC collaboration. The cooling section comprises of several different types of cooling cells therefore the choice of which one to build first to feel confident about the feasibility of the entire chain has been discussed at several occasions and is the object of a workpackage of the MuCol EU funded project. In this framework a workshop was organised in January 2024, leading to choices that have been reconfirmed during the annual meeting of the IMCC and MuCol, and that are listed in Tab. \ref{demo:tab:coolingcell}. The rationale behind those choices is to have a cell with challenging perfomances, but still within reach with a reasonable investment and in reasonable (~5 to 7 years) time. Such a long delay is justified by the fact that we will use a new technology for the coils of the solenoid, based on REBCO HTS tapes, for which a few solenoids have been built but none is operated in a regular basis. We decided therefore not to target the maximum field of \SI{20}{\tesla}, but an intermediate \SI{7}{\tesla}, which is still very challenging for this technology but considered within reach. Studies for more challenging solenoids will progress in parallel independently from the cooling cell. For RF, there are two main difficulties, reaching high average fields (40 MV/m) in a high magnetic field, that enhances breakdown phenomena. Mitigation measures are therefore studied and will be applied for a long term study in the cooling cell demonstrator, after dedicated studies in an RF test stand with magnetic field. Finally, the LiH (Lithium Hydride) technology has been chosen to start with because of the absence of major safety risks, but we plan to make sure that LH2 (Liquid Hydrogen) and GH2 (Gaseous Hydrogen) absorbers may be tested once all the technical and safety issues will be understood in a separate R\&D programme. 
MuCol will fund the design of the cell while its construction has still to be funded. The programme proposes to test the cell in a dedicated test stand at full power, and then test its features with a beam in the Ionisation Cooling demonstrator facility described in the next paragraph. 

\begin{table}[!h]
    \centering
    \begin{tabular}{lccc}
          Parameter & Unit & Value \\ \hline
         Cooling Cell Length&  mm& 800\\
         \multicolumn{3}{c}{\textbf{Beam Physics}}\\ \hline
         Momentum&  MeV/c& 200\\
         Twiss beta function&  mm& 107\\
         Dispersion in X&  mm& 38.5\\
         Dispersion in Y&  mm& 20.3\\
         Beam Pipe Radius&  mm& 81.6\\
         \multicolumn{3}{c}{\textbf{Solenoid Parameters}}\\ 
         &  Unit & Value & Tol\\\hline
         B0&  T& 8.75 & 0.25\\
         B0.5&  T& 0 & 0.02\\
         B1& T&1.25 & 0.025\\
         B2& T&0 & 0.5 \\
         \multicolumn{3}{c}{\textbf{Coil Geometry}}\\ \hline
         Inner Radius& mm&250\\
         Length& mm&140\\
         Radial Thickness& mm&169.3\\
         Z Centre Position& mm&100.7\\
         Current Density& A/mm$^2$&500\\
         \multicolumn{3}{c}{\textbf{RF Cavity}}\\ \hline
         Center-to-centre distance& mm&188.6\\
         Gradient E0& MV/m&30\\
         Iris Radius& mm&81.6\\
         Number of RF Cells& &3\\
         Frequency& GHz&0.704\\
         Synchronous Phase& degree&20\\
         Window Thickness& mm&0.1\\
         \multicolumn{3}{c}{\textbf{Wedge}}\\ \hline
         Material& &LiH\\
         Opening Angle& degree&10\\
         Thickness& mm&20\\
         Alignment& &Horizontal\\
         \multicolumn{3}{c}{\textbf{Dipole}}\\ \hline
         Length& mm&100\\
         Polarity& &+ - - +\\
         Field& T&0.2\\
         Z Centre Position& mm&160\\
         Field Direction & &Vertical\\
    \end{tabular}
    \caption{Cooling Cell Table}
    \label{demo:tab:coolingcell}
\end{table}

\subsection{TT7 Demonstrator}
\label{demo:sec:TT7}

\begin{figure} [h]
    \centering
    \includegraphics[width=1\textwidth]{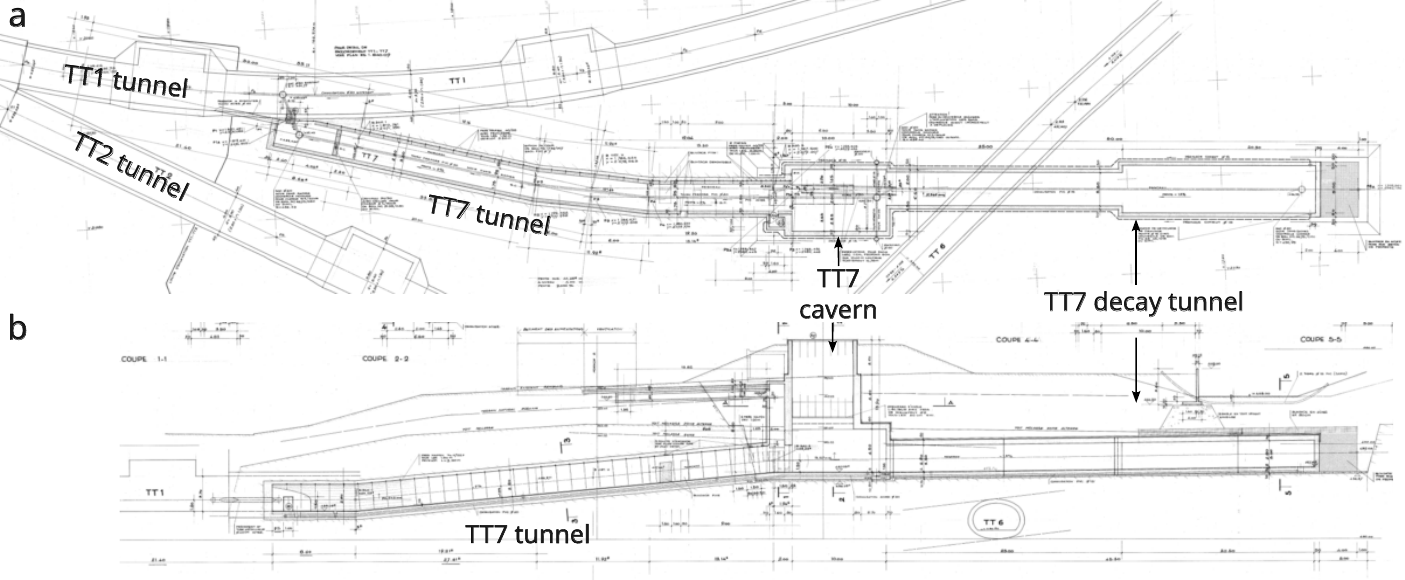}
    \caption{a) Top and b) side view of the TT7 tunnel.}
    \label{demo:fig:tt7schematic}
\end{figure}

\begin{figure} [h]
    \centering
    \includegraphics[width=1\textwidth]{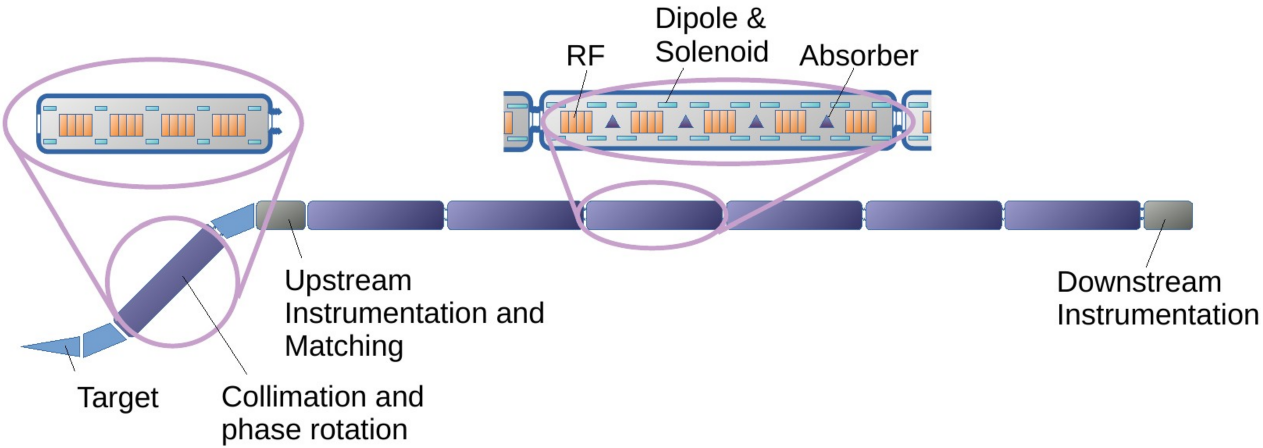}
    \caption{Cooling demonstrator conceptual layout.}
    \label{demo:fig:layoutschematic}
\end{figure}

Once the cooling cell integration and dedicated tests are concluded, it is proposed that the performance and efficiency of a cooling channel composed of series of cooling cells are characterised in a beam test. Two potential sites have been identified on the CERN premises that can host the demonstrator facility. Here we focus on the TT7 tunnel option, which would be suitable only for a low-power target (\SI{10}{\kilo\watt}) due to radiation protection restrictions but would reuse existing infrastructure. Fig. \ref{demo:fig:tt7schematic} shows engineering drawings of the TT7 tunnel, while a non-site specific conceptual layout of the demonstrator facility is shown in Fig. \ref{demo:fig:layoutschematic}. Parameters pertaining to various facility subsystems that have been produced in preliminary design studies or have been assumed (e.g. proton beam) are listed in Tab. \ref{demo:tab:demonstrator}. The ongoing beam physics design of the target and pion/muon transport line sections rests on the assumption that the target would be housed in the cavern in the middle region of the TT7 tunnel.

\begin{table}
    \centering
    \begin{tabular}{cccc}
         \textbf{Parameter}&  \textbf{Unit}&  \textbf{Baseline value}&  \textbf{Aspirational value}\\ \hline
         Transfer tunnel&  &  TT7&  --\\
         Tunnel ramp&  degrees&  5.18&  --\\
         Pion decay channel length&  m&  8-10&  20\\
         Chicane length&  m&  12.8&  --\\
         Muon bunch intensity&  \num{E6}&  1-10&  100\\
         \multicolumn{4}{c}{\textbf{Proton Beam}}\\ \hline
         Accelerator&  &  Proton Synchrotron&  SPS\\
         Cycle&  &  nTOF-like&  --\\
         Energy&  GeV&  14&  26, 100\\
         1-sigma beam size& mm& 2& <2\\
         1-sigma bunch length& ns& 10& <10\\
         Bunch intensity& \num{E13}& 1& >1\\
         \multicolumn{4}{c}{\textbf{Target System}}\\ \hline
         Target material& & Graphite& --\\
         Target length& cm& 80-100& --\\
         Target radius& mm& 6& --\\
         Pion capture& & Magnetic horn& Solenoid\\
         Horn length& m& 2& --\\
         Horn current& kA& 220& --\\
         \multicolumn{4}{c}{\textbf{Beam Preparation System}}\\ \hline
         Cell length& m& 1& --\\
         Peak solenoid field on-axis& T& 0.5& --\\
         Collimator radius& m& 0.05& --\\
         Dipole field& T& 0.67& --\\
         Dipole length& m& 1.04& --\\
         RF real estate gradient& MV/m& 7.5& --\\
         RF nominal phase& degrees& 0& --\\
         RF frequency& MHz& 704& --\\
         \multicolumn{4}{c}{\textbf{Cooling channel}}\\ \hline
         Length& m& 32& 48\\
         Number of cooling cells& & 40&60\\
         Number of vacuum vessel modules& & 8&12\\
         Number of cooling cells per vacuum vessel& & 5&5\\
    \end{tabular}
    \caption{Demonstrator design based off CERN PS TT7 geometry}
    \label{demo:tab:demonstrator}
\end{table}

Protons would be provided by the CERN Proton Synchrotron (PS) and extracted via the TT2 tunnel through part of the TT1 tunnel into the TT7 tunnel. The proton beam would then be directed towards a 80--100 cm-long graphite target situated inside a magnetic horn designed to focus the secondary pions with a central momentum of \SI{300}{MeV/c}. A nTOF-like proton beam is currently assumed, albeit with the beam energy reduced to \SI{14}{GeV} due to the beam power restrictions. Dedicated PS machine development studies are required to optimize the beam characteristics at this energy, in particular the bunch intensity and length. A magnetic horn was chosen over a solenoid-based pion capture solution due to its relatively compact build given the available space within the existing cavern, and due to cost-effectiveness. The target and horn parameters shown here are the outcome of a first-pass optimisation study, with further design work required \cite{demo:bib:jurj_targethorn_imcc_2024}. 

The design of the transport line downstream of the target is ongoing. It is forseen that a 8--10 m quadrupole-based lattice will be used as a pion decay channel, followed by a chicane used for muon momentum selection and for allowing the unspent protons and secondary particles to be dumped. A preliminary beam physics design for a beam preparation system, which would be used to tune the transverse emittance and length of the muon bunches prior to their delivery to the cooling section, has been established \cite{psf2023008037}. Due to the limited available space in the TT7, it is expected that this system will be integrated within the chicane and may undergo further alterations.

The TT7 cooling channel would be composed of a series of 40 cooling cells grouped into vacuum vessels, as shown in Fig \ref{demo:fig:layoutschematic}. The cooling performance for two different cooling channel lengths is listed in Tab. \ref{demo:tab:coolingperformance}.

\begin{table}
    \centering
    \begin{tabular}{ll|ccc} 
         \textbf{Simulated cooling performance}&  \textbf{Unit}&  \textbf{Start value}&  \textbf{End value (32 m)}& \textbf{End value (48 m)} 
\\ \hline
         Transverse emittance&  \si{\milli\meter} &  2.37&  1.61& 1.44
\\ 
         Longitudinal emittance&  \si{\milli\meter} &  4.99&  3.89& 3.58
\\ 
         6D emittance&  \si{\milli\meter\cubed} &  26.21&  9.77& 7.12\\ 
    \end{tabular}
    \caption{Simulated cooling performance.}
    \label{demo:tab:coolingperformance}
\end{table}
 \clearpage
\section{Site-Based Designs Considerations}
\label{site:sec}
Tenative parameter tables to guide future design efforts for the existing site options.
Different assumptions were made for the initial beam parameters and magnet technologies.
\subsection{RCS Layout at CERN}
\label{site:sec:RCS_cern}
The RCS layout on the CERN site is based on the usage of the existing Super Proton Synchrotron (SPS) and Large Hadron Collider (LHC) tunnels to host the high-energy acceleration chain. 
To make a comparison with the greenfield study possible, the same assumptions for the injection energy and the injection bunch population were chosen. 
The survival rate over the whole RCS chain is assumed as $70 \%$, only considering losses due to muon decay, while the individual survival rates of the rings were adjusted to achieve a high extraction energy from the last RCS. 
Table \ref{site:tab:RCS_key} and \ref{site:tab:RCS_add} for the site-based design correspond to Table \ref{high:tab:RCS_key} and \ref{high:tab:RCS_add} for the greenfield design respectively.

\begin{table}[!h]
    \centering
    \begin{tabular}{c|c|ccc} 
         Parameter&  Unit&  RCS SPS&  RCS LHC1&  RCS LHC2\\\hline
         Hybrid RCS&  -&  No&  No&  Yes\\
         Repetition rate&  Hz&  5&  5&  5\\
         Circumference& m& 6912& 26659& 26659\\
         Injection energy& GeV& 63& 350& 1600\\
         Extraction energy& GeV& 350& 1600& 3800\\
         Energy ratio&  - &  5.6&  4.6&  2.4\\
         Assumed survival rate& - &  0.88&  0.86&  0.92\\
         Total survival rate& - & 0.88& 0.76&0.70\\
         Acceleration time&  ms&  0.45&  2.60&  4.42\\
         Revolution period&  \textmu s&  23.0&  88.9&  88.9\\
         Number of turns& - & 19& 29& 50\\
         Required energy gain per turn& GeV& 15.1& 43.1& 44.4\\
         Average acel. gradient& MV/m& 2.15& 1.62& 1.68\\
         Number of bunches& - & 1& 1& 1\\
         Inj bunch population& $10^{12}$& 2.70& 2.38& 2.04\\
         Ext bunch population& $10^{12}$& 2.38& 2.04& 1.88\\
         Beam current per bunch& mA& 18.75& 4.29&3.68\\
         Beam power& MW& 803& 523&462\\
         Vert. norm. emittance& \si{\micro\meter} & 25& 25& 25\\
         Hor. norm. emittance& \si{\micro\meter} &  25&  25&  25\\
         Long. norm emittance&  eVs&  0.025&  0.025&  0.025\\
         Bunch length& ps& 33.2& 19.3&11.1\\ \hline
         Straight section length&  m&  2809&  8000&  8000\\
         Length with NC magnets& m& 4103& 18650& 12940\\
         Length with SC magnets& m& -& -& 5680\\
         Max NC dipole field& T& 1.8& 1.8& 1.8\\
         Max SC dipole field& T& -& -& 10\\
         Ramp rate& T/s& 3320& 1400& 810\\
         Main RF frequency& GHz& 1.3& 1.3& 1.3\\
    \end{tabular}
    \caption{Key acceleration Parameters for the CERN-site based RCS Acceleration Chain }
    \label{site:tab:RCS_key}
\end{table}

\begin{table}[!h]
    \centering
    \begin{tabular}{c|c|ccc} 
         Parameter&  Unit&  RCS SPS&  RCS LHC1&  RCS LHC2\\ \hline
         Harmonic number&  &  29900&  115345&  115345\\
         Packing Fraction& \%& 0.59& 0.70& 0.70\\
         Transition Gamma& - & 33& 45& 58\\
         Momentum compaction factor& $10^{-4}$& 9& 5& 3\\
    \end{tabular}
    \caption{Additional Parameters for the CERN-based RCS Acceleration Chain }
    \label{site:tab:RCS_add}
\end{table}

\pagebreak

\subsubsection{RF system for the RCS layout at CERN}
\label{site:sec:RCS_RF_CERN}
The design of the RF system for the RCS is based on the same assumptions as the RF system for the greenfield study. The assumptions are presented in \ref{rf:sec:high}.

\begin{table}[!h]
    \centering
    \begin{tabular}{lc|cccc}
         &  &  RCS SPS&  RCS LHC1&  RCS LHC2&  All\\ \hline
         Synchronous phase &  \textdegree&  135&  135&  135&  -\\
         Number of bunches/species &  - &  1&  1&  1&  -\\
         Combined beam current ($\mu^+$, $\mu^-$) &  mA&  37.5&  8.58&  7.35&  -\\
         Total RF voltage &  GV&  21.4&  61&  62.8&  145.2\\
         Total number of cavities &  - &  686&  1958&  2017&  4661\\
         Total number of cryomodules &  -  &  77&  218&  225&  520\\
         Total RF section length &  m&  974&  2760&  2850&  6584\\ \hline
         Combined peak beam power ($\mu^+$, $\mu^-$) &  MW&  803&  523&  462&  -\\
         External Q-factor &  $10^{6}$&  0.79&  3.49&  4.07&  -\\
         Cavity detuning for beam loading comp. & kHz& -1.16& -0.26& -0.23& -\\
         Beam acceleration time & ms& 0.45& 2.6& 4.42&-\\
         Cavity filling time & ms& 0.194& 0.854& 0.993&-\\
         RF pulse length & ms& 0.644& 3.454& 5.413&-\\
         RF duty factor & \%& 0.32& 1.73& 2.71& - \\
         Peak cavity power & kW& 987& 228& 195& - \\ \hline 
         Total peak RF power & MW & 905& 569& 529& -\\
         Total number of klystrons & - & 99& 60& 54& 213\\
         Cavities per klystron & - & 7& 33& 38& -\\
         Average RF power & MW& 2.91& 10.3& 14.3&27.51\\
         Average wall plug power for RF System & MW& 4.48& 15.8& 22&42.28\\
         HOM power losses per cavity per bunch& kW& 13.08& 4.54& 5.15& -\\
        Average HOM power per cavity& W& 58& 118& 227& -\\  
    \end{tabular}
    \caption[RF Parameters for the CERN-based RCS Acceleration Chain.]{RF Parameters for the CERN-based RCS Acceleration Chain. For the synchronous phase, \SI{90}{\degree} is defined as being on-crest}
    \label{site:tab:rf_rcs_cern}
\end{table}

\pagebreak

\subsection{RCS Layout at FNAL}
\label{site:sec:RCS_FNAL}
\begin{table}[!h]
    \centering
    \begin{tabular}{c|c|cccc} 
         Parameter&  Unit&  RCS1&  RCS2&  RCS3 &RCS4\\\hline
         Hybrid RCS&  -&  No&  Yes&   Yes&Yes\\
         Repetition rate&  Hz&  5&  5&   5&5\\
         Circumference& m& 6280& 10500&  15500&15500\\
         Injection energy& GeV& 173& 450&  1725&3035\\
         Extraction energy& GeV& 450& 1725&  3035&4059\\
         Energy ratio& - &  2.6&  3.83&   1.759&1.338\\
         Assumed survival rate& - &  0.85&  0.83&   0.946&0.972\\
 Total survival rate& -& 0.85& 0.71& 0.67&0.65\\
         Acceleration time&  ms&  0.97&  3.71&   3.22&2.44\\
         Revolution period&  \textmu s&  21&  35&   51.7&51.7\\
         Number of turns& -& 46& 106&  62&47\\
         Required energy gain per turn& GeV& 6& 12&  21&21.7\\
         Average acel. gradient& MV/m& 0.96& 1.15&  1.64&1.62\\
         Number of bunches& - & 1& 1&  1&1\\
         Inj bunch population& $10^{12}$& 3.3& 2.83&  2.35&2.22\\
         Ext bunch population& $10^{12}$& 2.83& 2.35&  2.22&2.16\\
         Vert. norm. emittance& mm& 0.025& 0.025&  0.025&0.025\\
         Hor. norm. emittance&  mm&  0.025&  0.025&   0.025&0.025\\
         Long. norm emittance&  eVs&  0.025&  0.025&   0.025&0.025\\
         Straight section length&  m&  1068&  1155&   1107&1107\\
         Length with NC magnets& m& 5233& 7448&  8495&6644\\
         Length with SC magnets& m& - & 1897&  1930&2876\\
         Max NC dipole field& T& 1.8& 1.8&  1.75&1.75\\
         Max SC dipole field& T& - & 12&  14&14\\
         Ramp rate& T/s& 1134& 970&  1087&1434\\
         Main RF frequency& GHz& 1.3& 1.3&  1.3&1.3\\
    \end{tabular}
    \caption{Key Parameters for the Fermilab-based RCS Acceleration Chain }
    \label{site:tab:FNAL_RCS_key}
\end{table}

\printbibliography

\end{document}